\magnification=1200
\baselineskip=6mm {\nopagenumbers
\noindent
\null\vskip .5truecm
\centerline {\bf OPEN QUANTUM SYSTEMS}
\centerline{A. Isar*, ~A. Sandulescu*, H. Scutaru*, E. Stefanescu*
and W. Scheid\#}
\centerline{*\it Department of Theoretical Physics,
Institute of Atomic Physics,}
\centerline{\it Bucharest-Magurele,
Romania} \centerline{\#\it Institut f\"ur Theoretische Physik der
Justus-Liebig- Universit\"at,}
\centerline{\it Giessen, Germany}
\centerline{ABSTRACT}
\baselineskip=4mm
The damping of the harmonic
oscillator is studied in the framework of the Lindblad theory for
open quantum systems. A generalization of the fundamental
constraints on quantum mechanical diffusion coefficients which
appear in the master equation for the damped quantum oscillator is
presented; the Schr\"odinger, Heisenberg and Weyl-Wigner-Moyal
representations of the Lindblad equation are given explicitly. On
the basis of these representations it is shown that various master
equations for the damped quantum oscillator used in the literature
are particular cases of the Lindblad equation and that not all of
these equations are satisfying the constraints on quantum mechanical
diffusion coefficients. Analytical expressions for the first two
moments of coordinate and momentum are obtained by using the
characteristic function of the Lindblad master equation. The master
equation is transformed into Fokker-Planck equations for
quasiprobability distributions and a comparative study is made for
the Glauber $P$ representation, the antinormal ordering $Q$
representation and the Wigner $W$ representation. The density matrix
is represented via a generating function, which is obtained by
solving a time-dependent linear partial differential equation
derived from the master equation. Illustrative examples for specific
initial conditions of the density matrix are provided. The solution
of the master equation in the Weyl-Wigner-Moyal representation is of
Gaussian type if the initial form of the Wigner function is taken to
be a Gaussian corresponding (for example) to a coherent
wavefunction. The damped harmonic oscillator is applied for the
description of the charge equilibration mode observed in deep
inelastic reactions. For a system consisting of two harmonic
oscillators the time dependence of expectation values, Wigner
function and Weyl operator are obtained and discussed. In addition
models for the damping of the angular momentum are studied. Using
this theory to the quantum tunneling through the nuclear barrier,
besides Gamow's transitions with energy conservation, additional
transitions with energy loss, are found. The tunneling spectrum is
obtained as a function of the barrier characteristics. When this
theory is used to the resonant atom-field interaction, new optical
equations describing the coupling through the environment of the
atomic observables are obtained. With these equations, some
characteristics of the laser radiation absorption spectrum and
optical bistability are described.
\vfil
\eject
}
\pageno=1
\baselineskip=6mm
{\bf 1. Introduction}

In the last two decades, more and more interest arose about the problem
of dissipation in quantum mechanics, i.e. the consistent description
of open quantum systems [1-7].The quantum description of dissipation is
important in many different areas of physics. In quantum optics we mention
the quantum theory of lasers and photon detection. There are some
directions in the theory of atomic nucleus in which dissipative processes
play a basic role. In this sense we mention the nuclear fission, giant
resonances and deep inelastic collisions of heavy ions. Dissipative
processes often occur also in many body or field-theoretical systems.

The irreversible, dissipative behaviour of the vast majority of physical
phenomena comes into an evident contradiction with the reversible nature
of our basic models. The very restrictive principles of conservative and
isolated systems are unable to deal with more complicated situations
which are determined by the features of open systems.

The fundamental quantum dynamical laws are of the reversible type. The
dynamics of a closed system is governed by the Hamiltonian that represents
 its total energy and which is a constant of motion. In this way
the paradox of irreversibility arises: the reversibility of microscopic
dynamics contrasting with the irreversibility of the macroscopic behaviour we
are trying to deduce from it.

One way to solve this paradox of irreversibility is to use models
to which Hamiltonian dynamics and Liouville's  theorem do not apply, but
the irreversible behaviour is clearly present even in the microscopic
dynamical description. The reason for replacing Hamiltonian dynamics
and Liouville's theorem is that no system is truly isolated, being subject
to uncontrollable random influences from outside. For this reason these
models are called open systems. There are two ways of treating
quantitatively their interaction with the outside. One way is to introduce
specific stochastic assumptions to simulate this interaction, the other one is
to treat the open systems with the usual laws of dynamics, by regarding them
as subsystems of larger systems which are closed (i.e. which
obey the usual laws of dynamics with a well-defined Hamiltonian).
The dissipation arises in general from the subsystem interactions
with a larger system, often referred to as the reservoir or bath.
The first of these two approaches has been used for the study of steady state
transport processes in systems obeying classical mechanics. The second of
the two approaches has been mainly used in quantum mechanics.
The main general result [2,8-10] is that under certain conditions the time
evolution of an open system can be described by a dynamical semigroup
$\Phi_t$($t\ge0$).
For a closed finite system with Liouville operator the evolution operator
is not restricted to nonnegative $t$. The importance of the dynamical
semigroup concept is that it generalizes the evolution operator to open
systems, for which there is no proper Liouville operator and no $\Phi_t$
for negative $t$. The mathematical theory of dynamical semigroups has
been developed in [2,11-15].

In Sect.2 the notion of  the quantum dynamical semigroup is
defined using the concept of a completely positive map [14]. The Lindblad
formalism replaces the dynamical group uniquely determined by its generator,
which is the Hamiltonian operator of the system, by the completely positive
dynamical semigroup with bounded generators. Then the general form of
Markovian quantum mechanical master equation is obtained.

The quantum mechanics of the one-dimensional damped harmonic oscillator
represents a fundamental theoretical problem with applications in
different domains of quantum optics, solid state physics, molecular and
nuclear physics. In the present review paper the quantum harmonic oscillator
is treated in the Lindblad axiomatic formalism of quantum dynamical
semigroups. In Sect.3 we give  the fundamental constraints on
quantum mechanical diffusion coefficients which appear in the corresponding
master equations [16,17]. On the basis of different representations
it is shown that various master equations for the damped quantum oscillator
used in the literature for the description of the damped collective modes in
deep inelastic collisions or in quantum optics are particular cases of the
Lindblad equation and that not all of these equations in literature are
satisfying the constraints on quantum mechanical diffusion coefficients. In
3.1, by using the Heisenberg representation, explicit expressions of the mean
values and variances are given [17].
In 3.2 we solve the master equation with the characteristic function
[18]. This function is found as a solution of a corresponding partial
differential equation. By this method one can derive explicit formulae for the
centroids and variances and, in general, for moments of any order.
In 3.3 we explore the applicability of quasiprobability distributions
to the Lindblad theory [19]. The methods of quasiprobabilities have provided
technical tools of great power for the statistical description of microscopic
systems formulated in terms of the density operator.
In the present review
the master equation of the one-dimensional damped harmonic oscillator
is transformed into Fokker-Planck equations for the Glauber $P$, antinormal
$Q$ and Wigner $W$ quasiprobability distributions associated with the density
operator. A comparative study of these representations is made.
The resulting equations are solved by standard methods and observables
directly calculated as correlations of these distribution functions. We solve
also the Fokker-Planck equations for the steady state and show that variances
found from the $P,Q$ and $W$ distributions are the same [19].
In 3.4 we study the time evolution of the density matrix that follows
from the master equation of the damped harmonic oscillator [20]. We
calculate the physically relevant solutions (including both
diagonal and off-diagonal matrix elements) of the master equation by applying
the method of generating function. This means that we represent the density
matrix with a generating function which is the solution of a time-dependent
partial differential equation of second order, derived from the master equation
of the damped harmonic oscillator. We discuss stationary solutions of the
generating function and derive the Bose-Einstein density matrix as example.
Then, formulas for the time evolution of the density matrix are presented
and illustrative examples for specific initial conditions provided. The same
method of generating function was already used by Jang [21] who studied the
damping of a collective degree of freedom coupled to a bosonic reservoir at
finite temperature with a second order RPA master equation in the collective
subspace. In 3.5 the master equation is given in the
Weyl-Wigner-Moyal representation [17]. We show that the solution of the
Lindblad
equation in this representation is of Gaussian type if the initial form of the
Wigner function is taken to be a Gaussian corresponding to a coherent
wavefunction.

In Sect.4 we give some applications to nuclear equilibration processes.
In 4.1 the charge
equilibration mode is treated by a damped harmonic oscillator
in the framework of the Lindblad theory [17]. It is shown that the
centroids and variances of the charge equilibration mode
observed in deep inelastic reactions can be well described by the
corresponding overdamped solutions [18,22].
In 4.2 we treat the damping of the proton and neutron
asymmetry degrees of freedom with the method of Lindblad
by studying the damping of two coupled oscillators [23]. We present
the equation of motion of the open quantum system of two
oscillators in the Heisenberg picture and we derive the time
dependence of the expectation values of the coordinates and
momenta and their variances. We discuss also the connection with
the Wigner function and Weyl operator. We demonstrate the time
dependence of the various quantities for a simplified version of
the model, where the decay constants can be calculated analytically.

In Sect.5 simple models for the damping of the angular momentum
are studied in the framework of the Lindblad theory of open
systems [24,25]. We assume that the system is opened by the
generators of the proper  Lorentz group and the Heisenberg group.

In Sect.6 we study the quantum tunneling [26]. Considering the tunneling
operator between the localized states, we solve Lindblad master equation in
the second order approximation of the theory of perturbations. We find that
diffusion leads to additional tunneling processes around the $Q$-value, while
friction determines only transitions at lower energies. Generally, this result
is contrary to the majority of the previous studies [27,28] where, usually,
it was concluded that the dissipation should decrease the tunneling rate.

In Sect.7 we study the resonant atom-field interaction [29]. From Lindblad's
theorem, instead of the conventional optical Bloch equations, where only decay
processes appear, we obtaine a more general system of equations where the
atomic observables are coupled through the environment. Based on the new
equations, experimental characteristics of the electromagnetic field absorbtion
spectrum and of the optical bistability are described. At the same time, we
predict that in some conditions, an energy transfer from the dissipative
environment to the coherent electromagnetic field appears.

The concluding summary is given in Sect.8.

{\bf 2. Lindblad theory for open quantum systems}

The standard quantum mechanics is Hamiltonian. The time evolution of a closed
physical system is given by a dynamical group $U_t$ which is uniquely
determined by its generator $H$, which is the Hamiltonian operator of the
system. The action of the dynamical group $U_t$ on any density matrix $\rho$
from the set $\cal D(H)$ of all density matrices of the quantum system, whose
corresponding Hilbert space is $\cal H$, is defined by
$$\rho(t)=U_t(\rho)=e^{-{i\over \hbar}Ht}\rho e^{{i\over \hbar}Ht}$$
for all $t\in(-\infty,\infty)$. We remind that, according to von Neumann,
density operators $\rho\in\cal D(H)$ are trace class (${\rm Tr}\rho<\infty$),
self-adjoint ($\rho^+=\rho$), positive ($\rho>0$) operators with
${\rm Tr}\rho=1$. All these properties are conserved by the time evolution
defined by $U_t$.

In the case of open quantum systems the main difficulty consists in finding
such time evolutions $\Phi_t$ for density operators $\rho(t)=\Phi_t(\rho)$
which preserve the von Neumann conditions for all times. From the last
requirement it follows that $\Phi_t$ must have the following properties:

$(i)~  \Phi_{t}(\lambda_{1}\rho_{1}+\lambda_{2}\rho_{2})=\lambda_{1}
\Phi_{t}(\rho_{1})+\lambda_{2}\Phi_{t}(\rho_{2}); \lambda_{1},\lambda_{2}
\ge 0$ with $\lambda_{1}+\lambda_2=1,$

$(ii)~  \Phi_{t}(\rho^+)=\Phi_{t}(\rho)^+,$

$(iii)~  \Phi_{t}(\rho)>0,$

$(iv)~  {\rm Tr}\Phi_{t}(\rho)=1.$

But these conditions are not restrictive enough in order to give a complete
description of the mappings $\Phi_{t}$ as in the case of the time evolutions
$U_{t}$ for closed systems. Even in this last case one has to impose other
restrictions to $U_t$, namely, it must be a group $U_{t+s}=U_tU_s.$ Also,
it is evident that in this case $U_{0}(\rho)=\rho$ and $U_{t}(\rho)\to \rho$
in the trace norm when $t\to 0$. For the dual group $\widetilde U_{t}$ acting
on the
observables $A\in \cal {B(H)},$ i.e. on the bounded operators on $\cal H$,
we have
$$\widetilde U_{t}(A)=e^{{i\over \hbar}Ht}Ae^{-{i\over \hbar}Ht}.$$
Then $\widetilde U_{t}(AB)=\widetilde U_{t}(A)\widetilde U_{t}(B)$ and
$\widetilde U_{t}(I)=I$, where $I$ denotes the identity operator on $\cal H.$
Also, $\widetilde U_{t}(A)\to A$
ultraweakly when $t\to 0$ and $\widetilde U_{t}$ is an ultraweakly continuous
mapping [2,9,13,14,30]. These mappings have a strong positivity property
called complete positivity:
$$\sum_{i,j} B_{i}^+\widetilde U_{t}(A_{i}^+ A_{j})B_{j} \ge 0,~~A_{i},
B_{i}\in \cal {B(H)}.$$
Because the detailed physically plausible conditions on the systems, which
correspond to these properties are not known, it is much more convenient to
adopt an axiomatic point of view which is based mainly on the simplicity and
the succes of physical applications. Accordingly [2,9,13,14,30] it is
convenient
to suppose that the time evolutions $\Phi _{t}$ for open systems are not very
different from the time evolutions for closed systems. The simplest dynamics
$\Phi _{t}$ which introduces a preferred direction in time,
characteristic for dissipative processes, is that in which the group condition
is replaced by the semigroup condition [9-12,31]
$$\Phi_{t+s}=\Phi_{t}\Phi_{s},~ t,s\ge 0.$$
The duality condition
$${\rm Tr}(\Phi_{t}(\rho)A)={\rm Tr}(\rho \widetilde \Phi_{t}(A)) \eqno(2.1)$$
defines $\widetilde \Phi_{t}$, the dual of $\Phi_{t}$ acting on $\cal B(H).$
Then the conditions $${\rm Tr}\Phi_{t}(\rho)=1$$and
$$\widetilde \Phi_{t}(I)=I \eqno (2.2)$$
are equivalent. Also the conditions
$$\widetilde \Phi_{t}(A)\to A \eqno (2.3)$$
ultraweakly when $t \to0$ and $$\Phi_{t}(\rho)\to \rho$$
in the trace norm when $t \to 0,$ are equivalent.
For the semigroups with the properties (2.2), (2.3) and
$$A\ge 0 \to \widetilde \Phi_{t}(A) \ge 0,$$
it is well known that there exists a (generally bounded) mapping
$\widetilde L$, the generator of $\widetilde \Phi_{t}.$
$\widetilde \Phi_{t}$ is uniquely determined by $\widetilde L.$
The dual generator of the dual semigroup $\Phi_{t}$ is denoted by $L$:
$${\rm Tr}(L(\rho)A)={\rm Tr}(\rho \widetilde L(A)).$$
The evolution equations by which $L$ and $\widetilde L$ determine uniquely
$\Phi_{t}$ and $\widetilde \Phi_{t}$, respectively, are
given in the Schr\"odinger and Heisenberg picture as
$${d\Phi_{t}(\rho)\over dt}=L(\Phi_{t}(\rho)) \eqno (2.4)$$
and
$${d\widetilde \Phi_{t}(A) \over dt}=\widetilde L(\widetilde \Phi_{t}(A)).
\eqno (2.5)$$
These equations replace in the case of open systems the von Neumann-Liouville
equations
$${dU_{t}(\rho)\over dt}=-{i\over \hbar}[H,U_{t}(\rho)]$$
and
$${d\widetilde U_{t}(A)\over dt}={i\over \hbar}[H,\widetilde U_{t}(A)].$$
For any applications Eqs. (2.4) and (2.5) are only useful if the detailed
structure of the generator $L(\widetilde L)$ is known and can be related to
the concrete properties of the open systems described by such equations.

Such a structural theorem was obtained by Lindblad [14] for the class of
dynamical semigroups $\widetilde \Phi_{t}$ which are completely positive and
norm continuous. For such semigroups the generator $\widetilde L$ is bounded.
In many applications the generator is unbounded.

A bounded mapping $\widetilde L:\cal {B(H)}\to \cal {B(H)}$ which satisfies
$\widetilde L(I)=0, ~\widetilde L(A^+)=\widetilde L(A)^+$ and
$$\widetilde L(A^+ A)-\widetilde L(A^+)A-A^+ \widetilde L(A)\ge 0$$
is called dissipative. The 2-positivity property of the completely positive
mapping $\widetilde \Phi_{t}$:
$$\widetilde \Phi_{t}(A^+ A)\ge \widetilde \Phi_{t}(A^+)\widetilde
 \Phi_{t}(A),\eqno(2.6)$$
with equality at $t=0$, implies that $\widetilde L$ is dissipative. Lindblad
[14] has shown that conversely, the dissipativity of $\widetilde L$ implies
that $\widetilde \Phi_{t}$ is 2-positive. $\widetilde L$ is called completely
dissipative if all trivial extensions of $\widetilde L$ are dissipative.
Lindblad has also shown that there exists a one-to-one correspondence between
the completely positive norm continuous semigroups $\widetilde \Phi_{t}$ and
completely dissipative generators $\widetilde L$. The structural theorem of
Lindblad gives the most general form of a completely dissipative mapping
$\widetilde L$ [14]:

{\it Theorem}: $\widetilde L$ is completely dissipative and ultraweakly
continuous if and only if it is of the form
$$\widetilde L(A)={i\over \hbar}[H,A]+{1\over 2\hbar}
\sum_{j} (V_{j}^+[A,V_{j}]+[V_{j}^+,A]V_{j}), \eqno (2.7)$$
where $V_{j}, \sum_{j} V_{j}^+ V_{j}\in{\cal B(H)}, ~H\in {\cal B(H)}_
{\rm s.a.}$.

The dual generator on the state space (Schr\"odinger picture) is of the form
$$L(\rho)=-{i\over\hbar}[H,\rho]+{1\over 2\hbar}\sum_{j}([V_{j}\rho,V_{j}^+]
+[V_{j},\rho V_{j}^+]). \eqno (2.8)$$
Eqs. (2.4) and (2.8) give an explicit form for the most general
time-homogeneous quantum mechanical Markovian master equation with a bounded
Liouville operator:
$${d\Phi_t(\rho)\over dt}=L(\Phi_t(\rho))=-{i\over \hbar}[H,\Phi_t(\rho)]+
{1\over 2\hbar}\sum_j([V_j\Phi_t(\rho),V_j^+]+[V_j,\Phi_t(\rho)V_j^+]).$$
Talkner [32] has shown that the assumption of a semigroup dynamics is only
applicable in the limit of weak coupling of the subsystem with its
environment, i.e. for long relaxation times.

We should like to mention that all Markovian master equations found in the
literature are of this form after some rearrangement of terms, even for
unbounded generators.

It is also an empirical fact for many physically interesting situations that
the time evolutions $\Phi_{t}$ drive the system towards a unique final state
$\rho (\infty)=\lim_{t\to \infty} \Phi_{t}(\rho(0))$ for all
$\rho (0)\in \cal D(H)$.

From the 2-positivity property (2.6) of $\widetilde\Phi_t$ it follows that
$$\widetilde\Phi(t)(\sum_jV_j^*V_j)\ge\sum_j\widetilde\Phi_t(V_j^*)\widetilde
\Phi_t(V_j)$$
or, by duality [17],
$${\rm Tr}(\widetilde\Phi_t(\rho)\sum_jV_j^*V_j)\ge\sum_j{\rm Tr}(\widetilde
\Phi_t(\rho)V_j^*){\rm Tr}(\widetilde\Phi_t(\rho)V_j).\eqno(2.9)$$
The evolution equations of Lindblad are operator equations. The problem of
finding their solutions is in general rather difficult. In cases when the
equations are exactly solvable, these solutions give complete information
about the studied problem and permit the calculation of expectation values
of the observables at any moment.

{\bf 3. The one-dimensional damped quantum harmonic oscillator}

In this Section the case of the damped quantum harmonic oscillator is
considered
in the spirit of the ideas presented in the previous Section. The basic
assumption is that the general form (2.8) of a bounded mapping $L$ given by
the Lindblad theorem [14] is also valid for an unbounded completely dissipative
mapping $L$:
$$L(\rho)=-{i\over \hbar}[H,\rho]+{1\over 2\hbar}\sum_{j} ([V_{j}\rho,V_{j}
^+]+[V_{j},\rho V_{j}^+]). \eqno (3.1)$$
This assumption gives one of the simplest way to construct an appropriate
model for a quantum dissipative system. Another simple condition imposed
to the operators $H,V_{j},V_{j}^+$ is that they are functions of the basic
observables $q$ and $p$ of the one-dimensional quantum mechanical system
(with $[q,p]=i\hbar I$, where $I$ is the identity operator on $\cal H$) of such
kind that the obtained model is exactly solvable. A precise version for this
last condition is that linear spaces spanned by the first degree (respectively
second degree) noncommutative polynomials in $p$ and $q$ are invariant under
the action of the completely dissipative mapping $L$. This condition implies
[16] that $V_{j}$ are at most first degree polynomials in $p$ and $q$ and
$H$ is at most a second degree polynomial in $p$ and $q$.

Because in the linear space of the first degree polynomials in $p$ and $q$
the operators $p$ and $q$ give a basis, there exist only two ${\bf C}$-linear
independent operators $V_{1},V_{2}$ which can be written in the form
$$V_{i}=a_{i}p+b_{i}q,~i=1,2,$$
with $a_{i},b_{i}=1,2$ complex numbers [16]. The constant term is omitted
because its contribution to the generator $L$ is equivalent to terms in $H$
linear in $p$ and $q$ which for simplicity are assumed to be zero. Then $H$ is
chosen of the form
$$H=H_{0}+{\mu \over 2}(pq+qp),~~~H_{0}={1\over 2m}p^2+{m\omega^2
\over 2}q^2. \eqno (3.2)$$
With these choices the Markovian master equation can be written:
$${d\rho \over dt}=-{i\over \hbar}[H_{0},\rho]-{i\over 2\hbar}(\lambda +\mu)
[q,\rho p+p\rho]+{i\over 2\hbar}(\lambda -\mu)[p,\rho q+q\rho]$$
$$-{D_{pp}\over {\hbar}^2}[q,[q,\rho]]-{D_{qq}\over {\hbar}^2}[p,[p,\rho]]+
{D_{pq}\over {\hbar}^2}([q,[p,\rho]]+[p,[q,\rho]]). \eqno (3.3)$$
Here we used the notations:
$$D_{qq}={\hbar\over 2}\sum_{j=1,2}{\vert a_{j}\vert}^2,
  D_{pp}={\hbar\over 2}\sum_{j=1,2}{\vert b_{j}\vert}^2,
D_{pq}=D_{qp}=-{\hbar\over 2}{\rm Re}\sum_{j=1,2}a_{j}^*b_{j},
\lambda=-{\rm Im}\sum_{j=1,2}a_{j}^*b_{j},$$
where $D_{pp},D_{qq}$ and $D_{pq}$ are the diffusion coefficients and
$\lambda$ the friction constant. They satisfy the following fundamental
constraints [17]:
$$i)~D_{pp}>0,$$
$$ii)~D_{qq}>0, \eqno (3.4)$$
$$iii)~D_{pp}D_{qq}-{D_{pq}}^2\ge {\lambda}^2{\hbar}^2/4.$$
Indeed, inequalities $i)$ and $ii)$ follow directly from the definitions of the
coefficients and $iii)$ from the Schwartz inequality.
The equality $\mu=\lambda$ is a necessary and sufficient condition for $L$ to
be translation invariant [16]: $[p,L(\rho)]=L([p,\rho]).$ In the
following general values for $\lambda$ and $\mu$ will be
considered as in [16].

Introducing the annihilation and creation operators
$$a={1\over \sqrt{2\hbar}}(\sqrt{m\omega}q+{i\over \sqrt{m\omega}}p),~
~a^+={1\over\sqrt{2\hbar}}(\sqrt{m\omega}q-{i\over\sqrt{m\omega}}p),\eqno
(3.5)$$
obeying the commutation relation $[a,a^+]=1$, we have
$$H_{0}=\hbar\omega(a^+a+{1\over 2}) \eqno (3.6)$$
and the master equation has the form
$${d\rho\over dt}={1\over 2}(D_{1}-\mu)(\rho a^+ a^+-a^+\rho a^+)+
{1\over 2}(D_{1}+\mu)(a^+ a^+\rho-a^+\rho a^+)$$
$$+{1\over 2}(D_{2}-\lambda-i\omega)(a^+\rho a-\rho aa^+)+
{1\over2}(D_{2}+\lambda+i\omega)(a\rho a^+-a^+ a\rho)+h.c.,\eqno (3.7)$$
where
$$D_{1}={1\over\hbar}(m\omega D_{qq}-{D_{pp}\over m\omega}+2iD_{pq}),
~~D_{2}={1\over\hbar}(m\omega D_{qq}+{D_{pp}\over m\omega}). \eqno (3.8)$$

In the literature, master equations of the type (3.3) or (3.7) are encountered
in models for the description of different physical phenomena in
quantum optics, in treatements of the damping of collective modes in deep
inelastic collisions of heavy ions or in the quantum mechanical
description of the dissipation for the one-dimensional harmonic
oscillator. In the following we show that some of these master
equations are particular cases of the Lindblad equation.

1) The master equation of Dekker for the damped quantum harmonic oscillator
[5,33-36] supplemented with the fundamental constraints (3.4) obtained in
[36] from the condition that the time evolution of this master equation does
not violate the generalized uncertainty principle [37] at any time, is a
particular case of
the Lindblad master equation (3.7) when $\mu=\lambda$. If the constraints
(3.4) are not fulfilled in the case $\mu=\lambda,$ then the uncertainty
principle is violated.

2) The quantum master equation considered in [38,39] by Hofmann et al. for
treating the charge equilibration process as a collective high frequency mode
is a particular case of the Lindblad master equation (3.3) if $\lambda=
\gamma(\omega)/2m=\mu, D_{qq}=D_{pq}=0, D_{pp}=\gamma(\omega)T^*(
\omega),$ but the fundamental constraints (3.4) are not satisfied.

3) For the quantum master equation considered in [40] for the description
of heavy ion collisions we have $\lambda=\mu=\gamma/2, D_{pp}=D, D_{qq}=0,
D_{pq}=D_{qp}=-d/2$ and consequently the fundamental constraints are not
fulfilled.

4) In [41], Spina and Weidenm\"uller considered two kinds of master equations
for describing the damping of collective modes in deep inelastic
collisions of heavy ions.

(i)The first equation can be obtained from Eq. (3.3) by replacing
$H_{0}$ by

$H_{0}-Am\omega q^2/2+f(t)q$ and setting
$\lambda=\mu=\Gamma/2, D_{pp}=D/2, D_{qq}=0$ and
$D_{pq}=D_{qp}=B/2$. Then the constraints (3.4) are not satisfied.

(ii)The second master equation is obtained from (3.3) by setting
$H_{0}-Am\omega q^2/2$
$-Ap^2/(2m\omega)+f(t)q$
instead of $H_0$ and $\Gamma_{R}=\Gamma_{p}=
\lambda, \mu=0, D_{pp}=D_{p}/2, D_{qq}=D_{R}/2,
D_{pq}=0$. Then last condition (3.4) is satisfied for all values of the
parameters.

5) The master equation for the density operator of the electromagnetic field
mode coupled to a squeezed bath [42,43] can be obtained from the master
equation (3.7) if we set
$\mu=0, \lambda=\gamma, D_1=2\gamma M, D_2=\gamma(2N+1)$
(in [42]: $\gamma/2$ instead of $\gamma$ and $\bar N$ instead of $N$).

6) The master equation for the density operator of  a harmonic oscillator
coupled to an environment of harmonic oscillators considered in [44-47] is a
particular case of the master equation (3.7) if we set
$\lambda=\mu=\gamma, D_{qq}=D_{pq}=0, D_{pp}=2\gamma(\bar n+1/2)m
\omega\hbar$
and the fundamental constraints (3.4) are not
fulfilled.

7) The master equation written in [48] for different models of correlated-
emission lasers can also be obtained from the master equation (3.7) by setting
$$D_{1}+\mu=2\Lambda_{4},~D_{1}-\mu=2\Lambda_{3},~
D_{2}+\lambda+i\omega=2 \Lambda_{2},~
D_{2}-\lambda-i\omega=2\Lambda_{1}.$$

8) Two master equations were introduced by Jang and Yannouleas in [21,49],
where the nuclear
dissipative pocess is described as the damping of a collective degree of
freedom coupled to a bosonic reservoir at finite temperature. The resulting
RPA master equation within the observed collective subspace was derived in a
purely dynamical way.

(i) The master equation written in [49] in the resonant
approximation (rotating-wave approximation) can be obtained as a particular
case of the Lindblad master equation (3.7). For this one has to set
$$D_{pp}=m^2\omega^2D_{qq},~D_{pq}=\mu=0,~
{4m\omega D_{qq}\over \hbar}=(2<n>+1)\Gamma,~\lambda={\Gamma\over 2},$$
where $<n>$ is the average number of the RPA collective phonons at thermal
equilibrium and $\Gamma$ the decay constant. The fundamental
constraints (3.4) are fulfilled in this case.

(ii)The master equation derived in [21] in order to extend the calculations
carried out in [49] with the before-mentioned master equation, can also be
obtained as a particular case of the master
equation (3.7) by setting
$$D_{qq}=D_{pq}=0,~D_{pp}={\hbar m\omega\over 2}(2<n>+1)\Gamma,~
\mu=\lambda={\Gamma\over 2}$$
or $D_{2}=-D_{1}=(2<n>+1)\Gamma/2$. In this case the fundamental
constraints (3.4) are not fulfilled.

As a conclusion, we can say that on various occasions the master equation for
$\lambda=\mu$ has been tested in the sense that it does not violate the
uncertainty principle in the final steady state, but a possible transient
violation of the uncertainty principle has been noted [5,36]. The above
mentioned models can be divided into three types. The first type are models
which fulfill the constraints (3.4). The corresponding master equations belong
to the cathegory of equations described by the Lindblad theory. These models
obey the uncertainty principle. The second type are models with $\lambda=\mu$
which do not fulfill the constraints (3.4) (for example model 2). According
to [36] these models show a dynamical violation of the uncertainty principle.
The third type are models with $\lambda\ne\mu$ which also do not fulfill
the constraints (3.4). These models and those of the second type do not
belong to the cathegory of master equations treated in the Lindblad theory.
For the models of the third type we can not prove that they do not conserve
the uncertainty principle in time.

{\bf 3.1 The master equation in the Heisenberg picture}

The following notations will be used:
$$\sigma_{q}(t)={\rm Tr}(\rho(t)q),$$
$$\sigma_{p}(t)={\rm Tr}(\rho(t)p),$$
$$\sigma_{qq}={\rm Tr}(\rho(t)q^2)-\sigma_{q}^2(t), \eqno (3.9)$$
$$\sigma_{pp}={\rm Tr}(\rho(t)p^2)-\sigma_{p}^2(t),$$
$$\sigma_{pq}(t)={\rm Tr}(\rho(t){pq+qp\over 2})-\sigma_{p}(t)\sigma_{q}(t).$$
In the Heisenberg picture the master equation has the following symmetric
form:
$${d\widetilde\Phi_{t}(A)\over dt}=\widetilde L(\widetilde\Phi_{t}(A))=
{i\over\hbar}[H_{0},\widetilde\Phi_{t}(A)]-{i\over 2\hbar}(\lambda+\mu)
([\widetilde\Phi_{t}(A),q]p+p[\widetilde\Phi_{t}(A),q])$$
$$+{i\over 2\hbar}(\lambda-\mu)(q[\widetilde\Phi_{t}(A),p]+[\widetilde\Phi_{t}
(A),p]q)-{D_{pp}\over\hbar^2}[q,[q,\widetilde\Phi_{t}(A)]]$$
$$-{D_{qq}\over\hbar^2}[p,[p,\widetilde\Phi_{t}(A)]]+{D_{pq}\over\hbar^2}
([p,[q,\widetilde\Phi_{t}(A)]]+[q,[p,\widetilde\Phi_{t}(A)]]).\eqno (3.10)$$
Denoting by $A$ a selfadjoint operator we have
$$\sigma_{A}(t)={\rm Tr}(\rho(t)A), \sigma_{AA}(t)={\rm Tr}(\rho(t)A^2)-\sigma_
{A}^2(t).$$
It follows that
$${d\sigma_{A}(t)\over dt}={\rm Tr}(L(\rho(t))A)={\rm Tr}(\rho(t)\widetilde
L(A)) \eqno (3.11)$$ and
$${d\sigma_{AA}(t)\over dt}={\rm Tr}(L(\rho(t)A^2)-2{d\sigma_{A}(t_)\over dt}
\sigma_{A}(t)={\rm Tr}(\rho(t)\widetilde L(A^2))-2\sigma_{A}(t){\rm Tr}((\rho(t
)\widetilde L(A)). \eqno (3.12)$$
An important consequence of the precise version of solvability condition
formulated at the beginning of the present Section is the fact that when
$A$ is put equal to $p$ or $q$ in (3.11) and (3.12), then
$d\sigma_{p}(t)/dt$ and $d\sigma_{q}(t)/dt$ are functions only
of $\sigma_{p}(t)$ and $\sigma_{q}(t)$ and $d\sigma_{pp}(t)/dt,
d\sigma_{qq}(t)/dt$ and $d\sigma_{pq}(t)/dt$ are functions only
of $\sigma_{pp}(t),\sigma_{qq}(t)$ and $\sigma_{pq}(t)$. This fact allows an
immediate determination of $\sigma_{p}(t),\sigma_{q}(t),
\sigma_{pp}(t),\sigma_{qq}(t)$ and $\sigma_{pq}(t)$ as functions of time.
The results are the following:
$${d\sigma_{q}(t)\over dt}=-(\lambda-\mu)\sigma_{q}(t)+{1\over m}\sigma_{p}
(t),$$
$${d\sigma_{p}(t)\over dt}=-m\omega^2\sigma_{q}(t)-(\lambda+\mu)\sigma_{p}
(t) \eqno(3.13)$$
and
$${d\sigma_{qq}(t)\over dt}=-2(\lambda-\mu)\sigma_{qq}(t)+{2\over m}
\sigma_{pq}(t)+2D_{qq},$$
$${d\sigma_{pp}\over dt}=-2(\lambda+\mu)\sigma_{pp}(t)-2m\omega^2\sigma_{pq}(t)
+2D_{pp}, \eqno(3.14)$$
$${d\sigma_{pq}(t)\over dt}=-m\omega^2\sigma_{qq}(t)+{1\over m}\sigma_{pp}(t)
-2\lambda\sigma_{pq}(t)+2D_{pq}.$$
All equations considered in various papers in
connection with damping of collective modes in deep inelastic collisions are
obtained as particular cases of Eqs. (3.13) and (3.14), as we already mentioned
before.

The integration of Eqs. (3.13) is straightforward. There are two cases:
{\it a)}$\mu>\omega$ (overdamped) and {\it b)} $\mu<\omega$ (underdamped).

In the case {\it a)} with the notation $\nu^2=\mu^2-\omega^2,$ we obtain [17]:
$$\sigma_q(t)=e^{-\lambda t}((\cosh\nu t+{\mu\over\nu}\sinh\nu t)\sigma_q(0)+
{1\over m\nu}\sinh\nu t\sigma_p(0)),$$
$$\sigma_p(t)=e^{-\lambda t}(-{m\omega^2\over\nu}\sinh\nu t\sigma_q(0)+
(\cosh\nu t-{\mu\over\nu}\sinh\nu t)\sigma_p(0)).\eqno(3.15)$$
If $\lambda>\nu$, then $\sigma_q(\infty)=\sigma_p(\infty)=0$. If $\lambda<\nu$,
then $\sigma_q(\infty)=\sigma_p(\infty)=\infty$.

In the case {\it b)} with
the notation $\Omega^2=\omega^2-\mu^2$, we obtain [17]:
$$\sigma_q(t)=e^{-\lambda t}((\cos\Omega t+{\mu\over\Omega}\sin\Omega t)
\sigma_q(0)+{1\over m\Omega}\sin\Omega t\sigma_p(0)),$$
$$\sigma_p(t)=e^{-\lambda t}(-{m\omega^2\over\Omega}\sin\Omega t\sigma_q(0)+
(\cos\Omega t-{\mu\over\Omega}\sin\Omega t)\sigma_p(0)) \eqno(3.16)$$
and $\sigma_q(\infty)=\sigma_p(\infty)=0.$
In order to integrate Eqs. (3.14), it is convenient to consider the vector
$$X(t)=\left(\matrix{m\omega\sigma_{qq}(t)\cr
\sigma_{pp}(t)/m\omega\cr
\sigma_{pq}(t)\cr}\right).$$
Then the system of equations (3.14) can be written in the form
$${dX(t)\over dt}=RX(t)+D,$$
where $R$ is the following $3\times 3$ matrix
$$R=\left(\matrix{-2(\lambda-\mu)&0&2\omega\cr
0&-2(\lambda+\mu)&-2\omega\cr
-\omega&\omega&-2\lambda\cr}\right)$$
and $D$ the following vector
$$D=\left(\matrix{2m\omega D_{qq}\cr
2D_{pp}/m\omega\cr
2D_{pq}\cr}\right).$$
Then there exists a matrix $T$ with the property $T^2=I$ where $I$ is the
identity matrix and a diagonal matrix $K$ such that $R=TKT$. From this it
follows that $$X(t)=(Te^{Kt}T)X(0)+T(e^{Kt}-I)K^{-1}TD.\eqno(3.17)$$
An interesting observation is that the time dependence of the variances
$\sigma_{qq}(t)$, $\sigma_{pp}(t)$, $\sigma_{pq}(t)$ decomposes in a classical
part given by $Te^{Kt}TX(0)$ and a quantum part given by $T(e^{Kt}-I)K^{-1}TD$.
Exactly this quantum part governs the asymptotic behaviour of the variances
when $t\to \infty$ [17].

In the overdamped case $(\mu>\omega)$ the matrices $T$ and $K$ are given by
$$T={1\over 2\nu}\left(\matrix{\mu+\nu&\mu-\nu&2\omega\cr
\mu-\nu&\mu+\nu&2\omega\cr
-\omega&-\omega&-2\mu\cr}\right)$$
and
$$K=\left(\matrix{-2(\lambda-\nu)&0&0\cr
0&-2(\lambda+\nu)&0\cr
0&0&-2\lambda\cr}\right)$$
with $\nu^2=\mu^2-\omega^2$.

In the underdamped case $(\mu<\omega)$ the matrices $T$ and $K$ are given by
$$T={1\over 2i\Omega}\left(\matrix{\mu+i\Omega&\mu-i\Omega&2\omega\cr
\mu-i\Omega&\mu+i\Omega&2\omega\cr
-\omega&-\omega&-2\mu\cr}\right)$$
and
$$K=\left(\matrix{-2(\lambda-i\Omega)&0&0\cr
0&-2(\lambda+i\Omega)&0\cr
0&0&-2\lambda\cr}\right)$$
with $\Omega^2=\omega^2-\mu^2$.
From (3.17) it follows that
$$X(\infty)=-(TK^{-1}T)D=-R^{-1}D \eqno(3.18)$$
(in the overdamped case the restriction $\lambda>\nu$ is necessary). Then
Eq.(3.17) can be written in the form
$$X(t)=(Te^{Kt}T)(X(0)-X(\infty))+X(\infty).\eqno(3.19)$$
Also
$${dX(t)\over dt}=(TKe^{Kt}T)(X(0)-X(\infty))=R(X(t)-X(\infty)).$$
The formula (3.18) is remarcable because it gives a very simple connection
between the asymptotic values $(t \to \infty)$ of $\sigma_{qq}(t),
\sigma_{pp}(t), \sigma_{pq}(t)$
and the diffusion coefficients $D_{qq},~D_{pp},~ $ $D_{pq}.$ As an immediate
consequence of (3.18) this connection is {\it the same} for both cases,
underdamped and overdamped, and has the following explicit form:
$$\sigma_{qq}(\infty)={1\over 2(m\omega)^2\lambda(\lambda^2+\omega^2-\mu^2)}
((m\omega)^2(2\lambda(\lambda+\mu)+\omega^2)D_{qq}$$
$$+\omega^2D_{pp}+2m\omega^2(\lambda+\mu)D_{pq}),$$
$$\sigma_{pp}(\infty)={1\over 2\lambda(\lambda^2+\omega^2-\mu^2)}((m\omega)^2
\omega^2D_{qq}+(2\lambda(\lambda-\mu)+\omega^2)D_{pp}-2m\omega^2(\lambda-
\mu)D_{pq}),\eqno(3.20)$$
$$\sigma_{pq}(\infty)={1\over 2m\lambda(\lambda^2+\omega^2-\mu^2)}(-(\lambda+
\mu)(m\omega)^2D_{qq}+(\lambda-\mu)D_{pp}+2m(\lambda^2-\mu^2)D_{pq}).$$
These relations show that the asymptotic values $\sigma_{qq}(\infty),
\sigma_{pp}(\infty),\sigma_{pq}(\infty)$ do not depend on the initial values
$\sigma_{qq}(0),\sigma_{pp}(0),\sigma_{pq}(0)$. In other words,
$$R^{-1}={-1\over 4\lambda(\lambda^2+\omega^2-\mu^2)}\left(\matrix{2\lambda
(\lambda+\mu)+\omega^2&\omega^2&2\omega(\lambda+\mu)\cr
\omega^2&2\lambda(\lambda-\mu)+\omega^2&-2\omega(\lambda-\mu)\cr
-(\lambda+\mu)\omega&(\lambda-\mu)\omega&2(\lambda^2-\mu^2)\cr}\right).$$
Conversely, if the relations $D=-RX(\infty)$ are considered,
then $$D_{qq}=(\lambda-\mu)\sigma_{qq}(\infty)-{1\over m}\sigma_{pq}(\infty),$$
$$D_{pp}=(\lambda+\mu)\sigma_{pp}(\infty)+m\omega^2\sigma_{pq}(\infty),\eqno
(3.21)$$
$$D_{pq}={1\over2}(m\omega^2\sigma_{qq}(\infty)-{1\over m}\sigma_{pp}(\infty)+
2\lambda\sigma_{pq}(\infty)).$$
Hence, from (3.4) fundamental constraints on $\sigma_{qq}(\infty),
\sigma_{pp}(\infty)$ and $\sigma_{pq}(\infty)$ follow:
$$D_{qq}=(\lambda-\mu)\sigma_{qq}(\infty)-{1\over m}\sigma_{pq}(\infty)>0,$$
$$D_{pp}=(\lambda+\mu)\sigma_{pp}(\infty)+m\omega^2\sigma_{pq}(\infty)>0,$$
$$D_{qq}D_{pp}-D_{pq}^2=(\lambda^2-\mu^2)\sigma_{qq}(\infty)\sigma_{pp}(\infty)
-\omega^2\sigma_{pq}^2(\infty)
+(\lambda-\mu)m\omega^2\sigma_{qq}(\infty)\sigma_{pq}(\infty)$$
$$-{(\lambda+\mu)
\over m}\sigma_{pp}(\infty)\sigma_{pq}(\infty)
-{1\over 4}(m\omega^2)^2\sigma_{qq}^2(\infty)-{1\over 4m^2}\sigma_{pp}^2
(\infty)-\lambda^2\sigma_{pq}^2(\infty)+{1\over 2}\omega^2\sigma_{qq}
(\infty)\sigma_{pp}(\infty)$$
$$-m\omega^2\lambda\sigma_{qq}(\infty)\sigma_{pq}(\infty)+{\lambda\over m}
\sigma_{pp}(\infty)\sigma_{pq}(\infty)\ge{\lambda^2\hbar^2\over 4}.
\eqno(3.22)$$
The constraint (3.22) can be put in a more clear form:
$$4(\lambda^2+\omega^2-\mu^2)(\sigma_{qq}(\infty)\sigma_{pp}(\infty)-
\sigma_{pq}(\infty)^2)$$
$$-(m\omega^2\sigma_{qq}(\infty)+{1\over m}\sigma_{pp}(\infty)+2\mu\sigma_{pq}
^2(\infty))\ge\hbar^2\lambda^2. \eqno(3.23)$$
If $\mu<\omega$ (the underdamped case), then $\lambda^2+\omega^2-\mu^2>
\lambda^2$. If $\mu>\omega$ (the overdamped case), then $0\le\lambda^2+
\omega^2-\mu^2<\lambda^2$ $(\lambda>\nu)$ and the constraint (3.23) is more
strong than the uncertainty inequality $\sigma_{qq}(\infty)\sigma_{pp}
(\infty)-\sigma_{pq}^2(\infty)\ge\hbar^2/4$. By using the fact that
the linear positive mapping ${\cal B(H)}\to {\bf C}$ defined by $A\to {\rm Tr}
(\rho A)$
is completely positive (hence 2-positive), in [17] the following inequality
was obtained:
$$D_{qq}\sigma_{pp}(t)+D_{pp}\sigma_{qq}(t)-2D_{pq}\sigma_{pq}(t)\ge
{\hbar^2\lambda\over 2}.$$
We found this inequality, which must be valid for all values of $t\in(0,
\infty),$ as the sufficient condition that the generalized
uncertainty inequality [37]
$$\sigma_{qq}(t)\sigma_{pp}(t)-\sigma_{pq}^2(t)\ge{\hbar^2
\over 4}$$
is fulfilled at any moment $t$, if the initial values $\sigma_{qq}(0),\sigma_
{pp}(0)$ and $\sigma_{pq}(0)$ for $t=0$ satisfy this inequality.
A restriction connecting the initial values $\sigma_{qq}(0),\sigma_{pp}(0),
\sigma_{pq}(0)$ with the asymptotic values $\sigma_{pp}(\infty),\sigma_{qq}(
\infty),\sigma_{pq}(\infty)$ is also obtained if the values $D_{qq}, D_{pp}$
and $ D_{pq}$ are expressed by Eq. (3.21) in terms of $\sigma_{pp}(\infty),
\sigma_{qq}(\infty), \sigma_{pq}(\infty):$
$$D_{qq}\sigma_{pp}(0)+D_{pp}\sigma_{qq}(0)-2D_{pq}\sigma_{pq}(0)\ge
{\hbar^2\lambda\over 2}.$$
More explicitly
$$\lambda(\sigma_{qq}(\infty)\sigma_{pp}(0)+\sigma_{pp}(\infty)\sigma_{qq}(0)-
2\sigma_{pq}(\infty)\sigma_{pq}(0))
-\mu(\sigma_{qq}(\infty)\sigma_{pp}(0)-\sigma_{pp}(\infty)\sigma_{qq}(0))$$$$-
{1\over m}(\sigma_{pq}(\infty)\sigma_{pp}(0)-\sigma_{pp}(\infty)\sigma_{pq}
(0))
+m\omega^2(\sigma_{pq}(\infty)\sigma_{qq}(0)-\sigma_{qq}(\infty)\sigma_{pq}
(0))\ge{\hbar^2\lambda\over 2}.\eqno(3.24)$$
If the asymptotic state is a Gibbs state
$$\rho_G(\infty)=e^{-H_0/kT}/{\rm Tr}(e^{-H_0/kT}),$$
then
$$\sigma_{qq}(\infty)={\hbar\over 2m\omega}\coth{\hbar\omega\over 2kT},~
\sigma_{pp}(\infty)={\hbar m\omega\over 2}\coth{\hbar\omega\over 2kT},~
 \sigma_{pq}(\infty)=0\eqno(3.25)$$
and
$$D_{pp}={\lambda+\mu\over 2}\hbar m\omega\coth{\hbar\omega\over 2kT},~
D_{qq}={\lambda-\mu\over 2}{\hbar\over m\omega}\coth{\hbar\omega\over 2kT},~
D_{pq}=0\eqno(3.26)$$
and the fundamental constraints (3.4) are satisfied only if $\lambda>\mu$ and
[16]:
$$(\lambda^2-\mu^2)(\coth{\hbar\omega\over 2kT})^2\ge\lambda^2.$$
If the initial state is the ground state of the harmonic oscillator, then
$$\sigma_{qq}(0)={\hbar\over 2m\omega},~\sigma_{pp}(0)={m\hbar\omega\over 2},~
\sigma_{pq}(0)=0.$$
Then (3.24) becomes
$$\lambda(\sigma_{qq}(\infty)m\omega+{\sigma_{pp}(\infty)\over m\omega})-
\mu(\sigma_{qq}(\infty)m\omega-{\sigma_{pp}(\infty)\over m\omega})\ge
\hbar\lambda.$$
For example, in the case (3.25), this implies $\coth(\hbar\omega/2kT)\ge1$
which is always valid.

Now, the explicit time dependence of $\sigma_{qq}(t),\sigma_{pp}(t),\sigma_{pq}
(t)$ will be given for both under- and overdamped cases. From Eq. (3.19) it
follows that in order to obtain the explicit time dependence it is necessary
to obtain the matrix elements of $Te^{Kt}T$. In the overdamped case
$(\mu>\omega, \nu^2=\mu^2-\omega^2)$ we have
$$Te^{Kt}T={e^{-2\lambda t}\over 2\nu^2}\left(\matrix{a_{11}&a_{12}&a_{13}\cr
a_{21}&a_{22}&a_{23}\cr
a_{31}&a_{32}&a_{33}\cr}\right),$$
with

$a_{11}=(\mu^2+\nu^2)\cosh 2\nu t+2\mu\nu\sinh 2\nu t-\omega^2,$

$a_{12}=(\mu^2-\nu^2)\cosh 2\nu t-\omega^2,$

$a_{13}=2\omega(\mu\cosh 2\nu t+\nu\sinh 2\nu t-\mu),$

$a_{21}=(\mu^2-\nu^2)\cosh 2\nu t-\omega^2,$

$a_{22}=(\mu^2+\nu^2)\cosh 2\nu t-2\mu\nu\sinh 2\nu t-\omega^2,$~~~~~~~~~~~~~~
~~~~~~~~~~~~~~~~~~~~~~~~~~~~~~~~~~(3.27)

$a_{23}=2\omega(\mu\cosh 2\nu t-\nu\sinh 2\nu t-\mu),$

$a_{31}=-\omega(\mu\cosh 2\nu t+\nu\sinh 2\nu t-\mu),$

$a_{32}=-\omega(\mu\cosh 2\nu t-\nu\sinh 2\nu t-\mu),$

$a_{33}=-2(\omega^2\cosh 2\nu t-\mu^2).$

In the underdamped case $(\mu<\omega, \Omega^2=\omega^2-\mu^2)$ we have
$$Te^{Kt}T=-{e^{-2\lambda t}\over 2\Omega^2}\left(\matrix{b_{11}&b_{12}&b_{13}
\cr
b_{21}&b_{22}&b_{23}\cr
b_{31}&b_{32}&b_{33}\cr}\right)$$
with

$b_{11}=(\mu^2-\Omega^2)\cos 2\Omega t-2\mu\Omega\sin 2\Omega t-\omega^2,$

$b_{12}=(\mu^2+\Omega^2)\cos 2\Omega t-\omega^2,$

$b_{13}=2\omega(\mu\cos 2\Omega t-\Omega\sin 2\Omega t-\mu),$

$b_{21}=(\mu^2+\Omega^2)\cos 2\Omega t-\omega^2,$

$b_{22}=(\mu^2-\Omega^2)\cos 2\Omega t+2\mu\Omega\sin 2\Omega t-\omega^2,$
~~~~~~~~~~~~~~~~~~~~~~~~~~~~~~~~~~~~~~~~~~~~~~~~~~~~~~(3.28)

$b_{23}=2\omega(\mu\cos 2\Omega t+\Omega\sin 2\Omega t-\mu),$

$b_{31}=-\omega(\mu\cos 2\Omega t-\Omega\sin 2\Omega t-\mu),$

$b_{32}=-\omega(\mu\cos 2\Omega t+\Omega\sin 2\Omega t-\mu),$

$b_{33}=-2(\omega^2\cos 2\Omega t-\mu^2).$

As was shown by Lindblad in [16] the equations of motion (3.10) written for
Weyl operators $(A=W(\xi,\eta)=e^{(i/\hbar)(\eta q-\xi p)})$ can be integrated
in a very simple and elegant way. This fact is important from both points of
view: practical and theoretical. From practical point of view, it gives in a
new way explicit formulas for $ \sigma_{qq}(t), \sigma_{pp}(t), \sigma_{pq}(t)$
and moreover, it gives the action  of the dynamical semigroup $ \widetilde
\Phi_t$ generated by (3.10) on any polynomial in the noncommutative variables
$p$ and $q$. Because $\widetilde \Phi_t(AB) \ne \widetilde \Phi_t(A) \widetilde
\Phi_t(B)$ it is not sufficient to know $\widetilde \Phi_t(p)$ and $ \widetilde
\Phi_t(q)$ as in the case of dynamical groups $\widetilde U_t(AB) = \widetilde
U_t(A) \widetilde U_t(B)$, where evidently it is sufficient to know the action
of $ \widetilde U_t$ on $p$ and $q$ in order to know this action on all
noncommutative polynomials. From theoretical point of view, the explicit
action of $\widetilde \Phi_t$ on Weyl operators $W(\xi,\eta)$ allows, as was
shown by Lindblad in the Appendix of [16], directly to prove
that the semigroup $\widetilde\Phi_t$ is indeed a semigroup of completely
positive mappings. This assertion cannot be considered as a consequence of the
structural theorem of Lindblad because $ \widetilde L$ is unbounded.

We can write as Lindblad [16],
$$ \widetilde \Phi_t(W(\xi,\eta))=W(\xi(t),\eta(t))e^{g(t)}$$
and then the master equation
$$ {d \widetilde \Phi_t(W(\xi,\eta)) \over dt}=\widetilde L(\widetilde \Phi_t
(W(\xi,\eta)))$$
is equivalent with the following differential equations
$$ {d\xi(t) \over dt}=-(\lambda+\mu)\xi(t)-{1 \over m} \eta(t),$$
$$ {d\eta(t) \over dt}=m \omega^2\xi(t)-(\lambda-\mu)\eta(t),$$
$$ {dg(t) \over dt}=-{1 \over \hbar^2}(D_{pp}\xi^2(t)+D_{qq}\eta^2(t)-
2D_{pq}\xi(t)\eta(t),$$
where $\xi(0)=\xi,~\eta(0)=\eta$ and $ g(0)=0.$
The explicit determination of $ \widetilde \Phi_t(W(\xi,\eta))$ as a function
of $t$ reduces to the integration of these differential equations. We obtain
[17]:
$$ \xi(t)=\alpha(t)\xi(0)+\beta(t)\eta(0),~
 \eta(t)=\gamma(t)\xi(0)+\delta(t)\eta(0),$$
where
$$ \left(\matrix{\alpha(t)&\beta(t) \cr
\gamma(t)&\delta(t) \cr} \right) =
e^{-\lambda t} \left(\matrix{\cosh \nu t-{\mu \over \nu} \sinh \nu t&-{1 \over
m \nu} \sinh \nu t \cr
{m \omega^2 \over \nu} \sinh \nu t&\cosh \nu t+{\mu \over \nu} \sinh \nu t \cr
} \right).$$
From
$$ \widetilde \Phi_t(W(\xi(0),~\eta(0)))=W(\xi(t),\eta(t))e^{g(t)}=
e^{(i/\hbar)(\eta(t)q-\xi(t)p)+g(t)}, \eqno (3.29)
$$
$$ {\partial W(\xi(0),\eta(0)) \over \partial \xi(0)} \vert_{\xi(0)=\eta(0)=0}=
-{i \over \hbar}p$$
and
$${\partial W(\xi(0),\eta(0)) \over \partial \eta(0)} \vert_{\xi(0)=\eta(0)=0}=
{i \over \hbar}q, $$
the action of $ \widetilde \Phi_t$ on $q$ and $p$, respectively, is obtained:
$$ \widetilde \Phi_t(q)=\delta(t)q-\beta(t)p,~~ \widetilde \Phi_t(p)=-\gamma(t)
q+\alpha(t)p. $$
As an immediate consequence of these relations we have $ [\widetilde \Phi_t(q),
\widetilde \Phi_t(p)]=i \hbar I e^{-2i\lambda t}. $
From (3.29) and
$$ {\partial^2 W(\xi(0),\eta(0)) \over \partial \xi^2(0)} \vert_{\xi(0)=\eta(0)
=0}=-{1 \over \hbar^2}p^2, $$
$${\partial^2 W(\xi(0),\eta(0)) \over \partial \eta^2(0)} \vert_{\xi(0)=\eta(0)
=0}=-{1 \over \hbar^2}q^2 $$
and
$$ {\partial^2 W(\xi(0),\eta(0)) \over \partial \xi(0) \partial \eta(0)}
\vert_{\xi(0)=\eta(0)=0}={1 \over 2 \hbar^2}(pq+qp),$$
it follows, respectively,
$$ \widetilde \Phi_t(q^2)= \widetilde \Phi_t^2(q)+2B(t), $$
$$ \widetilde \Phi_t(p^2)=\widetilde \Phi_t^2(p)+2A(t) $$
and
$$ \widetilde \Phi_t({pq+qp \over 2})=\widetilde \Phi_t(p)
\widetilde \Phi_t(q)-2C(t), $$
where
$$ \left(\matrix{2B(t)m\omega \cr
2A(t)/m\omega \cr
-2C(t) \cr} \right)=T(e^{Kt}-I)K^{-1}TD. $$
Let us denote by $\sigma(t)$ the correlation matrix
$$ \sigma(t)= \left(\matrix{m\omega\sigma_{qq}(t)&\sigma_{pq}(t) \cr
\sigma_{pq}(t)&\sigma_{pp}(t)/m\omega \cr} \right). $$
Now by definition, for any state $ \rho \in {\cal D}({\cal H}):$
$$ \sigma_{qq}(t)= {\rm Tr} (\rho \widetilde \Phi_t(q^2))-({\rm Tr}(\rho
\widetilde \Phi_t(q)))^2,$$
$$ \sigma_{pp}(t)= {\rm Tr} (\rho \widetilde \Phi_t(p^2))-({\rm Tr}(\rho
\widetilde \Phi_t(p)))^2 $$
and
$$ \sigma_{pq}(t)= {\rm Tr} (\rho \widetilde \Phi_t({pq+qp \over 2}))-
{\rm Tr}(\rho \widetilde \Phi_t(p)){\rm Tr}(\rho \widetilde \Phi_t(q)).$$
Using the relations (3.20) and (3.21) the following interesting formulas are
obtained
$$(\lambda^2+\omega^2-\mu^2)\det\sigma(\infty)={1\over 4}\det\widetilde D+{1
\over \lambda^2}({1 \over 2m}D_{pp}+{m\omega^2 \over 2}D_{qq}+\mu D_{pq})^2
\eqno(3.30) $$
and
$$(\lambda^2+\omega^2-\mu^2)\det\sigma(\infty)={1\over 4}\det \widetilde D+({1
\over 2m}\sigma_{pp}(\infty)+{m\omega^2 \over 2}\sigma_{qq}(\infty)+\mu \sigma
_{pq}(\infty))^2, \eqno (3.31) $$
where
$$\widetilde  D=2 \left(\matrix{m \omega D_{qq}&D_{pq} \cr
D_{pq}&D_{pp}/m\omega \cr} \right). $$
Comparison of (3.30) with (3.31) gives
$${1 \over 2m}\sigma_{pp}(\infty)+{m\omega^2 \over 2}\sigma_{qq}(\infty)+\mu
\sigma_{pq}(\infty)={1 \over \lambda}({1 \over 2m}D_{pp}+{m\omega^2 \over 2}
D_{qq}+\mu D_{pq}). \eqno (3.32)$$
But, the left-hand side of (3.32) is exactly the asymptotic mean value of the
energy of the open harmonic oscillator. Hence (3.32) gives the value of $E(
\infty)$ as a function of diffusion coefficients,
$$ E(\infty) = {1 \over \lambda}({1 \over 2m}D_{pp}+{m\omega^2 \over 2}
D_{qq}+\mu D_{pq}). $$
Another expression for $E(\infty)$ which also follows from (3.30) and (3.31) is
 $$ E(\infty)=((\lambda^2+\omega^2-\mu^2) \det \sigma(\infty)-{1 \over 4} \det
\widetilde D
)^{1/2}. $$
The reality condition for $E(\infty)$ implies
$$\det\sigma(\infty)\ge{1\over 4(\lambda^2+\omega^2-\mu^2)}\det\widetilde D \ge
{\lambda^2 \hbar^2 \over 4(\lambda^2+\omega^2-\mu^2)} $$
which is more restrictive for the overdamped case $(\omega < \mu)$ than
the generalized uncertainty inequality $\det\sigma(\infty) \ge \hbar^2/4.$

{\bf 3.2 The method of the characteristic function}

Instead of solving the master equation (3.7) directly, we first introduce the
normally ordered quantum characteristic function $\chi(\Lambda,\Lambda^*,t)$
defined in terms of the density operator $\rho$ by
$$\chi(\Lambda,\Lambda^*,t)={\rm Tr}[\rho(t)\exp(\Lambda a^+)\exp(-\Lambda^*a)]
,\eqno(3.33)$$
where $\Lambda$ is a complex variable and the trace is performed over the
states of system. Substituting Eq.(3.33) into the master equation (3.7) and
using the operator relations
$$a\exp(\Lambda a^+)=\exp(\Lambda a^+)(a+\Lambda),~
a^+\exp(-\Lambda^*a)=\exp(-\Lambda^*a)(a^++\Lambda^*),$$
$${\rm Tr}[\rho(t)\exp(\Lambda a^+)\exp(-\Lambda^*a)(a^++\Lambda^*)]=\partial_
\Lambda\chi,~
{\rm Tr}[\rho(t)\exp(\Lambda a^+)\exp(-\Lambda^*a)a]=-\partial_{\Lambda^*}
\chi
$$
or applying the rules:
$$\rho\leftrightarrow\chi,~
a\rho\leftrightarrow -{\partial\over\partial\Lambda^*}\chi,~
a^+\rho\leftrightarrow({\partial\over\partial\Lambda}-\Lambda^*)\chi,~
\rho a\leftrightarrow(-{\partial\over\partial\Lambda^*}+\Lambda)\chi,~
\rho a^+\leftrightarrow{\partial\over\partial\Lambda}\chi,$$
the following partial differential equation for $\chi$ is found [18]:
$$\{\partial_t+[(\lambda-i\omega)\Lambda+\mu\Lambda^*]\partial_\Lambda+
[(\lambda+i\omega)\Lambda^*+\mu\Lambda]\partial_{\Lambda^*}\}\chi(\Lambda,
\Lambda^*,t)$$
$$=\{L\vert\Lambda\vert^2+C\Lambda^2+C^*\Lambda^{*2}\}\chi(\Lambda,\Lambda^*,t)
,
\eqno(3.34)$$
where
$$L=\lambda-D_2,~C={1\over 2}(\mu+D_1^*).$$
We consider the state of the system initially to be a superposition of
coherent states. Coherent states $\vert\alpha>$ of the harmonic
oscillator are minimum uncertainty states having the mean coordinate $<q>$ and
mean momentum $<p>$ given by
$$<q>=<\alpha\vert q\vert\alpha>=\sqrt{{2\hbar\over m\omega}}{\rm Re}\alpha,~
 <p>=<\alpha\vert p\vert\alpha>=\sqrt{2\hbar m\omega}{\rm Im}\alpha.\eqno(3.35)
$$
Consequently, we take as the initial density operator
$$\rho(0)=\int d\alpha d\beta N(\alpha,\beta)\vert\alpha><\beta\vert.$$
The quantum characteristic function corresponding to the operator
$\vert\alpha><\beta\vert$ is given from Eq. (3.33) by
$$\chi=<\beta\vert\alpha>\exp(\lambda\beta^*-\lambda^*\alpha).\eqno(3.36)$$
We look for a solution of (3.34) having the exponential form
$$\chi(\Lambda,\Lambda^*,t)=
\int d\alpha d\beta N(\alpha,\beta)<\beta\vert\alpha>\exp[A(t)\Lambda+B(t)
\Lambda^*+f(t)\Lambda^2+f^*(t)\Lambda^{*2}+h(t)\vert\Lambda\vert^2].\eqno(3.37)
$$
The form of the solution (3.37) is suggested from the fact that the left-hand
side of Eq. (3.34) contains first-order derivatives with respect to the time
and variables $\Lambda$ and $\Lambda^*$ and is symmetric with
respect to complex conjugation.
The functions $A(t),B(t),f(t)$ and $h(t)$ depend only on time. Corresponding
to Eq. (3.36), these functions have to satisfy the
initial conditions
$$A(0)=\beta^*,~B(0)=\alpha,~f(0)=0,~h(0)=0.\eqno(3.38)$$
When we introduce the function (3.37) into Eq. (3.34) and equate the
coefficients
of equal powers of $\Lambda$ and $\Lambda^*$, we get the following two
systems of linear differential equations of first order with constant
coefficients:
$${dA(t)\over dt}+(\lambda-i\omega)A(t)+\mu B(t)=0$$
$${dB(t)\over dt}+\mu A(t)+(\lambda+i\omega)B(t)=0\eqno(3.39a)$$
and
$${dR(t)\over dt}+2\lambda R(t)+2\omega I(t)+\mu h(t)={\rm Re} C$$
$${dI(t)\over dt}+2\lambda I(t)-2\omega R(t)={\rm Im} C\eqno(3.39b)$$
$${dh(t)\over dt}+4\mu R(t)+2\lambda h(t)=L,$$
where $R(t)={\rm Re} f(t),I(t)={\rm Im} f(t)$ with the initial conditions $R(0)
=I(0)=h(0)=0$.
Subject to the initial conditions (3.38), the homogeneous system $(3.39a)$ has
the
solution [18]:
$$A(t)=u(t)\beta^*-v(t)\alpha,
B(t)=-u^*(t)\alpha+v(t)\beta^*,\eqno(3.40)$$
where
$$u(t)={1\over 2}[\exp(-\mu_-t)+\exp(-\mu_+t)+{2i\omega\over\mu_--\mu_+}
(\exp(-\mu_+t)-\exp(-\mu_-t))],$$
$$v(t)={\mu\over\mu_--\mu_+}(\exp(-\mu_-t)-\exp(-\mu_+t)).\eqno(3.41)$$
The eigenvalues $\mu_{\pm}$ are given by
$$\mu_{\pm}=\lambda\pm\sqrt{\mu^2-\omega^2},~\gamma\equiv\sqrt{\mu^2-\omega^2}.
\eqno(3.42)$$
The system $(3.39b)$ has the eigenvalues $-2\lambda, -2(\lambda\pm\sqrt{\mu^2-
\omega^2})=-2\mu_{\pm}$ and in order to integrate it we apply the
same method as for the system (3.14) in the preceding Subsection. We obtain:
$$f(t)={P\over 2\mu}\exp(-2\mu_+t)(\gamma-i\omega)-{N\over 2\mu}\exp(-2\mu_-t)
(\gamma+i\omega)-{i\mu M\over 2\omega}\exp(-2\lambda t)+f(\infty),$$
$$h(t)=M\exp(-2\lambda t)+N\exp(-2\mu_-t)+P\exp(-2\mu_+t)+h(\infty).\eqno(3.43)
$$
Here $M,N,P,f(\infty)$ and $h(\infty)$ are constants given by [18]:
$$M={\omega\over\lambda\gamma^2}(\mu {\rm Im} C+{\omega L\over 2}),~
N={\mu\over 2\gamma^2(\lambda-\gamma)}(\gamma {\rm Re}
C-\omega {\rm Im} C-{\mu L\over 2}),$$
$$ P=-{\mu\over 2\gamma^2(\lambda+\gamma)}(\gamma {\rm Re}
C+\omega {\rm Im} C+{\mu L\over 2})$$
and the asymptotic values connected with the diffusion coefficients $D_{qq},
D_{pp}$ and $D_{pq}$ are:
$$R(\infty)={2(\lambda {\rm Re} C-\omega {\rm Im} C)-L\mu\over 4(\lambda^2-
\gamma^2)},$$
$$I(\infty)={2\omega\lambda {\rm Re} C+2(\lambda^2-\mu^2){\rm Im} C-L\mu\omega
\over 4\lambda(\lambda^2-\gamma^2)},\eqno(3.44)$$
$$h(\infty)={L(\lambda^2+\omega^2)-2\mu(\lambda {\rm Re}
C-\omega {\rm Im} C)\over 2\lambda(\lambda^2-\gamma^2)}.$$
By knowing the characteristic function (3.37), (3.40)--(3.43) corresponding to
the initial density operator which represents a superposition of coherent
states, it is easy to obtain explicit formulae for the moments:
$$<a^{+m}(t)a^n(t)>={\rm Tr}[a^{+m}(t)a^n(t)\rho(t)]=(-1)^n{\partial^{n+m}\over
\partial\Lambda^{*n}\partial\Lambda^m}\chi(\Lambda,\Lambda^*,t)\vert_{
\Lambda=\Lambda^*=0}.$$
In the following, we take the density operator $\rho$ in the coherent state
representation
$$\rho(0)=\int P(\alpha)\vert\alpha><\alpha\vert d^2\alpha,$$
where $P(\alpha)$ is the diagonal or  Glauber $P$ distribution and $d^2\alpha=
d{\rm Re}\alpha d{\rm Im}\alpha$. The integration covers the entire complex
$\alpha$ plane. Then the characteristic function (3.37) becomes:
$$\chi(\Lambda,\Lambda^*,t)=\int d^2\alpha P(\alpha)\exp[(u\alpha^*-v\alpha)
\Lambda+(-u^*\alpha+v\alpha^*)\Lambda^*]\exp[f\Lambda^2+f^*\Lambda^{*2}+
h\vert\Lambda\vert^2].$$
Let us assume that the damped oscillator is at $t=0$ prepared in a pure
coherent state, say $\vert\alpha_0>$, corresponding to
$P(\alpha)=\delta({\rm Re}
\alpha-{\rm Re}\alpha_0)\delta({\rm Im}\alpha-{\rm Im}\alpha_0)$. One has
$$\chi^{(0)}(\Lambda,\Lambda^*,t)=\exp[(u\alpha_0^*-v\alpha_0)\Lambda+(-u^*
\alpha_0+v\alpha_0^*)\Lambda^*]\exp[f\Lambda^2+f^*\Lambda^{*2}+h\vert\Lambda
\vert^2].$$
The first moments are given by
$$<a^+(t)>={\partial\chi^{(0)}(t)\over\partial\Lambda}\vert_{\Lambda=\Lambda^*
=0}=u\alpha_0^*-v\alpha_0,~
<a(t)>=-{\partial\chi^{(0)}(t)\over\partial\Lambda^*}\vert_{\Lambda=\Lambda^*
=0}=u^*\alpha_0-v\alpha_0^*.$$
Then, with the notations (3.9), using (3.35) and the transformations
$$q(t)=\sqrt{{\hbar\over 2m\omega}}(a^+(t)+a(t)),~
p(t)=i\sqrt{{\hbar m\omega\over 2}}(a^+(t)-a(t))$$
for the displacement operator $q(t)$ and the momentum operator $p(t)$ of the
oscillator, we obtain the following mean values:
$$\sigma_q(t)=\sqrt{{\hbar\over 2m\omega}}((u-v)\alpha_0^*+(u^*-v)\alpha_0),~
\sigma_p(t)=i\sqrt{{\hbar m\omega\over 2}}((u+v)\alpha_0^*-(u^*+v)\alpha_0),
$$
with $u,v$ given by (3.41), (3.42).
There are two cases:

$a)$ the overdamped case: $\mu>\omega,\nu^2=\mu^2-\omega^2,\gamma\equiv\nu$;
then
$$u(t)=\exp(-\lambda t)(\cosh\nu t+{i\omega\over\nu}\sinh\nu t),~
v(t)=-{\mu\over\nu}\exp(-\lambda t)\sinh\nu t\eqno(3.45)$$
and $\sigma_q(t),\sigma_p(t)$ take the previous form (3.15);

$b)$ the underdamped case: $\mu<\omega,\Omega^2=\omega^2-\mu^2,\gamma\equiv
i\Omega$;
then
$$u(t)=\exp(-\lambda t)(\cosh\Omega t+{i\omega\over\Omega}\sinh\Omega t),~
v(t)=-{\mu\over\Omega}\exp(-\lambda t)\sinh\Omega t\eqno(3.46)$$
and $\sigma_q(t),\sigma_p(t)$ take the previous form (3.16).

For the variances one finds:
$$<a^2(t)>={\partial^2\chi^{(0)}(t)\over\partial\Lambda^{*2}}\vert_{\Lambda=
\Lambda^*
=0}=(u^*\alpha_0-v\alpha_0^*)^2+2f^*,$$
$$<a^{+2}(t)>={\partial^2\chi^{(0)}(t)\over\partial\Lambda^2}\vert_{\Lambda=
\Lambda^*
=0}=(u\alpha_0^*-v\alpha_0)^2+2f,\eqno(3.47)$$
$$<a^+(t)a(t)>=-{\partial^2\chi^{(0)}(t)\over\partial\Lambda\partial\Lambda^*}
\vert_
{\Lambda=\Lambda^*=0}=(u\alpha_0^*-v\alpha_0)(u^*\alpha_0-v\alpha_0^*)-h.$$
Then the relations (3.9) will give us the explicit time dependence of the
variances $\sigma_{qq}(t)$,
 $\sigma_{pp}(t)$, $\sigma_{pq}(t)$. The asymptotic
values of these variances are given by the following expresions [18]:
$$\sigma_{qq}(\infty)={\hbar\over m\omega}(f+f^*-h+{1\over 2})\vert_{t\to
\infty},$$
$$\sigma_{pp}(\infty)=-\hbar m\omega(f+f^*+h-{1\over 2})\vert_{t\to\infty},$$
$$\sigma_{pq}(\infty)=i\hbar(f-f^*)\vert_{t\to\infty}.$$
With $f(\infty)=R(\infty)+iI(\infty)$ and by using the formulas (3.44) for
$R(\infty),I(\infty),h(\infty)$, the asymptotic values of the variances take
the same form (3.20) as in the preceding subsection, as expected.

With the relations (3.47), the expectation value of the energy operator can be
calculated:
$$E(t)=\hbar\omega(<a^+a>+{1\over 2})+{i\hbar\mu\over 2}(<a^{+2}>-<a^2>).$$

{\bf 3.3 Quasiprobability distributions for the damped harmonic oscillator}

The methods of quasiprobabilities have provided technical tools of great power
for the statistical description  of microscopic systems formulated in terms of
the density operator [50-53]. The first quasiprobability distribution was
introduced by Wigner [54] in a quantum-mechanical context. In quantum optics
the $P$ representation introduced by Glauber [55,56] and Sudarshan [57]
provided many practical applications of quasiprobabilities. The development
of quantum-mechanical master equations was combined with the Glauber $P$
representation to give a Fokker-Planck equation for the laser [58,59]. One
useful way to study the consequences of the master equation for the
one-dimensional damped harmonic oscillator is to transform it into equations
for the $c$-number quasiprobability distributions associated with the density
operator. The resulting differential equations of the Fokker-Planck type for
the distribution functions can be solved by standard methods and observables
directly calculated as correlations of these distribution functions. However,
the Fokker-Planck equations do not always have positive-definite diffusion
coefficients. In this case one can treat the problem with the generalized
$P$ distribution [51].

First we present a short summary of the theory of quasiprobability
distributions [53,60]. For the master equation (3.7) of the harmonic
oscillator, physical observables can be derived from the expectation values of
polynomials of the annihilation and creation operators. The expectation values
are determined by using the quantum density operator $\rho$. Usually one
expands the density operator with the aid of coherent states, defined as
eigenstates of the annihilation operator: $a\vert\alpha>=\alpha\vert\alpha>$.
They are given in terms of the eigenstates of the harmonic oscillator as
$$\vert\alpha>=\exp(-{\vert\alpha\vert^2\over 2})\sum_{n=0}^\infty{1\over
\sqrt{n!}}\alpha^n\vert n>,$$
with the normalization $\vert<\beta\vert\alpha>\vert^2=\exp(-\vert\alpha-\beta
\vert^2)$. In order to solve the master equation (3.7) we represent the density
operator $\rho$ by a distribution function over a $c$-number phase space.
The chosen distribution function, introduced in [60], is defined as follows:
$$\Phi(\alpha,s)={1\over\pi^2}\int\chi(\Lambda,s)\exp(\alpha\Lambda^*-\alpha
^*\Lambda)d^2\Lambda, \eqno(3.48)$$
with the characteristic function
$$\chi(\Lambda,s)={\rm Tr}[\rho D(\Lambda,s)],$$
where $D(\Lambda,s)$ is the displacement operator
$$D(\Lambda,s)=\exp(\Lambda a^+-\Lambda^*a+{1\over 2}s\vert\Lambda\vert^2).$$
The interval of integration in Eq. (3.48) is the whole complex $\Lambda$ plane.
Because of
$$\delta^2(\alpha)={1\over\pi^2}\int\exp(\alpha\Lambda^*-\alpha^*\Lambda)d^2
\Lambda,$$
the distribution function $\Phi(\alpha,s)$ is the Fourier transform of the
characteristic function. Since the density operator is
normalized by ${\rm Tr}\rho=1$, one obtains the normalization of $\Phi$:
$$\int\Phi(\alpha,s)d^2\alpha=1.$$
In this paper we restrict ourselves to distribution functions with the
parameters $s=1,0$ and $-1$. These distribution functions can be used to
calculate expectation values of products of annihilation and creation
operators. For that purpose we first expand the displacement operator in a
power series of the operators $a$ and $a^+$:
$$D(\Lambda,s)=\sum_{m=0}^\infty\sum_{n=0}^\infty{\Lambda^m(-\Lambda^*)^n\over
 m!n!}\{a^{+m}a^n\}_s.\eqno(3.49)$$
The parentheses with the index $s$ indicate the special representations of the
polynomials depending on $s$. For example for $n=m=1$ we have the $s$-ordered
operators:
$$\{a^+a\}_{s=1}=a^+a,~
\{a^+a\}_{s=0}={1\over 2}(a^+a+aa^+),~
\{a^+a\}_{s=-1}=aa^+.\eqno(3.50)$$
Expectation values of the $s$-ordered operators can be calculated as follows:
$$<\{a^{+m}a^n\}_s>={\rm Tr}[\rho\{a^{+m}a^n\}_s]={\rm Tr}[\rho({\partial\over
\partial
\Lambda})^m(-{\partial\over\partial\Lambda^*})^nD(\Lambda,s)\vert_{\Lambda=0}]
$$
$$=({\partial\over\partial\Lambda})^m(-{\partial\over\partial\Lambda^*})^n\chi
(\Lambda,s)\vert_{\Lambda=0}=\int(\alpha^*)^m\alpha^n\Phi(\alpha,s)d^2\alpha.$$
For the last step we apply the inverse relation to Eq. (3.48):
$$\chi(\Lambda,s)=\int\Phi(\alpha,s)\exp(\Lambda\alpha^*-\Lambda^*\alpha)d^2
\alpha.$$
In the following we discuss the distribution functions for $s=1,0$ and $-1$ in
more detail. For $s=1$ we obtain the Glauber $P$ function [56,61], for $s=0$
the Wigner function [54] and for $s=-1$ the $Q$ function [60]. For $s=1$ we
have
$$D(\Lambda,1)=\exp(\Lambda a^+)\exp(-\Lambda^*a).$$
Then the $s$ ordering in Eq. (3.49) corresponds to normal ordering. Since the
Glauber $P$ function is the Fourier transform of the characteristic function
$$\chi_N(\Lambda)={\rm Tr}[\rho\exp(\Lambda a^+)\exp(-\Lambda^*a)]=\chi
(\Lambda,1),$$ it follows from Eq. (3.48) that the distribution $\Phi(\alpha,
1)$ is identical
to the $P$ function. This function is used for an expansion of the density
operator in diagonal coherent state projection operators [55-57,62]:
$$\rho=\int P(\alpha)d^2\alpha\vert\alpha><\alpha\vert.$$
Calculating the expectation values of normally ordered operator products we
obtain the relation
$$<a^{+m}a^n>=\int(\alpha^*)^m\alpha^n\Phi(\alpha,1)d^2
\alpha=\int(\alpha^*)^m\alpha^nP(\alpha)d^2\alpha,$$
from which we again derive $P(\alpha)=\Phi(\alpha,1)$. Despite the formal
similarity with averaging procedure with a classical probability distribution,
the function $P(\alpha)$ is not a true probability distribution. Because of the
overcompleteness of the coherent states, the $P$ function is not a unique,
well-behaved positive function for all density operators.

Cahill [63] studied the $P$ representation for density operators which
represent pure states and found a narrow class of states for which the $P$
representation exists. These states can be generated from a particular coherent
state
$\vert\alpha>$ by the application of a finite number of creation operators.
For example, for the ground state of the harmonic oscillator it is easy to
show that $\chi_N(\Lambda)=1$ for all $\Lambda$. In that case the $P$ function
becomes $P(\alpha)=\delta^2(\alpha)$. The delta function and its derivatives
are examples of a class of generalized functions known as tempered
distributions [61]. Also Cahill [64] introduced a representation of the
density operator of the electromagnetic field that is suitable for all density
operators and that reduces to the coherent state $P$ representation when the
latter exists. The representation has no singularities.

Sudarshan [57] offered a singular formula for the $P$ representation in terms
of an infinite series of derivatives of the delta function. From the
mathematical point of view, such a series is usually not considered to be a
distribution function [61,62].

For $s=-1$ we have
$$D(\Lambda,-1)=\exp(-\Lambda^*a)\exp(\Lambda a^+).$$
The $s$ ordering corresponds to antinormal ordering. Because the $Q$ function
is the Fourier transform of the characteristic function
$$\chi_A(\Lambda)={\rm Tr}[\rho\exp(-\Lambda^*a)\exp(\Lambda a^+)]=\chi
(\Lambda,-1),$$
it follows from Eq. (3.48) that the distribution $\Phi(\alpha,-1)$ is the $Q$
function. It is given by the diagonal matrix elements of the density operator
in terms of coherent states:
$$Q(\alpha)={1\over\pi}<\alpha\vert\rho\vert\alpha>.$$
Though for all density operators the $Q$ function is bounded, non-negative and
infinitely differentiable, it has the disadvantage that not every positive $Q$
function corresponds to a positive semidefinite Hermitian density operator.
Evaluating moments is only simple in the $Q$ representation for antinormally
ordered operator products.

For $s=0$, the distribution $\Phi(\Lambda,0)$ becomes the Wigner function $W$.
The latter function is defined as the Fourier transform of the characteristic
function
$$\chi_S(\Lambda)={\rm Tr}[\rho\exp(\Lambda a^+-\Lambda^*a)]=\chi(\Lambda,0).$$
Because this characteristic function is identical to $\chi(\Lambda,0)$, we
conclude that $\Phi(\alpha,0)$ is the Wigner function $W(\alpha)$. Therefore,
the Wigner function can be used to calculate expectation values of
symmetrically ordered operators:
$$<\{a^{+m}a^n\}_{s=0}>=\int(\alpha^*)^m\alpha^nW(\alpha)d^2\alpha.$$
The symmetrically ordered operators are the arithmetic average of $(m+n)!
/m!n!$ differently ordered products of $m$
factors of $a^+$ and $n$ factors of $a$. An example for $m=n=1$ is given in
Eq. (3.50).

The Wigner function is a nonsingular, uniformly continuous function of
$\alpha$ for all density operators and may in general assume negative values.
It is related to the density operator as follows:
$$W(\alpha)={1\over\pi^2}\int d^2\Lambda {\rm Tr}[\exp(\Lambda(a^+-\alpha^*)-
\Lambda^*(a-\alpha))\rho].$$
Also it can be obtained from the $P$ representation:
$$W(\alpha)={2\over\pi}\int P(\beta)\exp(-2\vert\alpha-\beta\vert^2)d^2\beta.$$
By using the standard transformations [65-67], the master equation (3.7) can be
transformed into a differential equation for a corresponding $c$-number
distribution. Here we apply these methods to derive Fokker-Planck
equations for the before mentioned distributions: Glauber $P$, $Q$ and
Wigner $W$ distributions. Using the relations
$${\partial D(\Lambda,s)\over\partial\Lambda}=[(s-1){\Lambda^*\over 2}+a^+]
D(\Lambda,s)=D(\Lambda,s)[(s+1){\Lambda^*\over 2}+a^+],$$
$${\partial D(\Lambda,s)\over\partial\Lambda^*}=[(s+1){\Lambda\over 2}-a]
D(\Lambda,s)=D(\Lambda,s)[(s-1){\Lambda\over 2}-a],$$
one can derive the following rules for transforming the master equation (3.7)
into Fokker-Planck equations in the Glauber $P(s=1)$, $Q(s=-1)$
and Wigner $W(s=0)$ representations:
$$a\rho\leftrightarrow (\alpha-{s-1\over 2}{\partial\over\partial\alpha^*})
\Phi,~~
a^+\rho\leftrightarrow (\alpha^*-{s+1\over 2}{\partial\over\partial\alpha})
\Phi,$$
$$\rho a\leftrightarrow (\alpha-{s+1\over 2}{\partial\over\partial\alpha^*})
\Phi,~~
\rho a^+\leftrightarrow (\alpha^*-{s-1\over 2}{\partial\over\partial\alpha})
\Phi.$$
Applying these operator correspondences (repeatedly, if necessary), we
find the following Fokker-Planck equations for the distributions
$\Phi(\alpha,s)$ [19]:
$${\partial\Phi(\alpha,s)\over\partial t}=-({\partial\over\partial\alpha}d_
\alpha+{\partial\over\partial\alpha^*}d_\alpha^*)\Phi(\alpha,s)
+{1\over 2}({\partial^2\over\partial\alpha^2}D_{\alpha\alpha}+{\partial^2
\over\partial \alpha^{*2}}D_{\alpha\alpha}^*+2{\partial^2\over\partial\alpha
\partial\alpha^*}D_{\alpha\alpha^*})\Phi(\alpha,s).\eqno(3.51)$$
Here, $\Phi(\alpha,s)$ is $P(s=1), Q(s=-1)$ and $W(s=0.)$ While the drift
coefficients are the same for the three distributions, the diffusion
coefficients are different:
$$d_\alpha=-(\lambda+i\omega)\alpha+\mu\alpha^*,~
 D_{\alpha\alpha}=D_1+s\mu,~
 D_{\alpha\alpha^*}=D_2-s\lambda.$$
The Fokker-Planck equation (3.51) can also be written in terms of real
coordinates $x_1$ and $x_2$ defined by
$$\alpha=x_1+ix_2\equiv\sqrt{{m\omega\over 2\hbar}}<q>+i{1\over\sqrt{2\hbar
 m\omega}}<p>,~
\alpha^*=x_1-ix_2\eqno(3.52).$$
as follows:
$${\partial\Phi(x_1,x_2)\over\partial t}=-({\partial\over\partial x_1}d_1+
{\partial\over\partial x_2}d_2)\Phi(x_1,x_2)
+{1\over 2}({\partial^2\over\partial x_1^2}D_{11}^{(s)}+{\partial^2\over
\partial x_2^2}D_{22}^{(s)}+2{\partial^2\over
\partial x_1\partial x_2}D_{12}^{(s)})\Phi(x_1,x_2),\eqno(3.53)$$
with the new drift and diffusion coefficients given by
$$d_1=-(\lambda-\mu)x_1+\omega x_2,~
d_2=-\omega x_1-(\lambda+\mu)x_2,$$
$$D_{11}^{(s)}={1\over\hbar}m\omega D_{qq}-{s\over 2}(\lambda-\mu),~
 D_{22}^{(s)}={1\over\hbar}{D_{pp}\over m\omega}-{s\over 2}(\lambda+\mu),~
 D_{12}^{(s)}={1\over\hbar}D_{pq}.$$
We note that the diffusion matrix
$$\left(\matrix{D_{11}^{(s)}&D_{12}^{(s)}\cr
D_{12}^{(s)}&D_{22}^{(s)}\cr}\right)$$
for the $P$ distribution ($s=1$) needs not to be positive definite.

Since the drift coefficients are linear in the variables $x_1$ and
$x_2 (i=1,2)$:
$$d_i=-\sum_{j=1}^2 A_{ij}x_j,~A_{ij}=-{\partial d_i\over\partial x_j},$$
with
$$A=\left(\matrix{\lambda-\mu&-\omega\cr
\omega&\lambda+\mu\cr}\right)\eqno(3.54)$$
and the diffusion coefficients are constant with respect to $x_1$ and $x_2$,
Eq. (3.53) describes an Ornstein-Uhlenbeck process [68,69].

The solution of the Fokker-Planck equation (3.53) can immediately be written
down provided that the diffusion matrix $D$ is positive definite. However,
the diffusion matrix in the Glauber $P$ representation is not, in general,
positive definite. For example, if
$$D_{11}^PD_{22}^P-(D_{12}^P)^2<0,$$
the $P$ distribution does not exist as a well-behaved function. In this
situation, the so-called generalized $P$ distributions can be taken which are
well-behaved, normal ordering functions [51]. The generalized
$P$ distributions are nondiagonal expansions of the density
operator in terms of coherent states projection operators. The $Q$ and $W$
distributions always exist; they are Gaussian functions if they are initially
of Gaussian type.

From Eq. (3.53) one can directly derive the equations of motion for the
expectation values of the variables $x_1$ and $x_2 (i=1,2)$:
$${d<x_i>\over dt}=-\sum_{j=1}^2 A_{ij}<x_j>.\eqno(3.55)$$
By using Eqs. (3.5), (3.52) and (3.55)
we obtain the equations of motion for the
expectation values $\sigma_q(t),\sigma_p(t)$  of the coordinate and momentum of
the harmonic oscillator which are identical with those derived in the preceding
two subsections by using the Heisenberg representation and the method of
characteristic function, respectively (see Eqs. (3.13)).

The variances of the variables $x_1$ and $x_2$ are defined by the expectation
values
$$\sigma_{ij}=<x_ix_j>-<x_i><x_j>,~i,j=1,2.$$
They are connected with the variances and covariance of the coordinate $q$ and
momentum $p$ by
$$\sigma_{qq}=(2\hbar/m\omega)\sigma_{11},~\sigma_{pp}=2\hbar m\omega\sigma_{22
},~
\sigma_{pq}=<{1\over 2}(pq+qp)>-<p><q>=2\hbar\sigma_{12}$$
and can be calculated with the help of the variances of the quasiprobability
distributions $(i,j=1,2)$:
$$\sigma_{ij}^{(s)}=\int x_ix_j\Phi(x_1,x_2,s)dx_1dx_2-\int x_i\Phi(x_1,x_2,s)
dx_1dx_2\int x_j\Phi(x_1,x_2,s)dx_1dx_2.$$
The following relations exist between the various variances:
$$\sigma_{ii}=\sigma_{ii}^P+{1\over 4}=\sigma_{ii}^Q-{1\over 4}=\sigma_{ii}^W,
 i=1,2,~~
\sigma_{12}=\sigma_{12}^P=\sigma_{12}^Q=\sigma_{12}^W.$$
The variances $\sigma_{ij}^{(s)}$ fulfill the following equations of motion:
$${d\sigma_{ij}^{(s)}\over dt}=-\sum_{l=1}^2(A_{il}\sigma_{lj}^{(s)}+\sigma
_{il}^{(s)}A_{lj}^{\rm T})+D_{ij}^{(s)}.\eqno(3.56)$$
They can be written explicitly in the form:
$${d\sigma_{11}^{(s)}\over dt}=-2A_{11}\sigma_{11}^{(s)}-2A_{12}\sigma_{12}
^{(s)}+D_{11}^{(s)},$$
$${d\sigma_{22}^{(s)}\over dt}=-2A_{21}\sigma_{12}^{(s)}-2A_{22}\sigma_{22}
^{(s)}+D_{22}^{(s)},$$
$${d\sigma_{12}^{(s)}\over dt}=-(A_{11}+A_{22})\sigma_{12}^{(s)}-A_{21}\sigma
_{11}^{(s)}-A_{12}\sigma_{22}^{(s)}+D_{12}^{(s)},$$
where the matrix elements $A_{ij}$ are defined in Eq. (3.54). These relations
are sufficient to prove  that the equations of motion for the variances
$\sigma_{11}$ and $\sigma_{22}$ and the covariance $\sigma_{12}$ are the same
for all representations as expected. The corresponding
equations of motion of the  variances and covariance of the coordinate and
momentum coincide with those obtained in the preceding two subsections by using
the Heisenberg representation and the method of characteristic function,
respectively (see Eqs. (3.14)).

In order that the system approaches a steady state, the condition $\lambda>\nu$
must be met. Thus the steady state solutions are
$$\Phi(x_1,x_2,s)={1\over 2\pi\sqrt{{\rm det}(\sigma(\infty))}}\exp[-{1\over 2}
\sum_{i,j=1,2}(\sigma^{-1})_{ij}(\infty)x_ix_j],\eqno(3.57)$$
where the stationary covariance matrix
$$\sigma^{(s)}(\infty)=\left(\matrix{\sigma_{11}^{(s)}(\infty)&
\sigma_{12}^{(s)}(\infty)\cr
\sigma_{12}^{(s)}(\infty)&\sigma_{22}^{(s)}(\infty)\cr}\right)$$
can be determined from the algebraic equation (see Eq. (3.56)):
$$\sum_{l=1}^2(A_{il}\sigma_{lj}^{(s)}(\infty)+\sigma_{il}^{(s)}(\infty)A_{lj}
^{\rm T})=D_{ij}^{(s)}.$$
With the matrix elements $A_{ij}$ given by (3.54), we obtain
$$\sigma_{11}^{(s)}(\infty)={(2\lambda(\lambda+\mu)+\omega^2)D_{11}^{(s)}+
\omega^2D_{22}^{(s)}+2\omega(\lambda+\mu)D_{12}^{(s)}\over 4\lambda(\lambda^2+
\omega^2-\mu^2)},$$
$$\sigma_{22}^{(s)}(\infty)={\omega^2D_{11}^{(s)}+(2\lambda(\lambda-\mu)+
\omega^2)D_{22}^{(s)}-2\omega(\lambda-\mu)D_{12}^{(s)}\over 4\lambda(\lambda^2+
\omega^2-\mu^2)},$$
$$\sigma_{12}^{(s)}(\infty)={-\omega(\lambda+\mu)D_{11}^{(s)}+\omega(\lambda-
\mu)D_{22}^{(s)}+2(\lambda^2-\mu^2)D_{12}^{(s)}\over 4\lambda(\lambda^2+\omega
^2-\mu^2)}.$$
The explicit matrix elements $\sigma_{ij}^{(s)}$ for the three representations
$P,Q$ and $W$ can be obtained by inserting the corresponding diffusion
coefficients.
The distribution functions (3.57) can be used to calculate the
expectation values of the coordinate and momentum and the variances by direct
integration. The following relations are noticed [48]:
$$\sigma_{ij}^W(\infty)={1\over 2}(\sigma_{ij}^P(\infty)+\sigma_{ij}^Q(\infty)
),~i,j=1,2.$$

{\bf 3.4 The master equation in the Weyl-Wigner-Moyal representation }

The Weyl-Wigner-Moyal representation is a remarkable phase-space representation
of the quantum mechanics. Roughly speaking, a phase-space representation of the
quantum mechanics is a mapping from the Hilbert space operators to the
functions
on the classical phase-space which is such that if $A$ is mapped onto $ f_A(x,
y)$ and $\rho$ is mapped onto $ f_{\rho}(x,y),$ then
$$ {\rm Tr}(\rho A)=\int_{-\infty}^{\infty} \int_{-\infty}^{\infty} f_{\rho}(
x,y)f_A(x,y) dx dy. \eqno (3.58) $$
In reality, it is not exactly so, because the Weyl mapping is a mapping from
the functions on the phase-space to the Hilbert space operators. This mapping
$W$ was defined by Weyl [70] in the following way:
$$W(f)={1 \over(2 \pi\hbar)^2}\int_{-\infty}^{\infty} \int_{-\infty}^{\infty}
(\int_{-\infty}^{\infty} \int_{-\infty}^{\infty} e^{-(i/\hbar)(x\eta-y\xi)}
f(x,y)dx dy) W(\xi,\eta)d\xi d\eta. $$
From this it follows very formally that for any $ \rho \in {\cal D}({\cal H}),$
$${\rm Tr}(\rho W(f))={1 \over(2 \pi\hbar)^2}\int_{-\infty}^{\infty} \int_
{-\infty}^{\infty}f(x,y)(\int_{-\infty}^{\infty} \int_{-\infty}^{\infty}
e^{-(i/ \hbar)(x\eta-y\xi)} {\rm Tr}(\rho W(\xi,\eta))d\xi d\eta)dx dy
\eqno (3.59) $$
and that (3.59) can be put in standard form (3.58) if the following function on
the phase space is associated to any $ \rho \in {\cal D}({\cal H}),$
$$ f_{\rho}(x,y)={1 \over(2 \pi\hbar)^2}\int_{-\infty}^{\infty} \int_
{-\infty}^{\infty}e^{-(i/ \hbar)(x\eta-y\xi)} {\rm Tr}(\rho W(\xi,\eta))
d\xi d\eta. \eqno (3.60) $$
The mapping $\rho\to f_{\rho}$ defined by (3.60) is exactly the Wigner mapping
[54] which is the dual of the Weyl mapping $f \to W(f)$ (hence it can be
denoted
by $ \widetilde W$) and $ f_{\rho} = \widetilde W(\rho)$ is the Wigner function
corresponding to the quantum state $\rho\in{\cal D}({\cal H}).$ But the quantum
nature of the expectation value is not lost because $ f_{\rho}(x,y)$ is not
a probability distribution on the phase space, taking positive and negative
values.

In the following, the phase space representation of the master equation (3.3)
is obtained [17] by using the Wigner mapping (3.60). Denoting
$$f(x,y,t)=f_{\rho(t)}(x,y)=f_{\Phi_t(\rho)}(x,y), $$
it follows from the definition (3.60) that
$$ f(x,y,t)={1 \over(2 \pi\hbar)^2}\int_{-\infty}^{\infty} \int_
{-\infty}^{\infty}e^{-(i/ \hbar)(x\eta-y\xi)} {\rm Tr}(\rho(t) W(\xi,\eta))
d\xi d\eta. \eqno (3.61)$$
Then
$${\partial f(x,y,t)\over\partial t}={1\over(2\pi\hbar)^2}\int_{-\infty}^
{\infty} \int_
{-\infty}^{\infty}e^{-(i/ \hbar)(x\eta-y\xi)} {\rm Tr}(L(\rho(t)) W(\xi,\eta))
d\xi d\eta $$
and by duality
$${\partial f(x,y,t)\over\partial t}={1\over(2\pi\hbar)^2}\int_{-\infty}^
{\infty} \int_
{-\infty}^{\infty}e^{-(i/ \hbar)(x\eta-y\xi)} {\rm Tr}(\rho(t)
\widetilde L( W(\xi,\eta)))d\xi d\eta. \eqno (3.62) $$
But from (3.10) in the case $ t=0, A=W(\xi,\eta)$ and by using the well-known
identities for the Fourier transformation, Eq. (3.62) is transformed into the
following evolution equation for the Wigner function [17]:
$$ {\partial f(x,y,t) \over \partial t}=-{y \over m}{\partial f(x,y,t) \over
\partial x}+m \omega^2x{\partial f(x,y,t) \over \partial y}
 + (\lambda-\mu){\partial \over \partial x}(xf(x,y,t))$$$$+(\lambda+\mu)
{\partial \over \partial y}(yf(x,y,t))
+ D_{qq}{\partial^2 f(x,y,t) \over \partial x^2}+D_{pp}{\partial^2 f(x,y,t)
\over \partial y^2}+2D_{pq}{\partial^2 f(x,y,t) \over \partial x \partial y}.
\eqno (3.63) $$
This equation looks classical and in fact is exactly an equation of the
Fokker-Planck type. But not every function $ f(x,y,0) $ on the phase-space is
the  Wigner transform of a density operator. Hence, the quantum mechanics
appears now in the restrictions imposed by this last condition on the initial
condition $f(x,y,0)$ for Eq. (3.63). Unfortunately, these restrictions are not
known explicitly.

Because the most frequently used choice for $ f(x,y,0)$ is a Gaussian function
and because Eq. (3.63) preserves this Gaussian type, i.e., $f(x,y,t)$ is also
a Gaussian function, the differences between the quantum mechanics and
classical
mechanics are completely lost in this representation of the master equation.
This is an explanation for the frequently occurring ambiguities on this
subject in the literature.

Now from (3.61) by duality it follows that
$$ f(x,y,t)={1 \over(2 \pi\hbar)^2}\int_{-\infty}^{\infty} \int_
{-\infty}^{\infty}e^{-(i/ \hbar)(x\eta-y\xi)} {\rm Tr}(\rho(0) \widetilde \Phi_
t(W(\xi,\eta))) d\xi d\eta $$
and from the results of Subsection (3.1) it follows that
$$ f(x,y,t)={1 \over(2 \pi\hbar)^2}\int_{-\infty}^{\infty} \int_
{-\infty}^{\infty}e^{-(i/ \hbar)(x\eta-y\xi)} {\rm Tr}(\rho(0)W(\xi(t),\eta
(t)) e^{g(t))} d\xi d\eta, $$
where
$$\xi(t)=\alpha(t)\xi+\beta(t) \eta,~~\eta(t)=\gamma(t)\xi+\delta(t)\eta, $$
$$\xi(0)=\xi,~\eta(0)=\eta,~g(t)=-{1 \over \hbar^2}(A(t) \xi^2+B(t)\eta^2+2C(t)
\xi\eta). $$
If the initial state $ \rho(0)$ is a pure state corresponding to a coherent
wave function
$$ \rho(0)\phi=(\psi_{\sigma_q(0),\sigma_p(0)}, \phi)\psi_{\sigma_q(0),\sigma_
p(0)}$$
centered in $ x=\sigma_q(0), y=\sigma_p(0),$ i.e., if
$$\psi_{\sigma_q(0),\sigma_p(0)}(x)=(W(\sigma_q(0),\sigma_p(0)) \psi_0)(x), $$
where
$$\psi_0(x)=(2 \pi\sigma_{qq}(0))^{-1/4} \exp(-x^2/4 \sigma_{qq}(0)), $$
the Wigner function can be analytically calculated because the integrand is an
exponential function with the exponent having a quadratic form in $\xi$ and
$\eta.$

With the help of the Wigner function the coordinate and momentum probability
distribution are defined, respectively, by
$$ P(x,t)=\int_{-\infty}^\infty f(x,y,t)dy $$
and
$$ P(y,t)=\int_{-\infty}^\infty f(x,y,t)dx. $$
It follows
$$P(x,t)={1 \over \sqrt{2 \pi \sigma_{qq}(t)}} \exp(-{(x-\sigma_q(t))^2 \over
2\sigma_{qq}(t)}) $$
and
$$P(y,t)={1 \over \sqrt{2 \pi \sigma_{pp}(t)}} \exp(-{(y-\sigma_p(t))^2 \over
2\sigma_{pp}(t)}), $$
where, with the notations defined in Subsection (3.1)
$$\sigma_q(t)=\delta(t)\sigma_q(0)-\beta(t)\sigma_p(0), $$
$$\sigma_p(t)=\alpha(t)\sigma_p(0)-\gamma(t)\sigma_q(0), $$
$$\sigma_{qq}(t)=\delta^2(t)\sigma_{qq}(0)+\beta^2(t)\sigma_{pp}(0)+2B(t), $$
$$\sigma_{pp}(t)=\alpha^2(t)\sigma_{pp}(0)+\gamma^2(t)\sigma_{qq}(0)+2A(t) $$
and $\sigma_{pp}(0)=\hbar^2/4\sigma_{qq}(0).$
With these notations the Wigner function takes the following form [17]:
$$ f(x,y,t)={1 \over 2\pi \sqrt{\sigma_{pp}(t)\sigma_{qq}(t)-\sigma^2_{pq}(t)}}
\exp \{-{1 \over 2(\sigma_{pp}(t)\sigma_{qq}(t)-\sigma^2_{pq}(t))} $$
$$ \times (\sigma_{pp}(t)(x-\sigma_q(t))^2+\sigma_{qq}(y-\sigma_p(t))^2
-2\sigma_{pq}(t)(x-\sigma_q(t))(y-\sigma_p(t))) \}, $$
where
$$ \sigma_{pq}(t)=\gamma(t)\delta(t)\sigma_{qq}(0)+\alpha(t)\beta(t)\sigma_{pp}
(0)+2C(t). $$
Evidently
$$ \int_{-\infty}^{\infty} \int_{-\infty}^{\infty} f(x,y,t)dx dy=1.$$

{\bf 3.5 Density matrix of the damped harmonic oscillator }

In this Subsection we explore the general results that follow from the master
equation of the one-dimensional damped harmonic oscillator. Namely, we discuss
the physically relevant solutions of the master equation, by using the method
of the generating function. In particular, we provide extended solutions
(including both diagonal and off-diagonal matrix elements) for different
initial conditions [20].

The method used in the following follows closely  the procedure of Jang [21].
Let us first rewrite the master equation (3.7) for the density matrix by means
of the number representation. Specificallly, we take the matrix elements of
each term between different number states denoted by $\vert n>$, and using
$a^+\vert n>=\sqrt{n+1}\vert n+1>$ and $a\vert n>=\sqrt n\vert n-1>$, we get
$${d\rho_{mn}\over dt}=-i\omega(m-n)\rho_{mn}+\lambda\rho_{mn}-(m+n+1)D_2\rho
_{mn}$$
$$-\sqrt{m+1)n}D_1^*\rho
_{m+1,n-1}-\sqrt{m(n+1)}D_1\rho_{m-1,n+1}$$
$$+{1\over 2}\sqrt{(n+1)(n+2)}(D_1-\mu)\rho
_{m,n+2}
+{1\over 2}\sqrt{(m+1)(m+2)}(D_1^*-\mu)\rho_{m+2,n}$$
$$+{1\over 2}\sqrt{m(m-1)}(D_1+\mu)\rho
_{m-2,n}
+{1\over 2}\sqrt{(n-1)n}(D_1^*+\mu)\rho_{m,n-2}$$
$$+\sqrt{(m+1)(n+1)}(D_2+
\lambda)\rho_{m+1,n+1}+\sqrt{mn}(D_2-\lambda)\rho_{m-1,n-1}.\eqno(3.64)$$
Here, we have used the abbreviated notation
$\rho_{mn}=<m\vert\rho(t)\vert n>.$
This master equation is complicated in form and in the indices involved. It
comprises not only the density matrix in symmetrical forms, such as
$\rho_{m\pm1,n\pm1}$, but also matrix elements in asymmetrical forms like
$\rho_{m\pm2,n}, \rho_{m,n\pm2}$ and $\rho_{m\mp1,n\pm1}$.
In order to solve Eq. (3.64) we use the method of a generating function which
allows us to
eliminate the variety of indices $m$ and $n$ implicated in the equation. When
we define the double-fold generating function by
$$G(x,y,t)=\sum_{m,n}{1\over\sqrt{m!n!}}x^my^n\rho_{mn}(t),\eqno(3.65)$$
the density matrix can be evaluated from the inverse relation of Eq. (3.65):
$$\rho_{mn}(t)={1\over\sqrt{m!n!}}({\partial\over\partial x})^m({\partial
\over\partial y})^nG(x,y,t)\vert_{x=y=0},\eqno(3.66)$$
provided that the generating function is calculated beforehand. When we
multiply both sides of Eq. (3.64) by $x^my^n/\sqrt{m!n!}$ and sum over the
result, we get the following linear partial differential equation of second
order for $G(x,y,t)$:
$${\partial\over\partial t}G(x,y,t)=\{[-(i\omega+D_2)x-D_1^*y]{\partial\over
\partial x}+[-D_1x+(i\omega-D_2)y]{\partial\over\partial y}
+(D_2+\lambda){\partial^2\over\partial x\partial y}$$$$+{1\over 2}[(D_1^*-\mu)
{\partial^2\over\partial x^2}+(D_1-\mu){\partial^2\over\partial y^2}]
+[{1\over 2}(D_1+\mu)x^2+{1\over 2}(D_1^*+\mu)y^2+(D_2-\lambda)(xy-1)]\}
G(x,y,t).\eqno(3.67)$$
The form of the solution of the Fokker-Planck equations treated in [17] and
[21] suggest us to look for the solution of Eq. (3.67) of the form
$$G(x,y,t)={1\over A}\exp\{xy-[B(x-C)^2+D(y-E)^2+F(x-C)(y-E)]/H\},\eqno(3.68)$$
where $A,B,C,D,E,F$ and $H$ are unknown functions of time which have to be
determined. When we first substitute the expression (3.68) for $G(x,y,t)$ into
Eq. (3.67) and  equate the coefficients of equal powers of $x,y$ and $xy$ on
both sides of the equation, we get the following differential equations
for the functions $A,B,D,F$ and $H$:
$$-{1\over A}{dA\over dt}=-(D_1^*-\mu){B\over H}-(D_1-\mu){D\over H}-(D_2+
\lambda){F\over H}+2\lambda,\eqno(3.69)$$
$${d\over dt}({B\over H})=2(\lambda-i\omega){B\over H}-\mu{F\over H}-{1
\over 2}(D_1-\mu){F^2\over H^2}-2(D_2+\lambda){FB\over H^2}-2(D_1^*-\mu){B^2
\over H^2},\eqno(3.70)$$
$${d\over dt}({D\over H})=2(\lambda+i\omega){D\over H}-\mu{F\over H}-{1\over 2}
(D_1^*-\mu){F^2\over H^2}-2(D_2+\lambda){DF\over H^2}-2(D_1-\mu){D^2\over H^2},
\eqno(3.71)$$
$${d\over dt}({F\over H})=2\lambda{F\over H}-(D_2+\lambda){F^2\over H^2}-
2\mu({B\over H}+{D\over H})-4(D_2+\lambda){DB\over H^2}-2(D_1^*-\mu){BF\over
H^2}-2(D_1-\mu){DF\over H^2}.\eqno(3.72)$$
In addition to these equations, we get for the functions $C$ and $E$
$$2B{dC\over dt}+F{dE\over dt}=(-2(\lambda-i\omega)B+\mu F)C+(2\mu B-(\lambda+i
\omega)F)E,\eqno(3.73)$$
$$2D{dE\over dt}+F{dC\over dt}=(-2(\lambda+i\omega)D+\mu F)E+(2\mu D-(\lambda-
i\omega)F)C.\eqno(3.74)$$
The equations (3.73) and (3.74) can be reformulated in order to eliminate the
functions $B,D$ and $F$, provided $BD-F^2/4\not=0$. We obtain
$${dC\over dt}=-(\lambda-i\omega)C+\mu E,\eqno(3.75)$$
$${dE\over dt}=-(\lambda+i\omega)E+\mu C.\eqno(3.76)$$
The functions $A,B,D,F$ and $H$ are connected by the auxiliary condition that
${\rm Tr}\rho$ is independent of time. The trace of $\rho$ can be evaluated by
summing the diagonal matrix elements $\rho_{nn}$ given in Eq.(3.66) or
directly by using the integral expression
$${\rm Tr}\rho=\sum_{n=0}^\infty \rho_{nn}={1\over(2\pi)^2}\int\exp(-k_1k_2)
\exp  (ik_1x+ik_2y)G(x,y,t)dk_1dk_2dxdy.$$
We obtain with the generating function (3.68)
$${\rm Tr}\rho=({4A^2\over H^2}({F^2\over 4}-BD))^{-1/2}.\eqno(3.77)$$
This quantity is time-independent which can be verified by constructing an
equation satisfied by the quantity $(F^2/4-BD)/H^2$. Combining Eqs.
(3.70)--(3.72) we get
$${d\over dt}({F^2/4-BD\over H^2})=2[2\lambda-(D_1^*-\mu){B\over H}-
(D_1-\mu){D\over H}-(D_2+\lambda){F\over H}]({F^2/4-BD\over H^2}).$$
We see immediately that the first factor on the right-hand side of this
equation is identical with the right-hand side of Eq. (3.69). Accordingly, we
find
$${d\over dt}(({F^2\over 4}-BD){A^2\over H^2})=0.$$
Since the scaling function $H$ is arbitrary, we simplify the following
equations by the choice
$${F^2\over 4}-BD=-H.\eqno (3.78)$$
Setting ${\rm Tr}\rho=1$, we obtain from Eqs. (3.77) and (3.78) the
normalization constant $A^2=-H/4$.
As a consequence, we can simplify Eqs. (3.70)--(3.72) by eliminating the
function $H$ from these equations. The resulting three equations are
$${dB\over dt}=-2(\lambda+i\omega)B-\mu F+2(D_1-\mu),$$
$${dD\over dt}=-2(\lambda-i\omega)D-\mu F+2(D_1^*-\mu),\eqno(3.79)$$
$${dF\over dt}=-2\mu(B+D)-2\lambda F-4(D_2+\lambda).$$
These equations imply that the function $D$ is complex conjugate to $B$,
provided that the function $F$ is real.

In order to integrate the equations for the time-dependent functions $B,C,D,E$
and $F$ we start with Eqs. (3.75) and (3.76). These equations imply that the
function $E$ is complex conjugate to the function $C$. By solving the coupled
equations we find:
$$C(t)=E^*(t)=u(t)C(0)-v(t)C^*(0),\eqno(3.80)$$
where $u(t)$ and $v(t)$ are given by (3.45) and (3.46) for the two considered
cases: overdamped and underdamped, respectively.
For integrating the system (3.79) we proceed in the same way as for integrating
the system (3.14). With the assumption that $F$ is real and
$$D(t)=B^*(t)=R(t)+iI(t),$$
we obtain explicitly:
$$R(t)={1\over 2}(e^{-2\mu_+t}+e^{-2\mu_-t})\widetilde R+{1\over 2\gamma}
(e^{-2\mu_+t}-e^{-2\mu_-t})(\omega\widetilde I+{\mu\over 2}\widetilde F)+
R(\infty),$$
$$I(t)=e^{-2\lambda t}{\mu\over\gamma^2}(\mu\widetilde I+{\omega\over 2}
\widetilde F)$$
$$-{\omega\over 2\gamma^2}(e^{-2\mu_+t}+e^{-2\mu_-t})(\omega\widetilde I+
{\mu\over 2}\widetilde F)-{\omega\over 2\gamma}(e^{-2\mu_+t}-e^{-2\mu_-t})
\widetilde R+I(\infty),$$
$$F(t)=-e^{-2\lambda t}{\omega\over\gamma^2}(2\mu\widetilde I+\omega
\widetilde F)$$
$$+{\mu\over\gamma^2} (e^{-2\mu_+t}+e^{-2\mu_-t})(\omega\widetilde I+{\mu
\over 2}\widetilde F)+{\mu\over\gamma}(e^{-2\mu_+t}-e^{-2\mu_-t})
\widetilde R+F(\infty),$$
where we used the notations:
$$\mu_\pm=\lambda\pm\gamma,~\gamma\equiv\sqrt{\mu^2-\omega^2},~
\widetilde R=R(0)-R(\infty),~\widetilde I=I(0)-I(\infty),~\widetilde F=F(0)-
F(\infty).$$
We can also obtain the connection between the asymptotic values of $B(t),D(t),
F(t)$ and the coefficients $D_1, D_2, \mu$ and $\lambda$:
$$R(\infty)={\rm Re}D(\infty)={\lambda({\rm Re}D_1-\mu)+\omega {\rm Im}D_1+\mu
(D_2+\lambda)\over\lambda^2-\gamma^2},$$
$$I(\infty)={\rm Im}D(\infty)={\omega\lambda({\rm Re}D_1-\mu)+(\mu^2-\lambda^2)
{\rm Im}D_1+\omega\mu(D_2+\lambda)\over\lambda(\lambda^2-\gamma^2)},$$
$$F(\infty)=-2{\mu[\lambda({\rm Re}D_1-\mu)+\omega {\rm Im}D_1]+(\lambda^2+
\omega^2)(D_2+\lambda)\over\lambda(\lambda^2-\gamma^2)}.$$
When all explicit expressions for $A,B,C,D,E,F$ and $H$ are introduced into
Eq. (3.68), we obtain an analytical form of the generating function $G(x,y,t)$
which allows us to evaluate the density matrix.

If the constants involved in the generating function satisfy the relations
$$C(0)=0,~R(0)=R(\infty),~I(0)=I(\infty),~F(0)=F(\infty),$$
we obtain the stationary solution
$$C(t)=E(t)=0,~R(t)=R(0),~I(t)=I(0),~F(t)=F(0)$$
and
$$D(t)=B^*(t)=R(0)+iI(0),~
H(t)=-4A^2(t)=R^2(0)+I^2(0)-F^2(0)/4.$$
Then the stationary solution of Eq. (3.67) is
$$G(x,y,t)={1\over A}\exp\{(1-{F\over H})xy-{1\over H}(Bx^2+B^*y^2)\}.
\eqno(3.81)$$
In addition, for a thermal bath [17] with
$$m\omega D_{qq}={D_{pp}\over m\omega},~D_{pq}=0,~\mu=0,$$
the stationary generating function is simply given by
$$G(x,y)={2\lambda\over D_2+\lambda}\exp({D_2-\lambda\over D_2+\lambda}xy).$$
The same generating function can be found for large times, if the asymptotic
state is a Gibbs state with $\mu=0$. In this case we obtain with Eq. (3.26) and
$\mu=0$
$$D_2=\lambda\coth{\hbar\omega\over 2kT}$$
and $$G(x,y)=(1-\exp(-{\hbar\omega\over kT}))\exp(\exp(-{\hbar\omega\over kT})
xy).$$
The density matrix can be calculated with Eq. (3.66) and yields the
Bose-Einstein distribution
$$\rho_{mn}(\infty)=(1-\exp(-{\hbar\omega\over kT}))\exp(-{n\hbar\omega\over
kT})\delta_{mn}.$$
  A formula for the density matrix can be written down by applying the relation
(3.66) to the generating function (3.68). We get
$$\rho_{mn}(t)={\sqrt{m!n!}\over A}\exp[-(BC^2+DE^2+FCE)/H]$$
$$\times\sum_{n_1,n_2,n_3=0}{(1-{F\over H})^{n_3}(-{B\over H})^{n_1}(-{D\over
H})^
{n_2}({2BC\over H}+{FE\over H})^{m-2n_1-n_3}({2DE\over H}+{FC\over H})^{n-2n_2
-n_3}\over n_1!n_2!n_3!(m-2n_1-n_3)!(n-2n_2-n_3)!}.\eqno(3.82)$$
In the case that the functions $C(t)$ and $E(t)$ vanish, the generating
function has the form of Eq. (3.81). Then the elements of the density matrix
with an odd sum $m+n$ are zero: $\rho_{mn}=0$ for $m+n=2k+1$ with $k=0,1,2,...$
The lowest non-vanishing elements are given with $\rho_{mn}=\rho_{nm}$ as
$$\rho_{00}={1\over A},~\rho_{20}=-{\sqrt 2 B\over AH},~\rho_{11}={1\over A}
(1-{F\over H}),$$
$$\rho_{22}={2BB^*\over AH^2}+{1\over A}(-{F\over H})^2,~\rho_{31}=-(1-{F
\over H}){\sqrt 6B\over AH},
~\rho_{40}={\sqrt 6B^2\over AH^2}.$$
It is also possible to choose the constants in such a way that the functions
$B$ and $D$ vanish at time $t=0$ and $F(0)=H(0)$. Then the density matrix
(3.82) becomes at $t=0 (E=C^*)$:
$$\rho_{mn}(0)={1\over\sqrt{m!n!}}(C^*(0))^m(C(0))^n\exp(-\vert C(0)
\vert^2).\eqno(3.83)$$
This is the initial Glauber
packet. The diagonal matrix elements of Eq. (3.83) represent a Poisson
distribution used also in the study of multi-phonon excitations in nuclear
physics. In the particular case when we assume
$$D_1=\mu=0,~D_2=\lambda,~
B(0)=D(0)=0,~F(0)=H(0)=-4,$$
the differential equations (3.79) yield
$B(t)=D(t)=0$ and $F(t)=H(t)=-4$.
Then the density matrix subject to the initial Glauber packet is (see also
[21])
$$\rho_{mn}(t)={1\over\sqrt{m!n!}}(C^*(t))^m(C(t))^n\exp(-\vert C(t)
\vert^2),$$
where $C(t)$ is given by Eq. (3.80).

{\bf 4. Applications to nuclear equilibration processes}

{\bf 4.1 Charge equilibration in deep inelastic reactions }

Over many years experimental data have been measured in the field of deep
inelastic heavy ion collisions [71,72]. The characteristic feature of these
collisions is the binary character of the system, i.e. the final fragments
have nearly the same masses as the initial nuclei. Deep inelastic collisions
are mainly described by the dynamics of selected collective degrees of freedom.
These are the relative motion of the nuclei, mass and charge exchange, the neck
degree of freedom and surface vibrations of the fragments [73]. Another
important
feature of these reactions is the dissipation of energy and angular momentum
out of the collective degrees of freedom into the intrinsic or single-particle
degrees of freedom.

In most of the approaches the collective degrees of freedom are chosen in an
{\it a priori } way, guided by measured macroscopic variables which do not
cover
the complete space associated with all of them. Also, intrinsic degrees of
freedom play a role. Thus, by taking into account the coupling between the
actually treated and intrinsic degrees of freedom and other collective degrees
of freedom, irreversible processes occur in the dynamics of the actually
treated
collective variables.

One of the most characteristic processes in deep inelastic reactions is the
fast
redistribution of neutrons and protons among the colliding nuclei in the early
stage of the reaction called charge equilibration, neutron excess mode or
isospin
relaxation. The conditional variance of the charge distribution at fixed mass
asymmetry has been largely used as a sensitive quantity to investigate the
nature
of the charge equilibration process.

There are two opposite approaches which have been developed in order to
describe
the charge equilibration mode, namely the quantum mechanical collective
treatment
and transport theories. The collective treatment stresses the coherent
properties
of the process and considers this mode as connected to the isovector dipole
giant
resonance of the composite system. An experimental indication for a quantum
process is that the charge variance reaches a saturation value at large
values of
the total kinetic energy loss. The other opposite view relies on stochastic
exchange processes between the colliding heavy nuclei.

In the following the charge equilibration mode is treated as a one-dimensional
collective mode in the charge asymmetry degree of freedom. The damping of this
collective mode is due to the coupling to the remaining collective and
intrinsic
degrees of freedom.

One method of introducing dissipation in a quantum mechanical description of
deep inelastic collisions is to assume that the energy dissipation is similar
to the loss
of energy of a harmonic oscillator coupled with a large number of other
harmonic
oscillators. This mechanism can be simulated by a friction term of the Kostin
type in the Schr\"odinger equation. According to this line, the role of
collective
motion and quantum fluctuations for charge and mass equilibration in deep
inelastic
reactions has been investigated [74,75].

Another method of treating dissipation in quantum mechanics,
especially
for the description of coupled collective modes, is the axiomatic method of
Lindblad. In this method, as we said before, a simple dynamics for the
subsystem of
the explicitly treated collective degrees of freedom is chosen, namely a
semigroup
of transformations which introduces a preferred direction in time and,
therefore,
can describe an irreversible process.

In the present Subsection, Lindblad's theory for open quantum systems is
applied
to the problem of the damping of a single collective degree of freedom in deep
inelastic collisions. We show the description of the charge equilibration
mode by a damped harmonic oscillator in the framework of the Lindblad theory
[17,18,22].

For a comparison of the theoretical results with experimental data we consider
the charge equilibration in deep inelastic collisions of heavy ions. We can
take
over the solutions for the centroids and variances of a damped oscillator given
by (3.15), (3.16), (3.19), (3.27), (3.28) for a comparison in order to
illustrate
our results. In [22] the charge equilibration of the systems
$^{56}{\rm Fe}+^{209}{\rm Bi}, ^{56}{\rm Fe}+^{238}{\rm U}$ and $ ^{98}{\rm Mo}
+^{98}{\rm Mo} $
was analysed within a collective
description of the charge asymmetry degree of freedom. The charge asymmetry
coordinate is defined as
$$ \xi = {Z_1-Z_2 \over Z_1+Z_2}$$
where $Z_1$ and $Z_2$ are the charge numbers of the fragments. The Hamiltonian
used for the description of this degree of freedom is a harmonic oscillator
Hamiltonian of the form
$$ H_0 = {P^2_\xi \over 2B}+{1 \over 2}K(\xi-\xi_0)^2.$$
The stiffness parameter $K$ and mass $B$ were calculated in [22] on the basis
of the liquid drop model for the potential energy and by use of a semiempirical
formula for the frequency $\omega=(K/B)^{1/2}.$ For the system $ ^{56}{\rm Fe}
+^{209}{\rm Bi}$ the following values were obtained [22]: $ B=1.6127 \times 10^
{-40}$ MeV
${\rm s}^2$ and $K=1.6363 \times 10^4$ MeV, corresponding to an oscillator
energy $ \hbar
\omega = 6.63$ MeV. Figs. 1--3 show the charge centroids and charge asymmetry
variances as functions of time for the reaction of $^{56}{\rm Fe} $ on $ ^{209}
{\rm Bi}$ at
$ E_{lab} = 465$ MeV. The full dots are the experimental data of Breuer et al.
[76]. Since the experimental data were obtained as functions of the total
energy
loss, Pop et al. [22] have related the data with the reaction times in order
to
compare them with calculations. As a model they used a parametrization of the
deflection function in terms of relaxation times for the dissipation of the
radial kinetic energy and relative angular momentum and for the development of
deformations [77-79]. These times were fixed by reproducing the ridge of the
double-differential cross-section $d^2 \sigma/d\Omega dE$ plotted as a contour
diagram in the total kinetic energy against scattering angle (Wilczynski plot).
For short reaction times the kinetic energy loss is proportional to the
reaction
time ($ \sim 30$ Mev/$10^{-22}$s). Our comparison does not depend much on the
specific assumptions of the model used to derive the reaction times from the
measured total energy losses.

The "experimental" values of the variance in Fig. 3 are constructed from the
experimental values of $ \sigma_{\xi\xi}$ by using the equation (see Eqs.
(3.14)) [22]
$${d\sigma_{\xi\xi} \over dt}=-2(\Lambda-\mu)\sigma_{\xi\xi}+{2 \over B}\sigma_
{\xi p_{\xi}}+2D_{\xi\xi}$$
and applying the corresponding fitted values for the parameters $ \Lambda, \mu$
and $D_{\xi\xi}.$ The curves are calculated with the parameters $ \xi_0,
\Lambda$
and $\mu$ fitted to the experimental data in Fig. 1 and in addition with
$\omega_1=-2m\omega D_{qq}/\hbar, \omega_2=-2D_{pp}/m\hbar\omega$ and $\omega
_3=2D_{qp}/\hbar$ fitted to the experimental variances in Figs. 2 and 3 (here
$q=\xi, p=p_{\xi}, m=B).$ The set of parameters $\omega_i$ is restricted by the
condition (3.4) which now reads as
$$\omega_1\omega_2-\omega^2_3 \ge \Lambda^2.$$
The full curves in Figs.1--3 represent the overdamped solution $(\omega=
1.0073\times 10^{22}s^{-1} < \mu=2.33 \times 10^{22} s^{-1})$ for the centroids
and variances. This solution is the only one for which a good description of
all
the experimental data could be obtained. In the case of the underdamped
solution
no set of parameters could be found fitting simultaneously the data for the
centroids and variances. Similar results were also obtained for the other
considered
systems $ ^{56}{\rm Fe}+^{238}{\rm U}$ and $^{98}{\rm Mo}+^{98}{\rm Mo}$ in
[22].

The parameters of the theory are accepted as free quantities, fitted to the
experimental
data of the charge distributions. The overdamped solution with imposed
fundamental
constraints for the damped quantum oscillator succeeded to describe the
experimental data within quite restricted ranges of the model parameters. Both
the underdamped and the equilibrium solutions can not describe the data for
the charge centroids and the isobaric variances.

The given treatment has the advantage that analytical solutions can be
obtained
for the damping of a harmonic oscillator. This is in contrast to the method
when one introduces a friction term of the Kostin type into the Schr\"odinger
equation and solves the resulting equation numerically. However, future
investigations have to be carried out to interpret the fitted parameters of
the theory from a microscopical point of view. For example, the parameters
can be related to those obtained from a microscopical derivation of
Fokker-Planck equations.

{\bf 4.2 Charge and neutron equilibration and the open quantum system of two
harmonic oscillators}

In the present Subsection we extend the work of Gupta et al. [80] on the
dynamics of the charge equilibration process in deep inelastic collisions and
treat the damping of the proton and neutron asymmetry degrees of freedom with
the method of Lindblad [23]. The charge and mass distribution in di-nuclear
systems
can be described with continuous coordinates of the charge and neutron
asymmetries
defined by
$$ \eta_Z = {Z_1-Z_2 \over Z_1+Z_2},~~\eta_N = {N_1-N_2 \over N_1+N_2}.$$
Here, $ Z_1(N_1)$ and $Z_2(N_2)$ are the total charge numbers (neutron numbers)
on the left-hand and right-hand side of a plane through the neck of the
di-nuclear
system [73]. Without damping, the charge and neutron asymmetry degrees of
freedom
are described by a wavefunction $\psi (\eta_Z,\eta_N,t),$ which is the solution
of the following time-dependent Schr\"odinger equation:
$$H(\eta_Z,\eta_N)\psi(\eta_Z,\eta_N,t)=i\hbar \partial \psi(\eta_Z,\eta_N,t)/
\partial t, $$
where the Hamiltonian of the model is given as $ (\vert k_{ZN} \vert \le (k_Z
k_N)^{1/2}):$
$$ H(\eta_Z,\eta_N)=-{\hbar \over 2B_{ZZ}}{\partial^2 \over \partial \eta^2_Z}-
{\hbar^2 \over 2B_{NN}}{\partial^2 \over\partial \eta^2_N}+{1 \over2}k_Z \eta^2
_Z+{1 \over 2}k_N \eta^2_N-k_{ZN} \eta_Z \eta_N. \eqno (4.1) $$
The Hamiltonian has the simple structure of two coupled oscillators in the
coordinates $\eta_Z$ and $\eta_N$ in order to keep the time development of the
wavefunction analytically solvable. The string constants of the potential can
be
calculated with the liquid drop model for the sticking configuration of the
nuclei, i.e. for the relative distance $ R=R_1+R_2,$ where $R_1$ and $R_2$ are
the radii of the two colliding nuclei (for more details see [80]).

Hofmann et al. [39] have considered the coupling of the mass asymmetry
coordinate
to the charge asymmetry coordinate on the basis of a density operator formalism
using a quantum master equation in a perturbative treatment. The coupling of
the
neutron and charge asymmetry coordinates has also been studied in the framework
of a two-dimensional Fokker-Planck equation by Gross and Hartmann [81],
Schr\"oder et al. [82], Birkelund et al. [83] and Merchant and N\"orenberg
[84].

With the Hamiltonian (4.1) as an example, we study the damping of two coupled
oscillators in the framework of the Lindblad theory. In order to have a
formalism
which is generally applicable, we give the following formulae in terms of the
two
general coordinates $q_1$ and $q_2$ instead of $\eta_Z$ and $ \eta_N.$ We
present
the equation of motion of the open quantum system of two oscillators in the
Heisenberg picture. With this equation we derive the time dependence of the
expectation values of the coordinates and momenta and their variances. The
connection with the Wigner function and Weyl operator is also discussed.
Finally,
we demonstrate the time dependence of the various quantities for a simplified
version of the model, where the decay costants can be calculated analytically.

If $ \widetilde \Phi_t $ is the dynamical semigroup describing the time
evolution of the open quantum system in the Heisenberg picture, then the master
equation is given for an operator A as follows [14,16,17]:
$${d\widetilde\Phi_t(A)\over dt}=\widetilde L(\widetilde\Phi_t(A))={i\over
\hbar}[H,\widetilde\Phi_t(A)]+{1\over 2\hbar}\sum_j(V_j^+[\widetilde\Phi_t(A),
V_j]+[V_j^+,\widetilde\Phi_t(A)]V_j).\eqno(4.2)$$

The operators H, $ V_j , V^+_j (j = 1,2,3,4)$ are taken to be functions of the
basic observables of the two quantum oscillators.
The coordinates $q_k$ and the momenta $p_k$ obey the usual commutation
relations $(k,l=1,2)$:
$$[q_k,p_l]=i\hbar\delta_{kl},~~[q_k,q_l]=[p_k,p_l]=0.$$
In order to obtain an analytically solvable model, $H$ is taken to be a
polynomial
of second degree in these basic observables and $V_j,V_j^+$ are
taken to be polynomials of only first degree.
Then in the linear space spanned by $q_k,p_k (k=1,2)$, there exist four
linearly independent operators $V_{j=1,2,3,4}$:
$$V_j=\sum_{k=1}^2 a_{jk}p_k+\sum_{k=1}^2 b_{jk}q_k,$$
where $a_{jk},b_{jk}\in {\bf C}$ with $j=1,2,3,4$.
It yields
$$V_j^+=\sum_{k=1}^2 a_{jk}^*p_k+\sum_{k=1}^2 b_{jk}^*q_k,$$
where $a_{jk}^*,b_{jk}^*$ are the complex conjugates of $a_{jk},b_{jk}$.

The Hamiltonian $H$ is chosen in the form of two coupled oscillators
$$H=\sum_{k=1}^2({1\over 2m_k}p_k^2+{m_k\omega_k^2\over 2}q_k^2)
+ k_{12} p_1 p_2 + {1 \over 2} \sum^2_{k_1,k_2=1} \mu_{k_1,k_2}
(p_{k_1} q_{k_2} + q_{k_2} p_{k_1} ) + \nu_{12} q_1 q_2.$$
Here we use the following abbreviations $(k = 1,2):$
$$D_{q_kq_l}=D_{q_lq_k}={\hbar\over 2}{\rm Re}({\bf a}_k^*{\bf a}_l),~
D_{p_kp_l}=D_{p_lp_k}={\hbar\over 2}{\rm Re}({\bf b}_k^*{\bf b}_l),~
D_{q_kp_l}=D_{p_lq_k}=-{\hbar\over 2}{\rm Re}({\bf a}_k^*{\bf b}_l),$$
$$\alpha_{12}=-\alpha_{21}=-{\rm Im}({\bf a}_1^*{\bf a}_2),~
\beta_{12}=-\beta_{21}=-{\rm Im}({\bf b}_1^*{\bf b}_2),~
\lambda_{kl}=-{\rm Im}({\bf a}_k^*{\bf b}_l).\eqno(4.3)$$
The scalar products are formed with the vectors ${\bf a}_k,{\bf b}_k$ and their
complex conjugates ${\bf a}_k^*,{\bf b}_k^*$. The vectors have the components
$${\bf a}_k=(a_{1k},a_{2k},a_{3k},a_{4k}),~
{\bf b}_k=(b_{1k},b_{2k},b_{3k},b_{4k}).$$
Now, as a consequence of the definitions (4.3) of the phenomenological
constants
which appear in $\widetilde L(A)$ and of the positivity of the matrix formed by
the four vectors $ {\bf a}_1, {\bf a}_2, {\bf b}_1, {\bf b}_2,$ it follows that
the principal minors of this matrix are positive or zero. This matrix is given
by:
$${1 \over 2} \hbar \left( \matrix{ {\bf a}^*_1 {\bf a}_1&{\bf a}^*_1 {\bf a}_2
&{\bf a}^*_1 {\bf b}_1&{\bf a}^*_1 {\bf b}_2 \cr
{\bf a}^*_2 {\bf a}_1&{\bf a}^*_2 {\bf a}_2
&{\bf a}^*_2 {\bf b}_1&{\bf a}^*_2 {\bf b}_2 \cr
{\bf b}^*_1 {\bf a}_1&{\bf b}^*_1 {\bf a}_2
&{\bf b}^*_1 {\bf b}_1&{\bf b}^*_1 {\bf b}_2 \cr
{\bf b}^*_2 {\bf a}_1&{\bf b}^*_2 {\bf a}_2
&{\bf b}^*_2 {\bf b}_1&{\bf b}^*_2 {\bf b}_2 \cr} \right) $$
$$ = \left( \matrix{ D_{q_1 q_1}&D_{q_1 q_2} - i \hbar\alpha_{12}/2&
- D_{q_1 p_1} - i \hbar \lambda_{11}/2&- D_{q_1 p_2} -i\hbar\lambda_{12}/2 \cr
D_{q_2 q_1} - i \hbar \alpha_{21}/2&D_{q_2 q_2}&- D_{q_2 p_1} -
i \hbar \lambda_{21}/2&- D_{q_2 p_2} - i \hbar \lambda_{22}/2 \cr
- D_{p_1 q_1} + i \hbar \lambda_{11}/2&- D_{p_1 q_2} + i \hbar
\lambda_{21}/2&D_{p_1 p_1}&D_{p_1 p_2} - i \hbar  \beta_{12}/2 \cr
- D_{p_2 q_1} + i \hbar \lambda_{12}/2&- D_{p_2 q_2} + i \hbar
\lambda_{22}/2&D_{p_2 p_1} - i \hbar \beta_{21}/2&D_{p_2 p_
2} \cr} \right). \eqno (4.4) $$
For example, we can write one of the conditions obtained from the positivity
of (4.4):
$$ D_{q_1 q_1} D_{q_2 q_2} - D^2_{q_1 q_2} \ge {1 \over 4} \hbar^2 \alpha^2_
{12}. $$
This inequality and the corresponding ones derived from (4.4) are constraints
imposed on the phenomenological constants by the fact that $ \widetilde \Phi_t
$ is a dynamical semigroup [14,16,17].

The time-dependent expectation values of self-adjoint operators $A$ and $B$
can be written with the density operator $\rho$, describing the initial state
of the quantum system, as follows:
$$m_A(t)={\rm Tr}(\rho\widetilde\Phi_t(A)),~
\sigma_{AB}(t)={1\over 2}{\rm Tr}(\rho\widetilde\Phi_t(AB+BA)).$$

In the following we denote the vector with the four components $m_{q_i}(t),
m_{p_i}(t), i=1,2$, by ${\bf m}(t)$ and the following $4\times 4$ matrix by
$\sigma(t)$:
$$\sigma(t)=\left(\matrix{\sigma_{q_1q_1}&\sigma_{q_1q_2}&\sigma_{q_1p_1}&
\sigma_{q_1p_2}\cr
\sigma_{q_2q_1}&\sigma_{q_2q_2}&\sigma_{q_2p_1}&\sigma_{q_2p_2}\cr
\sigma_{p_1q_1}&\sigma_{p_1q_2}&\sigma_{p_1p_1}&\sigma_{p_1p_2}\cr
\sigma_{p_2q_1}&\sigma_{p_2q_2}&\sigma_{p_2p_1}&\sigma_{p_2p_2}\cr}\right).$$
Then via direct calculation of $\widetilde L(q_k)$ and $\widetilde L(p_k)$ we
obtain
$${d{\bf m}\over dt}= Y{\bf m},\eqno(4.5)$$
where
$$ Y=\left(\matrix{-\lambda_{11} + \mu_{11}&-\lambda_{12} + \mu_{12}&1/m_1
&-\alpha_{12}+k_{12}\cr
-\lambda_{21} + \mu_{21}&-\lambda_{22} + \mu_{22}&\alpha_{12} + k_{12}&
1/m_2 \cr
-m_1\omega_1^2&\beta_{12} - \nu_{12}&-\lambda_{11} - \mu_{11}&-\lambda_{21}-
\mu_{21} \cr
- \beta_{12} - \nu_{12}&-m_2\omega_2^2&-\lambda_{12} - \mu_{12}&-\lambda_{22}
- \mu_{22}\cr}\right).\eqno(4.6)$$
From (4.5) it follows that
$${\bf m}(t) = M(t){\bf m}(0) = \exp(tY){\bf m}(0),\eqno(4.7)$$
where ${\bf m}(0)$ is given by the initial conditions. The matrix $M(t)$ has
to fulfil the condition:
$$\lim_{t\to\infty} M(t)=0.\eqno(4.8)$$
In order that this limit exists, $Y$ must have only eigenvalues with
negative real parts.

By direct calculation of $\widetilde L(q_kq_l),\widetilde L(p_kp_l)$ and
$\widetilde L(q_kp_l+p_lq_k),(k,l=1,2),$ we obtain
$${d \sigma\over dt} = Y \sigma + \sigma Y^T+2 D,\eqno(4.9)$$
where $D$ is the matrix of the diffusion coefficients
$$ D=\left(\matrix{D_{q_1q_1}&D_{q_1q_2}&D_{q_1p_1}&D_{q_1p_2}\cr
D_{q_2q_1}&D_{q_2q_2}&D_{q_2p_1}&D_{q_2p_2}\cr
D_{p_1q_1}&D_{p_1q_2}&D_{p_1p_1}&D_{p_1p_2}\cr
D_{p_2q_1}&D_{p_2q_2}&D_{p_2p_1}&D_{p_2p_2}\cr}\right)$$
and $Y^T$ the transposed matrix of $Y$. The time-dependent
solution of (4.9) can be written as
$$ \sigma(t)= M(t)( \sigma(0)- \Sigma) M^T(t)+\Sigma,\eqno(4.10)$$
where $ M(t)$ is defined in (4.7). The matrix $\Sigma$ is time
independent and solves the static problem (4.9) $(d\sigma/dt=0)$:
$$ Y \Sigma+\Sigma Y^T+2 D=0.\eqno(4.11)$$
Now we assume that the following limit exists for $t\to\infty$:
$$\sigma(\infty)=\lim_{t\to\infty}\sigma(t).\eqno(4.12)$$
In this case it follows from (4.10) and (4.8):
$$\sigma(\infty)=\Sigma.\eqno(4.13)$$
Inserting (4.13) into (4.10) we obtain the basic equations for our purposes:
$$\sigma(t)= M(t)(\sigma(0)-\sigma(\infty)) M^T(t)+\sigma(\infty),\eqno(4.14)$$
where
$$Y\sigma(\infty)+\sigma(\infty) Y^T=-2 D.\eqno(4.15)$$
Now we want to discuss the time dependence of the Wigner function. This
function is defined as:
$$ f(x_1,x_2,y_1,y_2,t) = {1 \over (2 \pi \hbar)^4} \int_{-\infty}^\infty
\int_{-\infty}^\infty \int_{-\infty}^\infty \int_{-\infty}^\infty \exp
\left(-{i \over \hbar}(x_1\eta_1 + x_2\eta_2 - y_1\xi_1 - y_2\xi_2)\right)$$
$$\times {\rm Tr}[\rho \widetilde \Phi_t(W(\xi_1,\xi_2;\eta_1,\eta_2))]
d\xi_1d\xi_2d\eta_1d\eta_2, \eqno (4.16)$$
where the Weyl operator $W$ is defined by $(\xi_1,\xi_2,\eta_1,\eta_2$ real)
$$W(\xi_1,\xi_2;\eta_1,\eta_2) = \exp[i \hbar^{-1}(\eta_1q_1 + \eta_2q_2 -
\xi_1p_1 - \xi_2p_2)].$$
Using the method developed by Lindblad [14,16,17] for the one-dimensional case,
we find the following relation for the time development of the Weyl operator:
$$\widetilde \Phi_t(W(\xi_1,\xi_2;\eta_1,\eta_2)) = W(\xi_1(t),\xi_2(t);
\eta_1(t),\eta_2(t)) \exp g(t). \eqno (4.17)$$
The real functions $ {\bf \xi}(t) = (\xi_1(t),\xi_2(t);\eta_1(t),\eta_2(t))$
and $ g(t) $ satisfy the equations of motion:
$$ d{\bf \xi}(t)/dt = JY^TJ^{-1}{\bf \xi}(t) \eqno (4.18)$$
$$ dg(t)/dt = - \hbar^{-2} {\bf \xi}(t)JDJ^{-1}{\bf \xi}(t), \eqno (4.19)$$
where
$$ J = \left(\matrix{0&0&-1&0 \cr
0&0&0&-1 \cr
1&0&0&0 \cr
0&1&0&0 \cr }\right).$$
Eqs. (4.18) and (4.19) are obtained by inserting the Weyl operator
$W(\xi_1,\xi_2;\eta_1,\eta_2)$ into the equation of motion (4.2). The initial
conditions for the coordinates $\xi_1(t),\xi_2(t),$ $\eta_1(t)$ and $\eta_2(t)$
are determined by $ \xi_1(0) = \xi_1, \xi_2(0) = \xi_2, \eta_1(0) = \eta_1$
and $ \eta_2(0) = \eta_2$, respectively, and $ g(t)$ by $ g(0) = 0.$  From
(4.18) and (4.19) we find that $ {\bf \xi}(t)$ is a linear function in the
coordinates
$ \xi_1,\xi_2,\eta_1$ and $\eta_2$ and $ g(t)$ a quadratic function.

The Weyl operator can be used to calculate the time-dependent expectation
values
$ {\bf m}(t)$ and $ \sigma(t)$ (see (4.7) and (4.10)), since this operator is
connected with the coordinates and momenta via the derivatives
$${\partial W\over \partial \xi_i} \vert_{{\bf \xi}=0} = - {i \over\hbar}p_i,$$
$$ { \partial W\over \partial \eta_i} \vert_{{\bf\xi}=0} = {i \over\hbar}q_i,$$
$$ { \partial^2 W\over\partial \xi_i \partial \xi_j} \vert_{{\bf\xi}=0} = - {1
\over \hbar^2}p_ip_j, \eqno (4.20)$$
$$ { \partial^2 W \over \partial \xi_i \partial \eta_j} \vert_{{\bf\xi}=0} = {1
\over 2\hbar^2}(p_iq_j + q_jp_i),$$
$$ { \partial^2 W \over\partial\eta_i\partial\eta_j} \vert_{{\bf\xi}=0} = - {1
\over \hbar^2}q_iq_j.$$
For example, one obtains by using (4.20)
$$ \sigma_{p_ip_j}(t) = - \hbar^2 {\rm Tr} (\rho { \partial^2 \widetilde \Phi_t
(W) \over \partial \xi_i(0) \partial \xi_j(0)} \vert_{{\bf\xi}(t=0)=0}).$$
Equations of this type can be evaluated with the help of (4.17)--(4.19) and
lead
to the same results for ${\bf m}(t) $ and $\sigma(t)$ as given before. With the
Weyl operator (4.17) we can calculate the time development of the Wigner
function.
For this purpose we use the Fourier transform of the Wigner function at $t=0$:
$$ {\rm Tr}\{ \rho \exp[i \hbar^{-1}(\eta'_1q_1 + \eta'_2q_2 -
\xi'_1p_1 - \xi'_2p_2)]\}$$
$$=\int_{-\infty}^\infty \int_{-\infty}^\infty \int_{-\infty}^\infty \int_{-
\infty}^\infty
\exp[i\hbar^{-1}(x_1\eta'_1 + x_2\eta'_2 - y_1\xi'_1
-y_2\xi'_2)]$$
$$ \times f(x_1,x_2,y_1,y_2,t=0)dx_1dx_2dy_1dy_2.$$
When this relation is inserted into (4.16) after the Weyl operator $ \widetilde
\Phi_t (W)$ is expressed by (4.17), one can integrate over the coordinates
$ \xi_1,\xi_2,\eta_1$ and $\eta_2$ with the following result for the Wigner
function:
$$ f(x_1,x_2,y_1,y_2,t) = {1 \over[{\rm det}(4 \pi Z)]^{1/2}} \int_{- \infty}^
\infty\int_{- \infty}^
\infty\int_{- \infty}^
\infty\int_{- \infty}^
\infty \exp[ -{1 \over 4}({\bf x} - {\bf x}'M^T)Z^{-1}({\bf x} - M{\bf x}')]$$
$$ \times f(x'_1,x'_2,y'_1,y'_2,t=0)dx'_1dx'_2dy'_1dy'_2, \eqno (4.21)
$$
where ${\bf x} =(x_1,x_2,y_1,y_2)$ and the matrix $Z(t)$ is given by
$$Z(t)=\int_0^t M(t')DM^T(t')dt'.$$
This definition can be applied in order to rewrite (4.14):
$$ \sigma(t)=M(t)\sigma(0)M^T(t)+2Z(t). \eqno (4.22)$$
In the particular case when we set
$$f(x_1,x_2,y_1,y_2,t=0)={1 \over[{\rm det}(2 \pi \sigma(0))]^{1/2}} \exp
[-{1 \over 2}({\bf x} - {\bf m}(0)) \sigma(0)^{-1} ({\bf x} - {\bf m}(0))],$$
we obtain from (4.21):
$$f(x_1,x_2,y_1,y_2,t)={1 \over[{\rm det}(2 \pi \sigma(t))]^{1/2}} \exp
[-{1 \over 2}({\bf x} - {\bf m}(t)) \sigma(t)^{-1} ({\bf x} - {\bf m}(t)], $$
which is the well-known result for Wigner functions [45,69,85].

In order to illustrate the developed formalism we present an example of two
oscillators, which are coupled by a potential of the form as used for the
proton
and neutron degrees of freedom in (4.1), i.e. $ k_{12} = 0, \mu_{ij} = 0, \nu _
{12} \not= 0.$ In this case the matrix $Y$, governing the time development of
the expectation values ${\bf m} (t)$ and $\sigma (t),$ becomes
$$ Y=\left(\matrix{-\lambda_{11}&-\lambda_{12}&1/m_1&-\alpha_{12}\cr
-\lambda_{21}&-\lambda_{22}&\alpha_{12}&1/m_2 \cr
-m_1\omega_1^2&\beta_{12} - \nu_{12}&-\lambda_{11}&-\lambda_{21}\cr
- \beta_{12} - \nu_{12}&-m_2\omega_2^2&-\lambda_{12}&-\lambda_{22}
\cr}\right).\eqno(4.23)$$
For the calculation of the matrix $ M(t)$ we must diagonalize the matrix
$Y$ by solving the corresponding secular equation, i.e. $\det(Y-zI)=0$,
where $z$ is the eigenvalue and $I$ is the unit matrix. According to (4.23)
one obtains an equation of fourth order for the eigenvalues $z$, which can be
simply solved only for special examples. In the particular case with
$\alpha_{12}=0,\beta_{12}=0,\lambda_{12}=0$ and $ \lambda_{21} = 0,$ the
secular equation is obtained as
$$[(z+\lambda_{11})^2+\omega_1^2][(z+\lambda_{22})^2+\omega_2^2]=
 \nu_{12}^2 / m_1m_2 .$$
The roots of this equation have the general structure
$$z_1 = -\gamma_+ + i\omega_+,~z_2 = -\gamma_+ - i\omega_+,~z_3 = -\gamma_-
+ i\omega_-,~z_4 = -\gamma_- - i\omega_-.$$
The constants $\gamma_{\pm} $ and $\omega_{\pm}$ can be easily calculated for
the case $ \nu_{12} = 0:$
$$ \gamma_+ = \lambda_{11},~\gamma_- = \lambda_{22},~\omega_+ = \omega_1,~
\omega_- = \omega_2 \eqno (4.24) $$
or for $ \lambda_{11} = \lambda_{22} = \lambda, \omega_1 = \omega_2 =\omega(
\omega^2 > \nu_{12}/(m_1m_2)^{1/2})$
$$ \gamma_+ = \gamma_- = \lambda,~\omega^2_\pm = \omega^2 \pm \nu_{12}/
(m_1 m_2)^{1/2}.$$
Only positive values of $\gamma_+$ and $ \gamma_-$ fulfil (4.8).
Applying the eigenvalues $z_i$ of $Y$ we can write the time-dependent
matrix
$M(t)$ as follows:
$$M_{mn}(t)=\sum_i N_{mi}\exp(z_it)N_{in}^{-1},$$
where the matrix $N$ represents the eigenvectors of $Y$:
$$\sum_n Y_{mn}N_{ni}=z_iN_{mi}.$$
With the relations $M_{mn}(t=0)=\delta_{mn}$ and $dM_{mn}(t)/dt\vert_{t=0}=
Y_{mn}$ and using (4.7), (4.14), we conclude that the expectation values of
the
coordinates and momenta decay with the exponential factors $\exp(-\gamma_
+t)$ and $\exp(-\gamma_-t)$ and the matrix
elements $\sigma_{mn}$ with the combined factors $\exp(-2\gamma_+t)$,
$\exp(-\gamma_-t)$ and $\exp[-(\gamma_+ +\gamma_-)t].$

Since the matrix elements $ M_{mn}$ are in general lengthy expressions, we
present here the matrix $M(t)$ only for the special and simple case
that the oscillators are uncoupled. With the roots given in (4.24) we obtain

$M(t)=$
$$\left(\matrix{e^{-\lambda_{11}t}\cos\omega_1t&0&{1\over m_1\omega_1}
e^{-\lambda_{11}t}\sin\omega_1t&0 \cr
0&e^{-\lambda_{22}t}\cos\omega_2t&0&{1\over m_2\omega_2}e^{-\lambda_{22}t}
\sin\omega_2t \cr
-m_1\omega_1e^{-\lambda_{11}t}\sin\omega_1t&0&e^{-\lambda_{11}t}\cos\omega_1t
&0 \cr
0&-m_2\omega_2e^{-\lambda_{22}t}\sin\omega_2t&0&e^{-\lambda_{22}t}\cos
\omega_2t}\right). \eqno(4.25)$$
This matrix can be used to evaluate $\sigma(t)$ defined by (4.14) or (4.22).
For example, we find the following expression for $\sigma_{12} = \sigma_
{q_1q_2}$ with $M(t)$ of (4.25):
$$\sigma_{q_1q_2}(t)=\exp[-(\lambda_{11}+\lambda_{22})t]((\sigma_{q_1q_2}(0)-
\sigma_{q_1q_2}(\infty))\cos\omega_1t\cos\omega_2t$$
$$+{1\over m_1\omega_1}(\sigma_{q_2p_1}(0)-\sigma_{q_2p_1}(\infty))\sin\omega_
1t\cos\omega_2t
+{1 \over m_2\omega_2}(\sigma_{q_1p_2}(0)-\sigma_{q_1p_2}(\infty))\cos\omega_
1t\sin\omega_2t$$
$$+{1 \over m_1m_2\omega_1\omega_2}(\sigma_{p_1p_2}(0)-\sigma_{p_1p_2}(\infty))
\sin\omega_1t\sin\omega_2t)+\sigma_{q_1q_2}(\infty).$$

Similar expressions are found for the other matrix elements of $\sigma(t)$.
The matrix elements of $\sigma(\infty)$ depend on $Y$ and $D$
and must be evaluated with (4.15) or by the relation:
$$\sigma(\infty)=2\int_0^\infty  M(t') D M^T(t')dt'.$$
As an example we present the value of $ \sigma_{q_1q_2}(\infty):$
$$\sigma_{q_1q_2} (\infty) = 2 \{[(\lambda_{11} + \lambda_{22})^2 + (\omega_1 +
\omega_2)^2][(\lambda_{11}+ \lambda_{22})^2 + (\omega_1 - \omega_2)^2]\}^{-1}$$
$$\times\{(\lambda_{11}+\lambda_{22})[(\lambda_{11}+ \lambda_{22})^2 +\omega^2_
1 + \omega^2_2]D_{q_1q_2}
 + [(\lambda_{11} + \lambda_{22})^2 +\omega^2_1 - \omega^2_2]D_{q_2p_1}/m_1$$
$$ + [(\lambda_{11} + \lambda_{22})^2 +\omega^2_2 - \omega^2_1]D_{q_1p_2}/m_2 +
2(\lambda_{11} + \lambda_{22})D_{p_1p_2}/m_1m_2 \}.$$
Similar expressions are obtained for the other matrix elements of $ \sigma
(\infty).$ The diffusion coefficients $ D_{q_1q_2}, D_{q_1p_2}, D_{q_2p_1}$
and $D_{p_1p_2}$ are in general zero for uncoupled oscillators interacting
with
an usual environment. This has the consequence that the expectation values $
\sigma_{q_1q_2}, \sigma_{q_1p_2}, \sigma_{q_2p_1}$ and $\sigma_{p_1p_2}$ vanish
for $ t \to \infty.$ It is a very interesting point that the general theory
of Lindblad allows couplings via the environment between uncoupled oscillators
with $ k_{12} = 0, \mu_{ij} = 0,\nu_{12} = 0.$ According to the definitions of
the parameters in terms of the vectors ${\bf a}_k$ and ${\bf b}_k,$ the
diffusion
coefficients above can be different from zero and can simulate an interaction
between
the "uncoupled" oscillators. In this case a structure of the environment is
reflected in the motion of the oscillators.

{\bf 5. Damping of angular momentum in open quantum systems}

In heavy-ion collisions, especially in deep inelastic reactions, one has
experimentally studied the loss of angular momentum in the relative motion of
the nuclei [72]. Measurements of the $ \gamma$-multiplicities yielded
information on the spin of the excited nuclei, which means information on the
loss of angular momentum in the relative motion of the nuclei. If the energy
loss in the relative motion is simultaneously measured one can determine a time
scale of the loss of angular momentum, i.e. one can explore the damping process
as a function of time. This example motivated us to study the damping of
angular momentum.
Another example for the damping of angular momentum is the spin-relaxation in
the presence of a magnetic field which was theoretically studied in [86].

The damping of angular momentum can be described at different levels of
approximations: one may use quantum-mechanical or classical methods and
microscopical or phenomenological models. In this paper we describe the
damping
of angular momentum with the quantum-mechanical theory of Lindblad [24,25].
We choose the operators, which open the system, proportional to the components
of the angular-momentum operator, to generators of the proper Lorentz group
and
proportional to a linear combination of momenta and coordinates.

Let us assume a finite set of Hermitian operators $A_i$ depending on the basic
variables of the physical problem and given in the Heisenberg picture. Then
their time evolution can be determined within the theory of Lindblad by the
following Markovian master equations:
$${dA_i(t) \over dt}=L(A_i(t))= \sum^N_{k=1} M_{ik}A_k(t)+D_i,~~i=1,...,N.$$
The matrix elements $M_{ik}$ and components $D_i$ are time-independent numbers.
They are calculated by the equations $(A_i=A_i(t=0)):$
$$L(A_i)={i \over \hbar}[H,A_i]+{1 \over 2 \hbar} \sum_j(V^+_j[A_i,V_j]+[V^+_j,
A_i]V_j)\eqno(5.1).$$
Here, $H$ is the Hamiltonian, $V_j$ and $V^+_j$ are operators which are
functions
of the basic variables. $V^+_j$ is the Hermitian adjoint operator to $V_j.$ The
expectation values of the operators $A_i(t)$ are solutions of the differential
equations $(i=1,...,N):$
$${d \langle A_i(t) \rangle \over dt}= \sum^N_{k=1} M_{ik} \langle A_k(t)
\rangle +D_i.$$
The following considerations are restricted to three spatial degrees of
freedom
described by the coordinates $ q_1,q_2,q_3$ and linear momenta $p_1,p_2,p_3.$
We assume that the Hamiltonian conserves angular momentum:
$$ [H,L_i] = 0,~~i=1,2,3.$$
Further we assume for simplicity that the opening of the quantum-mechanical
system fulfils rotational symmetry. If we restrict this study to three
operators
$V_{j=1,2,3},$ these operators have to transform like the components of a
vector
under spatial rotations. They fulfil the commutators
$$ [ \sum^3_{j=1} V^+_jV_j,L_i ] = 0,~~i=1,2,3.$$
The following three examples for $V_j$ will be considered $ (j=1,2,3):$
$$(1)~~~~~~ V_j=\alpha L_j,  $$
$$(2) ~~~~~~~~~~~~~~~~~~~~~~~~V_j=\alpha N_j=\alpha(sp_j+q_jp_s), $$
$$(3) ~~~~~~~~~~~~~V_j=ap_j+bq_j.  $$
The operators $N_j$ contain a fourth coordinate $s$ and the corresponding
momentum $p_s=-i\hbar \partial / \partial s.$ This coordinate can be thought
to describe an intrinsic degree of freedom not affected by spatial rotations.
The operators $L_j$ and $N_j$ have commutation relations as the generators of
the proper Lorentz group [87]. The free parameters $\alpha, a$ and $b$ are
complex numbers.

(1) For the case $V_j$ proportional to $L_j,$ we obtain from Eq. (5.1)
$$L(A)={i \over \hbar}[H,A]+{\vert \alpha \vert^2 \over 2 \hbar} \sum^3_{j=1}
[L_j,[A,L_j]].$$
If we set the operator $A$ equal to the square of the angular momentum, $A=L^2,
$ we find conservation of the total angular momentum:
$${dL^2(t) \over dt}=0,~~{d \langle L^2(t) \rangle \over dt}=0 \eqno (5.2) $$
Here $ \langle L^2(t) \rangle $ denotes the expectation value of $L^2(t).$ The
squares of the components of $L^2$ get equalized:
$$ {dL^2_i(t) \over dt}=- \vert \alpha \vert^2 \hbar(3L^2_i(t)-L^2),~~i=1,2,3.
\eqno (5.3) $$
The expectation value of $L^2_i$ is calculated as
$$ \langle L^2_i(t) \rangle = \langle L^2_i \rangle_{t=0} \exp (-3 \hbar \vert
\alpha \vert^2t)+{ \langle L^2 \rangle \over 3} (1- \exp (-3 \hbar \vert \alpha
\vert^2t)), $$
where $ \langle L^2 \rangle$ is time independent as shown in Eq. (5.2). These
results prove that the ansatz (1) can be used in applications when the square
of the angular momentum is conserved, but the squares of the individual
components
are equalized with time. Examples are depolarization phenomena.

We note the fact that in the derivation of Eqs. (5.2) and (5.3) we only used
the
commutators between angular-momentum operators. Therefore, the results in this
case are also true for systems with many degrees of freedom.

(2) For the case $V_j$ proportional to $N_j$ we obtain for $L(A)$ according to
Eq. (5.1):
$$L(A)={i \over \hbar}[H,A]+{\vert \alpha \vert^2 \over 2 \hbar} \sum^3_{j=1}
[N_j,[A,N_j]].$$
If we choose $A=L^2,$ we find the following time derivative of $L^2(t):$
$$ {dL^2(t) \over dt}=2 \vert \alpha \vert^2 \hbar(L^2(t)+N^2(t)).\eqno(5.4)$$
Here we used the commutation relations
$$[L_i,N_j]=i \hbar \sum^3_{l=1} \varepsilon_{ijl}N_l,~
~[N_i,N_j]=-i \hbar \sum^3_{l=1} \varepsilon_{ijl}L_l,$$
where $\varepsilon_{ijl}$ is the Levi-Civita antisymmetric tensor. Since the
right-hand side of Eq. (5.4) contains the square $N^2= \sum^3_{j=1} N^2_j,$
we have also to calculate the time derivative of $N^2(t):$
$${dN^2(t) \over dt}={i \over \hbar}[H,N^2(t)]+2 \vert \alpha \vert^2
\hbar(L^2(t)+N^2(t)). \eqno (5.5) $$
In order to simplify the discussion of Eqs. (5.4) and (5.5) let us assume that
the commutator of $N^2$ with $H$ vanishes. Then the solution for the
expectation
values can be written as
$$ \langle L^2(t) \rangle = \langle L^2 \rangle_{t=0} +{1\over 2} ( \langle L^2
\rangle_{t=0}+ \langle N^2 \rangle_{t=0}) (\exp(4 \vert \alpha \vert^2 \hbar t)
-1), \eqno (5.6) $$
$$ \langle N^2(t) \rangle = \langle N^2 \rangle_{t=0} +{1\over 2} ( \langle L^2
\rangle_{t=0}+ \langle N^2 \rangle_{t=0}) (\exp(4 \vert \alpha \vert^2 \hbar t)
-1). \eqno (5.7) $$
This result shows that the expectation values of $L^2(t)$ and $N^2(t)$ increase
exponentially with time. If the Hamiltonian $H$ is a possitive function of
$L^2$ and $N^2$ alone, also the energy would grow in time. In this case the
master equation describes a physical system coupled to a reservoir of infinite
temperature which increases the energy and angular momentum exponentially.
However, such a system is not applicable for a phenomenological description
of nuclear processes. But it is thinkable that realistic applications in
solid-state physics can be found.

Let us assume that the system has the smallest possible value of $L^2$, namely
$ \langle L^2 \rangle_{t=0} = 0.$ Then the increase of $L^2$ and $N^2$ is given
according to Eqs. (5.6) and (5.7):
$$ \langle L^2(t) \rangle ={1 \over 2} \langle N^2 \rangle_{t=0}(\exp(4 \vert
\alpha \vert^2 \hbar t)-1),~\langle N^2(t) \rangle ={1\over 2} \langle N^2
\rangle_{t=0}(\exp(4 \vert \alpha \vert^2 \hbar t)+1). $$
From these equations it is interesting to note that both $ \langle L^2(t)
\rangle$ and $ \langle N^2(t) \rangle$ are directly proportional to $ \langle
N^2 \rangle_{t=0}.$  Both expectation values are in general related by the
equation
$$ \langle L^2(t) \rangle - \langle L^2(t) \rangle = \langle N^2(t) \rangle -
\langle N^2(t) \rangle_{t=0}.$$

(3) In the case $V_j=ap_j+bq_j,$ we obtain
$$ L(A)={i \over \hbar}[H,A]+{1 \over \hbar^2}D_{pp} \sum^3_{j=1}[q_j,[A,q_j]]
+{1 \over \hbar^2}D_{qq} \sum^3_{j=1}[p_j,[A,p_j]] $$
$$- {1 \over \hbar^2}D_{pq} \sum^3_{j=1}([q_j,[A,p_j]]+[p_j,[A,q_j]])+
{i \lambda \over 2 \hbar} \sum^3_{j=1}([q_j,[A,p_j]+[A,p_j]q_j-p_j[A,q_j]-
[A,q_j]p_j). \eqno (5.8) $$
In this equation we used the following abbreviations:
$$ D_{qq}={\hbar \over 2} a^*a,~D_{pp}={\hbar \over 2} b^*b,~
D_{pq}=-{\hbar \over 2} {\rm Re} (a^*b),~\lambda=-{\rm Im} (a^*b).$$
The quantities $D_{qq}, D_{pp}$ and $D_{pq}$ are denoted as the diffusion
coefficients and $ \lambda$ as the friction constant. With the complex numbers
$a$ and $b$ we can form the following determinant which is zero:
$$ 0={\hbar \over 2} \left | \matrix{a^*a&a^*b \cr
b^*a&b^*b \cr } \right | = \left | \matrix{D_{qq}&-D_{pq}-i \hbar \lambda/2 \cr
-D_{pq}+i \hbar \lambda/2&D_{pp} \cr} \right | .$$
This determinant can be used to write down an equality serving as a constraint
for the phenomenological parameters:
$$D_{qq}D_{pp}-D^2_{pq}=\hbar^2 \lambda^2/4. \eqno (5.9) $$
If we set $A=L^2$ in Eq. (5.8), we find
$$L(L^2)=4(D_{pp}q^2+D_{qq}p^2-D_{pq}({\bf qp+pq}))-4 \lambda(L^2+{3 \over 2}
\hbar^2), \eqno (5.10) $$
with $q^2=q^2_1+q^2_2+q^2_3, p^2=p^2_1+p^2_2+p^2_3 $ and ${\bf qp}=q_1p_1+
q_2p_2+q_3p_3.$ Since Eq. (5.10) contains other operators in addition to $L^2$
we have to calculate the corresponding equations of motion of these opertors,
too. We obtain
$${dq^2(t) \over dt} ={i \over \hbar}[H,q^2(t)]-2 \lambda q^2(t)+6D_{qq}, \eqno
(5.11) $$
$${dp^2(t) \over dt} ={i \over \hbar}[H,p^2(t)]-2 \lambda p^2(t)+6D_{pp}, \eqno
(5.12) $$
$${d({\bf qp}+{\bf pq})_t \over dt} ={i \over \hbar}[H,({\bf qp}+{\bf pq})_t]
-2 \lambda ({\bf qp}+{\bf pq})_t+12D_{pq}. \eqno(5.13) $$
Let us first assume that the Hamiltonian depends on $L^2$ only, e.g., $H=L^2/
2\Theta$, where $\Theta$ is the moment of inertia. Then all commutators with
$H$
vanish in Eqs. (5.11)--(5.13). As function of time the expectation values of
$q^2,p^2$ and ${\bf qp}+{\bf pq}$ drop exponentially down $(\sim\exp(-2 \lambda
t)).$ For large times they approach the final values
$$\langle q^2 \rangle_{t \to \infty}=3D_{qq}/ \lambda,~
\langle p^2 \rangle_{t \to \infty}=3D_{pp}/ \lambda,~
{1 \over 2} \langle {\bf qp}+{\bf pq} \rangle_{t \to \infty}=3D_{pq}/
\lambda. $$
If we insert these values into the equation for $d \langle L^2(t) \rangle /dt$
using Eq. (5.10), we find for large times
$$ \langle L^2 \rangle_{t \to \infty}=0.$$
Solving the expectation values we obtain
$$\langle q^2(t) \rangle=A \exp(-2 \lambda t)+3D_{qq}/ \lambda, $$
$$\langle p^2(t) \rangle=B \exp(-2 \lambda t)+3D_{pp}/ \lambda, $$
$${1 \over 2} \langle ({\bf qp}+{\bf pq})_t \rangle=C \exp(-2 \lambda t)+3D_
{pq}/ \lambda, $$
$$\langle L^2(t) \rangle=D \exp(-4 \lambda t)+{2 \over \lambda}(AD_{pp}+BD_{qq}
-2CD_{pq}) \exp(-2 \lambda t). $$
The parameters $A,B,C$ and $D$ have to be chosen by initial conditions.
A second example is the choice of the Hamiltonian of a spherical oscillator:
$$H={1 \over 2m}p^2+{m \over 2} \omega^2q^2.$$
If we calculate the expectation value of $L^2(t)$ for large times using the
equations (5.10)--(5.13) we obtain
$$ \langle L^2 \rangle_{t \to \infty} ={6 \over \lambda^2(\lambda^2 + \omega^2)
}\{ {1 \over 4} ({D_{pp} \over m} - m \omega^2 D_{qq})^2 + \omega^2D^2_{pq} \}.
\eqno (5.14) $$
This relation can be rewritten by using the expectation value of the energy at
large times:
$$ \langle H \rangle_{t \to \infty} ={3 \over 2 \lambda}({D_{pp} \over m} +
m \omega^2 D_{qq}). \eqno (5.15) $$
Inserting (5.15) into (5.14) and applying (5.9) we obtain
$$ \langle L^2 \rangle_{t \to \infty} ={2 \over 3(\lambda^2 + \omega^2)}
(\langle H \rangle^2_{t \to \infty} - ({3 \over 2} \hbar \omega)^2). $$
If the diffusion coefficients are chosen in accordance with Eq. (5.9),
$$ D_{pp}={\lambda \hbar \over 2m \omega},~~D_{pp}={\lambda \hbar \over 2}
m \omega,~~D_{pq}=0, $$
we reach the lowest value of $ \langle H \rangle_{t \to \infty}=(3/2) \hbar
\omega $ and $ \langle L^2 \rangle_{t \to \infty} = 0. $

In simple examples we have found operators $V_j$ needed to open a system in
which
angular momentum is damped. The difficulty in choosing these operators is the
fact that in general new and usually complex operators appear on the
right-hand
sides of the Lindblad equations after evaluating the commutators. For these new
operators one has to determine their master equations. Only if the number of
coupled master equations remains small, one can hope to find analytical
solutions
for the damping processes.

As shown in Eqs. (5.10)--(5.14) the damping depends on the parameter $ \lambda$
which results from a product of two free parameters. If only a single free
parameter is contained in the operators $V_j,$ as for example given in the case
(2), we obtain no damping of the square of angular momentum in general but an
exponential increase. Systems with an energy and angular momentum
increasing in time are not suitable for application in nuclear physiscs.

A damping of angular momentum may be easier reached, if the rotational
symmetry
of the dissipative operators is given up. But the equations which have
to be solved get more complex and cannot analytically be evaluated without
approximations [88].

One can apply the damped three-dimensional harmonic oscillator presented here
to heavy-ion collisions. For deep inelastic reactions of not too heavy nuclei
one may approximate the internuclear potential near the barrier by a reverse
parabola:
$$V(r)=V_B-{m \over 2} \kappa^2(r-r_B)^2. \eqno (5.16) $$
Here, $r$ is the internuclear distance and $r_B$ the location of the barrier
with
height $V_B.$ Equation (5.16) means a replacement of the frequency $ \omega$ by
$ i \kappa.$ A damping of angular momentum of the relative motion can be gained
if the friction constant $\lambda$ is larger than $ \kappa.$

{\bf 6.Quantum tunneling in open systems}

Tunneling is a remarkable, essentially quantum phenomenon,
consisting in the propagation of a particle through a classically
impenetrable potential barrier. Standing at the basis of important
applications in electronics, chemistry and nuclear physics, it has
continuously been investigated from the beginning of the quantum
mechanics [27],[89],[90], especially due to the difficulties raised
by the effects of the dissipation [91-94], always present in
practical cases.

In this Section, the master equation (3.1) is used for the study of dissipative
phenomena on tunneling processes. We consider a fission potential barrier $U(q)
$ [26], of the form represented in Fig. 4, where the first well
corresponds to a compound nucleus and the second well to a fission channel.
In order to generalize Gamow's formula of the tunneling rate, for the case when
a dissipative environment in present we use the theory of perturbations.
With the potential $U(q)$, the Hamiltonian is
$$ H = {p^2 \over 2m} + U(q) \eqno(6.1) $$
and the master equation (3.1) becomes
$$ {d \rho \over d t} = - {i \over \hbar} [H, \rho] + \Lambda (\rho),\eqno(6.2)
$$
where
$$ \Lambda(\rho) = - {i \lambda \over 2 \hbar} ([q, p \rho + \rho p] -
[p, q \rho + \rho q])  $$
$$ - {1 \over \hbar^2} D_{qq} [p,[p,\rho]] - {1 \over \hbar^2} D_{pp} [q,[q,
\rho]]
 + {1 \over \hbar^2} D_{pq} ([q,[p,\rho]] + [p,[q,\rho]]). \eqno (6.3) $$

Using the commutation relation $ [q, p] = i \hbar,$ from Eqs. (6.2)--(6.3)
we obtain the following evolution equations for the expectation values of the
canonical coordinates $q$ and $p$:
$$ {d<q> \over dt} + \lambda <q> = {<p> \over m},$$
$$ {d<p> \over dt } + \lambda <p> = < - {dU \over dq} > \eqno (6.4) $$
and for their variances:
$$ {d \over dt} \sigma_{qq} + 2 \lambda \sigma_{qq} - {2 \over m} \sigma
_{pq} = 2D_{qq},$$
$$ {d \over dt} \sigma_{pp} + 2 \lambda \sigma_{pp} -
2 \sigma_{pU} = 2D_{pp},\eqno(6.5)$$
$$ {d \over dt} \sigma_{pq} + 2 \lambda \sigma_{pq} -
{1 \over m} \sigma_{pp} - \sigma_{qU} = 2 D_{pq},$$
where
$$  \sigma_{qq} = <q^{2}> - <q>^{2},~
\sigma_{pp} = <p^{2}> - <p>^{2},~
 \sigma_{qq} = <{ qp + pq \over 2}> - <q> <p>  \eqno (6.6) $$
and
$$ \sigma_{pU} = - {1 \over 2} < {dU \over dq} p + p {dU \over dq} > +
< {dU \over dq} > <p>,~
 \sigma_{qU} = - < {dU \over dq} q > + < {dU \over dq} > <q>. \eqno (6.7) $$
In this way we see that the quantum master equation (6.2)--(6.3) leads to the
Newtonian equations (6.4) with additional friction terms and to the similar
equations (6.5) for the variances, with $D_{qq} , D_{pp} , D_{pq} $ as
zero-point values and $ \sigma_{pq} , \sigma_{pU} , \sigma_{qU} $ as
generalized forces.

Quantum tunneling is described as a transition
from some "localized state" of the first well to some "localized state" of the
second well. As we shall see later, this can be done by defining a Hamiltonian
$H_{0} $ with approximate solutions in the first and in the second well, which
disregard the tails corresponding to the other regions. The difference $V$
between
the two Hamiltonians defines a tunneling operator. Consequently, the
Hamiltonian of the system takes the form:
$$H = H_{0} + V,  \eqno (6.8) $$
where $H_{0} $ describes the localized states and $V$ the transitions between
them.

From Eqs. (6.4), with the expression $ U(q) = - \omega^{2}_{b} q^{2} /2
, m = 1 $ for the potential barrier, one obtains:
$$ { d < p > \over d < q > } = { - \lambda < p > + \omega^{2}_{b} < q > \over
- \lambda < q > + < p > }, \eqno (6.9) $$
which can be compared with the similar equation standing at the basis of the
Dekker's theory of the quantum tuuneling  with dissipation [28],
where the majority of the previous results are incorporated:
$${d< p > \over d < q > } = - 2 \lambda + { \omega^{2}_{b} < q > \over < p > }.
\eqno (6.10) $$
This equation has the solutions
$$ < p >_{1} = ( \sqrt{ \omega^{2}_{b} + \lambda^{2}} - \lambda) < q >,~
 < p >_{2} = - ( \sqrt{ \omega^{2}_{b} + \lambda^{2}} + \lambda) < q >.
\eqno (6.11) $$
In this theory, the tunneling rate is considered of the form:
$$ \Gamma (E) = \kappa \nu (E) P (E), \eqno (6.12) $$
where $P(E)$ is the penetrability and the coefficient $\kappa $ describing the
effect of the dissipation is defined by the expression:
$$ \kappa = { < p ( \lambda) > \over < p (0) >}. \eqno (6.13) $$
With the first solution (6.11) describing the particle passing through the
barrier,
the expression (6.13) takes the explicit form:
$$ \kappa = \sqrt{ 1 + { \lambda^{2} \over \omega^{2}_{b}}} - { \lambda \over
\omega_{b}}, \eqno (6.14) $$
which, in accordance to all previous results shows that due to the dissipation,
the tunneling rate decreases.
At the same time, with our expression (6.9) we obtain $ \kappa = 1.$ We
conclude
that the result (6.14) is a consequence of the particular way of introducing
the
dissipation in the equation (6.10), which is only empirically justified. When
one considers the more accurate equations (6.4), resulting from the quantum
theory
of open systems and which can be easily understood from a physical point of
view,
the effect described by the expression (6.14) disappears.
As we shall see in the following, the dissipative environment generates new
transitions increasing the barrier penetrability.

By considering a basis of localized states, $\vert\Psi_{0}> $ for the ground
state of the first well (the compound nucleus) and
 $ \vert \Psi_{i} >, ~~i = 1,2,...$ for the second well (the reaction channel)
the transition operator $V$ takes the form:
$$V=\pmatrix{0&V_{01}&V_{02}\ldots&V_{0n}\cr V_{10}&0&0\ldots&0\cr
V_{20}&0&0\ldots&0\cr \vdots&\vdots&\ddots&\vdots\cr
V_{n0}&0&0\ldots&0\cr}, \eqno (6.15) $$
$$ V_{i0} = V_{0i},~~~ i = 1,2,...,n, $$
where the matrix elements have the approximate expressions [26]:
$$ V_{0i} = V_{i0} = { \hbar^{2} \over 2 m} [ \Psi^{'}_{0} (r_{0} ) \Psi_{i}
(r_{0} ) - \Psi_{0} (r_{0} ) \Psi^{'}_{i} (r_{0} ) ]. \eqno (6.16) $$

We consider that the operators $V$ and $\Lambda(\rho) $ are small perturbations
of the Hamiltonian $ H_{0} $ defining the basis of the localized states. With
the notations:
$$\rho^{'}(t)=e^{{i\over \hbar} H_{0} t} \rho(t) e^{-{i \over \hbar} H_{0} t},~
V^{'}(t) = e^{{i \over \hbar} H_{0} t} V(t) e^{-{i \over \hbar} H_{0} t},~
\Lambda^{'} ( \rho^{'}(t) ; t) = e^{{i \over \hbar}H_{0}t}\Lambda (\rho)e^{-{i
\over\hbar} H_{0} t} \eqno (6.17) $$
in the interaction picture the quantum master equation (6.2) becomes:
$$ {d \rho^{'} \over dt} = - {i \over \hbar} [ V^{'}(t) , \rho^{'}(t) ] +
\Lambda^{'}( \rho^{'}(t) ; t ). \eqno (6.18) $$
Considering the zero-order density operator $\rho^{'(0)}=\vert 0> < 0 \vert $ ,
for the initial state the density operator in the second order
approximation with respect to the operators $V$ and $\Lambda,$ can be written
in the form:
$$\rho^{'} (t) = \rho^{'(0)} + \rho^{'(1)}(t) + \rho^{'(2)}(t). \eqno (6.19) $$
Introducing this expression in Eq. (6.18) and integrating, we obtain:
$$\rho^{'}(t) = \rho^{'(0)} + \rho^{'}_{V}(t) + \rho^{'}_{\Lambda}(t) + \rho^
{'}_{VV}
(t) + \rho^{'}_{V \Lambda}(t) + \rho^{'}_{LV}(t) + \rho^{'}_{\Lambda \Lambda}
(t), \eqno (6.20) $$
where:
$$ \rho^{'}_{V}(t) = -{i \over \hbar} \int^{t}_{0} dt^{'} [V^{'}(t^{'}) ,
\rho^{'(0)} ], $$
$$\rho^{'}_{\Lambda}(t)=\int^{t}_{0}dt^{'} \Lambda^{'} ( \rho^{'(0)};t^{'}),$$
$$ \rho^{'}_{VV}(t) = -{1 \over \hbar^{2}} \int^{t}_{0} dt^{'} \int^{t^{'}}_{0}
dt^{''} [V^{'}(t^{'}) , [V^{'}(t^{''}) , \rho^{'(0)}]],$$
$$ \rho^{'}_{V\Lambda}(t) = -{i \over \hbar} \int^{t}_{0}dt^{'}\int^{t^{'}}_{0}
dt^{''} [V^{'}(t^{'}) , \Lambda^{'}( \rho^{'(0)} ; t^{''})], $$
$$\rho^{'}_{\Lambda V}(t)=-{i\over \hbar} \int^{t}_{0} dt^{'}\int^{t^{'}}_{0}
dt^{''}\Lambda^{'} ([V^{'}(t^{''}) , \rho^{'(0)} ] ; t^{'}), $$
$$ \rho^{'}_{\Lambda\Lambda}(t) =\int^{t}_{0} dt^{'} \int^{t^{'}}_{0} dt^{''}
\Lambda^{'} (\Lambda^{'} ( \rho^{'(0)} ; t^{''}) ; t^{'}). \eqno (6.21) $$
In principle we have to calculate the diagonal matrix elements $\rho_{ii}(t) $
in the representation of the localized states $ \vert \psi_{i}(t) > = \exp(-i
H_{0}t/\hbar) \vert i >, i=1,2,... $
In this case, $ \rho_{ii}(t)=<\psi_{i}(t)\vert\rho\vert\psi_{i}(t)>=
<i\vert \rho^{'}(t) \vert i > $ and the transition rates have the expressiones
$ \rho_{ii}(t)/t $ .
The functions $ \rho_{ii}(t)/t $ describe the dependence of the transition
probabilities on the frequency $\omega_{i}=(E_{i}- E_{0} )/ \hbar $ due to the
energy uncertainty corresponding to the time $t$. In this way we can see the
influence of the openess on the tunneling spectrum and on the total tunneling
rate
$ \Gamma = \sum_{i} \rho_{ii}(t)/t $.
Using the expressions (6.3) and (6.15) of the openess and tunneling operators,
neglecting the sums of the rapidly varying terms $\sum_{j}e^{i\omega_{j}t}$ and
making the integrals (6.21), from the expression (6.20) we obtain:
$$ \rho_{ii}(t) = \rho^{G}_{ii}(t) + \rho^{L}_{ii}(t) + \rho^{ \lambda}_{ii}(t)
+ \rho^{D}_{ii} +\rho^{N}_{ii}+\rho^{A}_{ii}+\rho^{B}_{ii}(t)+\rho^{C}_{ii}(t)
+ \rho^{F}_{ii}(t) \eqno (6.22) $$
with
$$ \rho^{G}_{ii}(t) = \Omega^{2}_{i0} { \sin^{2}( \omega_{i}t/2) \over ( \omega
_{i}/2)^{2}}, \eqno (6.23) $$
$$ \rho^{L}_{ii}(t)=\left[\lambda C_{i0} \left( \lambda C_{0i} + 2D_{qq} u_{0i}
-2{D_{pp} \over \hbar^{2}} w_{0i}\right) - 2 {D_{pq} \over \hbar} \Omega
_{i0} (C_{0i} + 2v_{0i})\right]{ \sin^{2} ( \omega_{i} t/2) \over 2( \omega_{i}
/2)^{2}}, \eqno (6.24) $$
$$ \rho^{ \lambda}_{ii}(t)= \lambda C_{i0} \Omega_{0i} { \sin^{2} ( \omega_{i}t
/2) \over \omega_{i}/2} t, \eqno (6.25) $$
$$ \rho^{D}_{ii}(t)=-2( D_{qq} u_{00} + {D_{pp} \over \hbar^2} w_{00} ) C_{i0}
\Omega_{0i} {t\over\omega_{i}} ({ \sin \omega_{i} t \over \omega_{i} t } - \cos
 \omega_{i} t),\eqno (6.26) $$
$$ \rho^{N}_{ii}(t)= -2{D_{pq} \over \hbar} \lambda C_{i0} ( C_{0i} + 2v_{0i} )
{t \over \omega_{i}}(1-{\sin \omega_{i} t \over \omega_{i} t} ),\eqno (6.27) $$
$$ \rho^{A}_{ii}(t) = \left( - 2 \lambda q_{i0} s_{0i} + 2D_{qq} s^{2}_{0i} +
2 {D_{pp} \over \hbar^{2}} q^{2}_{i0} \right) t , \eqno (6.28) $$
$$ \rho^{B}_{ii}(t) = \left( -  \lambda q_{i0} s_{0i} + D_{qq} s^{2}_{0i} +
{D_{pp} \over \hbar^{2}} q^{2}_{i0} \right) \lambda t^{2} , \eqno (6.29) $$
$$ \rho^{C}_{ii} = C^{2}_{i0}, \eqno (6.30) $$
$$ \rho^{F}_{ii}(t) = C_{0i} \{ 2 \left[ \Omega_{0i} - {D_{pq} \over \hbar}
(C_{0i} + 2v_{0i}) \right] { \sin^{2} ( \omega_{i} t/2) \over \omega_{i}/2}
  $$
$$+\left( \lambda C_{0i} + 2 D_{qq} u_{0i} - 2 {D_{pp} \over \hbar^{2}} w_{0i}
\right) { \sin \omega_{i}t \over \omega_{i}} \} \eqno (6.31) $$
depending on the following overlap integrals or matrix elements:
$$ C_{0i} = C_{i0} = \int \Psi_{0} (q) \Psi_{i} (q) dq,
 \Omega_{0i} = \Omega_{i0} = {V_{i0} \over \hbar},$$
$$ q_{i0} = \int \Psi_{0} (q) q \Psi_{i} (q) dq,~
s_{0i} = \int \Psi_{0} (q) d \Psi_{i} (q), $$
$$ u_{0i} = - \int {d \Psi_{0} \over dq} {d \Psi_{i} \over dq} dq,~
u_{00} = \int ({d \Psi_{0} \over dq})^{2} dq, $$
$$ v_{0i} =  \int \Psi_{0} (q) q d \Psi_{i} (q),
 w_{0i} = \int \Psi_{0} (q) q^{2} \Psi_{i} (q) dq,
 w_{00} = \int \Psi_{0} (q) q^{2} \Psi_{0} (q) dq. \eqno (6.32) $$
First of all, we notice that although the matrix elements given by the
expressions
(6.22)--(6.32) depend on the very small arbitrary distance $ \delta \omega$
between
the energy levels in the fission channel, the sum:
$$\sum_{i} \rho_{ii} (t) \simeq \int \rho_{ii} ( \omega_{i} ; t ) {d \omega_{i}
\over \delta \omega } \eqno (6.33) $$
of the probabilities on every span of energy  do not depend on it, because
every
term contains two times the wave functions $\Psi_{i} (q) $ which, according to
the expression (6.24) is proportional to $ \sqrt{ \delta \omega} $ .

Secondly, one notice that, using the relation (6.33), the first term (6.23)
leads to the Fermi's golden rule (6.32):
$$ \Gamma^{G} = \sum_{i} { \rho^{G}_{ii} (t) \over t} = 2 \pi { \Omega^{2}_{i0}
\over \delta \omega}, \eqno (6.34) $$
where we have neglected the dependence on $ { \it i}$ of the matrix element
$ \Omega_{i0} $, in the very narrow span of energy $ \Delta \omega = 1/t. $
This term is symmetric with the energy $ \hbar \omega_{i} = E_{i} - E_{0} $ and
has a very small width of the spectrum, representing the energy uncertainty
for the time $t$.

The second term (6.24), having the same time dependence as (6.23) is merely a
correction to the Gamow's term, due to the openess parameters.

The third term (6.25), being asymmetric with $ \omega_{i},$ describes a
tunneling energy shift proportional with the coefficient $ \lambda$. Because
the matrix
element $ \Omega_{i0} $ is negative, this term increases the transition rates
for energies $ E_{i} $ smaller than the ground energy of the nucleus $ E_{0}
(\omega_{i}<0 ), $ while for $ E_{i}> E_0 $ the transition rates are decreased.

The fourth term (6.26), rapidly varying with $ \omega_{i},$ describes
transition rate modifications due to the diffusion coefficients $ D_{pp} $ and
$ D_{qq} $ which, on the other hand, do not change the energy expectation
values, the new introduced positive terms being compensated by negative ones
in the summation.

The fifth term (6.27), depending on $ D_{pq},$  introduces an energy shift
similar to that introduced by $ \lambda.$ Because we have previously shown that
at thermal equilibrium $ D_{pq} = 0$ [17], this term describes
nonequilibrium processes.

The sixth term (6.28) leads to a transition rate $ \rho^{A}_{ii} (t)/t $ also
not depending on time, but with a very large spectrum, having a dependence on
$ \omega_{i} $ much weaker
than that of the functions multiplying the matrix elements in the expressions
(6.23)--(6.27).

The seventh term (6.29) leads to a transition rate $ \rho^{B}_{ii} (t)/t $
also with the very large spectrum given by the matrix elements $ q_{i0} ,
s_{0i}, $ but proportional with time. From the fundamental constraints (3.4)
we find that both terms (6.28) and (6.29) are positive.

The terms (6.30) and (6.31) lead to a time-independent desintegration
probability due to the overlap of the initial state $ \vert 0 > $ with
the states $ \vert i > $.

For a long interval of time, the evolution of the system can be described by
the
expressions (6.22)--(6.31). In this case, the validity condition of the
perturbation theory $\rho_{ii}(t)<< \rho_{00} (0) $ holds also for large values
of time,
when $ \sum_{i} \rho_{ii} (t) > > \rho_{00} (0) $, the number of energy levels
$ \it i $ being very large.
Making the summation of the expressions (6.22)--(6.31), by using the formula
(6.33), one obtains an expression of the form
$$ \sum_{i} \rho_{ii} (t)=\chi+\Gamma_{0} t + \Gamma_{1} t^{2}, \eqno (6.35) $$
where:
$$ \Gamma_{0} t = \sum_{i} \left[ \rho^{G}_{ii} (t) +
\rho^{L}_{ii} (t) + \rho^{\lambda}_{ii} (t) + \rho^{D}_{ii} (t) + \rho^{N}_{ii}
(t) + \rho^{A}_{ii} (t) \right], $$
$$ \Gamma_{1} t^{2} = \sum_{i} \rho^{B}_{ii} (t),
 \chi = \sum_{i} \left[ C^{2}_{i0} + \rho^{F}_{ii} (t)
\right]. \eqno (6.36) $$

If $N(t)$ is the number of nondesintegrated nuclei, we can consider the
equality
$$\sum_{i} \rho_{ii} (t) = - \int^{t}_{0} { dN (t^{'}) \over N (t^{'})} , \eqno
(6.37) $$
which leads to the following desintegration law :
$$ N (t) = N_{0} e^{- \chi - \Gamma_{0} t - \Gamma_{1} t^{2}} =
N(0) e^{- \Gamma_{0} t - \Gamma_{1} t^{2}}. \eqno (6.38) $$
In this expression appear two additional parameters: $\Gamma_{1}$ describing
the
irreversibility of the desintegration process in a dissipative environment and
$ \chi $ describing the nonorthogonality of the initial state $ \vert 0 > $
of the nucleus with the fission channel states $ \vert i > $. Estimating the
time $ t_{0} = \chi/ \Gamma_{0} \sim m \Delta q^{2}/ \hbar \sim 10^{-23} s $,
where $ \Delta q $ is the width of the barrier, we find that $ \chi$ can be
neglected.

In the tunneling spectrum described by the expressions (6.22)--(6.31) we
distinguish the Gamow's spectral line:
$$f( \omega_{i} t ) = { \sin^{2} ( \omega_{i} t/2) \over ( \omega_{i} t/2)^{2}}
\eqno (6.39) $$
represented in Fig. 5 and additional lines due to the openess, described by
other
functions:
$$g^{\lambda}(\omega_{i} t ) = - { \sin^{2} ( \omega_{i} t/2 ) \over \omega_{i}
t/2 }  \eqno (6.40) $$
$$g^{D}( \omega_{i} t) = {1 \over \omega_{i} t} \left({ \sin \omega_{i} t \over
\omega_{i} t} - \cos \omega_{i} t \right), \eqno (6.41) $$
$$g^{N}( \omega_{i} t) = -{1 \over \omega_{i} t} \left( 1 - { \sin \omega_{i} t
\over \omega_{i} t } \right) \eqno (6.42) $$
represented in Figs. 6, 7 and respectively 8. In this case, the transition rate
spectrum can be written in the form:
$$ \Gamma ( \omega_{i} ) = { \rho_{ii}(t) \over t} = W^{2}_{i0} t F( \omega_{
i} t) + \varphi( \omega_{i}) (1+ { \lambda \over 2} t), \eqno (6.43) $$
where:
$$ W^{2}_{i0} = \Omega^{2}_{i0} + \lambda C_{i0} \left( \lambda C_{0i} +
2D_{qq} u_{0i} - 2{D_{pp} \over \hbar^{2}} w_{0i} \right )
-2{D_{pq} \over \hbar} \Omega_{i0} (C_{0i} + 2v_{0i} ) \simeq \Omega^{2}_{i0},
\eqno (6.44) $$
$$ \varphi ( \omega_{i}) = -2 \lambda q_{i0} s_{0i} + 2D_{qq} s^{2}_{0i} +
2{D_{pp} \over \hbar^{2}} q^{2}_{i0} \ge 0, \eqno (6.45) $$
$$ F ( \omega_{i} t ) = f ( \omega_{i} t ) + \eta^{ \lambda} g^{ \lambda}
( \omega _{i} t) + \eta^{D} g^{D} ( \omega_{i} t) + \eta^{N} g^{N} ( \omega_{i}
t), \eqno (6.46) $$
the quantities
$$ \eta^{ \lambda} = - \lambda { C_{0i} \Omega_{0i} \over W^{2}_{i0}}
\simeq - \lambda { C_{i0} \over \Omega_{i0}} \ge 0 , $$
$$\eta^{D} = -2 \left( D_{qq} u_{00} + { D_{pp} \over \hbar^{2}} w_{00} \right)
{ C_{i0} \Omega_{0i} \over W^{2}_{i0}} \simeq
 - 2 \left( D_{qq} u_{00} + { D_{pp} \over \hbar^{2}} w_{00} \right)
{ C_{i0} \over \Omega_{i0}} \ge 0 , $$
$$ \eta^{N} = 2 { D_{pq} \over \hbar} \lambda { C_{i0} ( C_{0i} + 2 v_{0i} )
\over W^{2}_{i0}} \ge 0 \eqno (6.47) $$
being practically independent, positive parameters, describing the four
processes
of the interaction of the system with the environment: the friction, the
diffusion
of $q$, the diffusion of $p$ and respectively the non-equilibrium. The function
$ \varphi ( \omega_{i} ) $ is also positive due to the fundamental constraints.

{\bf 7. Open quantum systems and the atom-field interaction}

We consider a linearly polarized single mode electromagnetic
field $  \vec {E}$,
of the frequency $ \omega$ and the wavevector $ { \vec {k}}_{1} $
propagating through an absorbing medium of two level atoms with the transition
frequency $ \omega_{0} $, the electric dipole $ \mu $ and the density {\it N};
every atom {\it n} is described by the Pauli operators $\sigma^{n}_{x},
\sigma^{n}_{y} , \sigma^{n}_{z}$. Due to the interaction between the radiation
and the atoms, higher order harmonics $ {\vec {E}}_{\nu}$ of the frequency
$ \nu \omega$ and wavevector ${\vec{k}}_{\nu} $ are generated. For the electric
dipole interaction, the Hamiltonian of the system is of the form
$$ H = - {1 \over 2} \hbar\omega_{0} \sum_{n} \sigma^{n}_{z} + {1 \over 2} \sum
_\nu ( p^{2}_\nu + \nu^{2} \omega^{2} q^{2}_\nu )
 +\bar\mu \sum_{n \nu} \sigma^{n}_{x} ( \nu \omega q_\nu \sin {\vec {k}}
_\nu{\vec{r}}_{n}+ p_\nu \cos {\vec {k}}_\nu \vec {r}_{n} ),\eqno (7.1) $$
where $\bar{\mu}=\mu/\sqrt{\varepsilon_{0}V}$, $V$ is the volume of
quantization,
$ p_{\nu}, q_{ \nu} $ are the canonical variables of the harmonic $ \nu$ , and
$ {\vec {r}}_{n} $ the position vector of the atom $n$.
In this case, we consider the operator $V_{j}$ from the equation (3.1),
of the form [29]:
$$ V_{j}= \sum_{\nu} (a_{j \nu} p_{\nu} + b_{j \nu} q_{\nu}) + \sum_{n} (A_{jn}
\sigma^{n}_{x}+ B_{jn} \sigma^{n}_{y} + C_{jn} \sigma^{n}_{z} ). \eqno (7.2) $$
With the expressions (7.1) and (7.2) we obtain the master equation:
$$ {d \rho \over dt}=-{i\over \hbar} [H_{0} , \rho] - i{ \bar{\mu} \over \hbar}
\sum _{n \nu} [( \nu \omega q_{\nu} \sin \vec{ k}_{\nu} \vec{r}_n
+ p_{\nu} \cos \vec {k}_\nu \vec {r}_{n}) \sigma^{n}_{x}, \rho]  $$
$$+\sum_{\nu}\{i{\lambda_{\nu}\over 2 \hbar} \left([ p_{ \nu} , \rho q_{ \nu} +
q_{ \nu} \rho] - [ q_{ \nu} , \rho p_{ \nu} + p_{\nu} \rho] \right)
 - {D_{qq \nu} \over \hbar^{2}} [p_{ \nu} , [p_{ \nu} , \rho]] -
{D_{pp \nu} \over \hbar^{2}} [q_{ \nu} , [q_{\nu} , \rho]]  $$
$$ + {D_{\rho q \nu} \over \hbar^{2}}\left([q_{\nu},[p_{\nu},\rho]] + [p_{\nu}
, [q_{ \nu} , \rho]] \right) \} +\sum_{n}\{-\Lambda^{n}_{xy} [ \sigma^{n}_{z} ,
[ \sigma^{n}_{z} , \rho]]
 - \Lambda^{n}_{yz} [ \sigma^{n}_x, [ \sigma^{n}_{x} , \rho]]$$
$$ - \Lambda^{n}_{zx}
[ \sigma^{n}_{y} , [ \sigma^{n}_{y} , \rho]]
 + \Gamma^{n}_{xy} ([ \sigma^{n}_{x} , [ \sigma^{n}_{y} , \rho]] + [ \sigma^{n}
_{y} , [ \sigma^{n}_{x} , \rho]] )
 + \Gamma^{n}_{yz} ([ \sigma^{n}_{y} , [ \sigma^{n}_{z} , \rho]] + [ \sigma^{n}
_{z} , [ \sigma^{n}_{y} , \rho]] )  $$
$$ + \Gamma^{n}_{zx} ([ \sigma^{n}_{z} , [ \sigma^{n}_{x},\rho]] + [ \sigma^{n}
_{x} , [ \sigma^{n}_{z} , \rho]] )
 - {i \over 2} [ D^{n}_{x} ([ \sigma^{n}_{y} ,\sigma^{n}_{z} \rho + \rho \sigma
^{n}_{z} ] - [ \sigma^{n}_{z} , \sigma^{n}_{y} \rho + \rho \sigma^{n}_{y}
])  $$
$$ +D^{n}_{y} ([ \sigma^{n}_{z} , \sigma^{n}_{x} \rho + \rho \sigma
^{n}_{x} ] - [ \sigma^{n}_{x} , \sigma^{n}_{z} \rho + \rho \sigma^{n}_{z}
])
 +D^{n}_{z} ([ \sigma^{n}_{x} , \sigma^{n}_{y} \rho + \rho \sigma
^{n}_{y} ] - [ \sigma^{n}_{y} , \sigma^{n}_{x} \rho + \rho \sigma^{n}_{x}
]) ] \}, \eqno (7.3) $$
where $ H_{0} $ is the Hamiltonian of the system without interaction, and the
phenomenological parameters are defined by the expression:
$$ D_{qq \nu} = {\hbar \over 2} \sum_{j} a^{*}_{j \nu} a_{j \nu},~
D_{pp \nu} = {\hbar \over 2} \sum_{j} b^{*}_{j \nu} b_{j \nu},$$
$$D_{pq\nu} = - { \hbar \over 4} \sum_{j} ( a^{*}_{j \nu} b_{j \nu} + a_{j \nu}
b^{*}_{j \nu}),~
 \lambda_{\nu} = {i \over 2} \sum_{j} (a^{*}_{j \nu} b_{j \nu} - a_{j \nu}
b^{*}_{j \nu}), $$
$$ \Lambda^{n}_{xy} = {1 \over 2 \hbar} \sum_{j} C^{*}_{jn} C_{jn},~
\Lambda^{n}_{yz} = {1 \over 2 \hbar} \sum_{j} A^{*}_{jn} A_{jn},~
\Lambda ^{n}_{zx} = {1 \over 2 \hbar} \sum_{j} B^{*}_{jn} B_{jn}, $$
$$ \Gamma^{n}_{xy} = - {1 \over 4 \hbar} \sum_{j} (A^{*}_{jn} B_{jn} + A_{jn}
B^{*}_{jn}),~D^{n}_{z} = - {i \over 2 \hbar} \sum_{j} (A^{*}_{jn} B_{jn} -
A_{jn} B^{*}_{jn}), $$
$$ \Gamma^{n}_{yz} = - {1 \over 4 \hbar} \sum_{j} (B^{*}_{jn} C_{jn} + B_{jn}
C^{*}_{jn}),~D^{n}_{x} = - {i \over 2 \hbar} \sum_{j} (B^{*}_{jn} C_{jn} -
B_{jn} C^{*}_{jn}), $$
$$ \Gamma^{n}_{zx} = - {1 \over 4 \hbar} \sum_{j} (C^{*}_{jn} A_{jn} + C_{jn}
A^{*}_{jn}),~D^{n}_{y} = - {i \over 2 \hbar} \sum_{j} (C^{*}_{jn} A_{jn} -
C_{jn} A^{*}_{jn}). \eqno (7.4) $$
Separating the field from the atomic observables by the "mean-field
approximation",
for the atomic expectation values we obtain the equations
$$ {d \over dt} < \sigma_{x} > + \gamma^{'}_{\perp} < \sigma_{x} > -
(\omega_{0} - s) < \sigma_{y} > + \gamma_{1} < \sigma_{z} > - D_{1} = 0, $$
$$ {d \over dt} < \sigma_{y} > + (\omega_{0} - s) < \sigma_{x} >
+ \gamma^{''}_{\perp} <\sigma_{y} > + (\gamma_{2} - \bar {\chi})
<\sigma_{z} > - D_{2} = 0, $$
$$ {d \over dt}<\sigma_{z}>+\gamma_{1}<\sigma_{x} > + (\gamma_{2} + \bar{\chi})
+ \gamma_{\parallel} < \sigma_{z} > - D_{3} = 0, \eqno (7.5) $$
where
$$ \gamma^{'}_{\perp} = 4( \Lambda_{xy} + \Lambda_{zx} ) ,~~\gamma^{"}_{
\perp} = 4(\Lambda_{xy} + \Lambda_{yz} ),~~\gamma_{\parallel} = 4( \Lambda_{zx}
+ \Lambda_{yz} ), $$
$$ s = 4 \Gamma_{xy},~~\gamma_{1} = 4 \Gamma_{zx},~~\gamma_{2}= 4 \Gamma_{yz},~
 D_{1} = 4 D_{x},~~D_{2} = 4 D_{y},~~D_{3} = 4 D_{z} \eqno (7.6) $$
and
$$ \bar {\chi} = {2 \bar{\mu} \over \hbar}\sum_{\nu}{\rm Tr}[(\nu\omega q_{\nu}
+ p_{\nu} ) \rho] \eqno (7.7) $$
is a normalized field variable. From these equations one can obtain the
classical
Bloch-Feynman equations when $ s = \gamma_{1} = \gamma_{2} = 0,~~D_{1} = D_{2}
= 0,~~\gamma^{'}_{\perp} = \gamma^{"}_{\perp}$ [29].

Having in view the equation (3.3), in the master equation (7.3), we distinguish
two additional parts describing the interaction with the environment:~~one part
for the harmonic oscillators coresponding to the electromagnetic field and
another
part for the ensemble of atoms. This latter part describes three processes of
the
atom interaction with the environment: 1) the friction, described by the
parameters
$ \Lambda $ or $ \gamma^{'}_{\perp}, \gamma^{"}_{\perp}, \gamma_{\parallel} $ ;
2) the diffusion, leading to zero-point values of the observables, described
by
the parameters $D_i $; 3) the coupling between the observables described by the
parameters
$\Gamma$ or s, $ \gamma_{1}, \gamma_{2} $. These parameters satisfy fundamental
constraints, similar to the expressions (3.4) and (3.24):
$$D_{qq\nu} \ge 0 ,~~D_{pp \nu} \ge 0 ,~~D_{pp \nu} D_{qq \nu} - D^{2}_{pq \nu}
\ge { \hbar^{2} \lambda^{2}_{\nu} \over 4}, $$
$$ D_{qq\nu} \Delta_{pp\nu} (t) + D_{pp\nu} \Delta_{qq\nu}(t)-2D_{pq\nu}
\Delta_{pq\nu} (t)
\ge {h^{2} \lambda^{2}_{\nu} \over 4}, $$
$$\Lambda^{n}_{xy} \ge 0 ,~~\Lambda^{n}_{yz} \ge 0 ,~~\Lambda^{n}_{zx} \ge 0,$$
$$ \Lambda^{n}_{xy} \Lambda^{n}_{zx} - \Gamma^{n^{2}}_{yz} \ge {1 \over 4}
D^{n^{2}}_{x} ,~~\Lambda^{n}_{yz} \Lambda^{n}_{xy} - \Gamma^{n^{2}}_{zx} \ge
{1 \over 4} D^{n^{2}}_{y},~
\Lambda^{n}_{zx}\Lambda^{n}_{yz}- \Gamma^{n^{2}}_{xy} \ge {1 \over 4} D^{n^{2}}
_{z}, $$
$$\Lambda^{n}_{xy} \Delta^{n}_{zz} (t) + \Lambda^{n}_{zx} \Delta^{n}_{yy} (t)
- 2 \Gamma^{n}_{yz} \Delta^{n}_{yz} (t) \ge D^{n}_{x} < \sigma^{n}_{x} (t) >,$$
$$ \Lambda^{n}_{yz} \Delta^{n}_{xx} (t) + \Lambda^{n}_{xy} \Delta^{n}_{zz} (t)
- 2 \Gamma^{n}_{zx} \Delta^{n}_{zx} (t) \ge D^{n}_{y} < \sigma^{n}_{y} (t) >,$$
$$ \Lambda^{n}_{zx} \Delta^{n}_{yy} (t) + \Lambda^{n}_{yz} \Delta^{n}_{xx} (t)
- 2 \Gamma^{n}_{xy} \Delta^{n}_{xy} (t) \ge D^{n}_{z} < \sigma^{n}_{z} (t) >,
\eqno (7.8) $$
where $ \Delta_{pp}, \Delta_{qq}, \Delta_{xx}, \Delta_{yy}, \Delta_{zz} $ are
the variances of the observables and the quantities $ \Delta_{pq}, \Delta_{xy},
\Delta_{yz}, \Delta_{zx} $ are expressions of the form:
$$ \Delta_{pq} = < {pq + qp \over 2} > - <p> <q>. \eqno (7.9) $$
Following the model of the geometrical representation originally
adapted by Feynman, Vernon and Hellwarth [95], we consider the
atomic observables with the "amplitudes" $u,v,w,$ as components of
the Bloch vector in a "rotating frame" of the frequency $ \omega$ :
$$ < \sigma_{x} > = u \cos \omega t - v \sin \omega t,~
 < \sigma_{y} > = - u \sin \omega t - v \cos \omega t,~
 <\sigma_{z} > = - w. \eqno (7.10) $$
For a system of atoms with the density { \bf N}, using (7.10) one
obtains
the following expressions of the macroscopic polarization ${\sl S} = {\bf N}
< \sigma_{x}
> $ and of the population $ N = {\bf N } < \sigma_{z} > $:
$$ S = {1 \over 2} ( {\bf S} e^{-i \omega t} + {\bf S}^{*} e^{i \omega t} ),~
   N = - {\bf N} w, \eqno (7.11) $$
where
$$ {\bf S = N } (u - iv). \eqno (7.12) $$
With the expressions (7.10)-- (7.12), the equations (7.5) become:
$$ {d {\bf S} \over dt} + ( \gamma_{\perp} +  \Delta ) {\bf S} + \gamma_{a}
{\bf S}^{*} e^{2i \omega t} = (i \chi - \gamma e^{i \omega t} + i \chi^{*}
e^{2 i \omega t} ) N + D e^{i \omega t}, $$
$$ {d N \over dt} + \gamma_{\parallel} (N - N_{3} ) = {1 \over 2} \left[
( i \chi^{*} - \gamma^{*} e^{i \omega t} + i \chi e^{-2i \omega t} ) {\bf S} -
(i \chi - \gamma e^{i \omega t} + i \chi^{*} e^{2i \omega t} ) {\bf S}^{*}
\right], \eqno (7.13) $$
where
$$\gamma_{\perp} = ( \gamma^{'}_{\perp} + \gamma^{"}_{\perp}) / 2,~~ \gamma_{a}
= ( \gamma^{'}_{\perp} - \gamma^{"}_{\perp} ) / 2,~~ \Delta = \omega_{0} - s -
\omega, $$
$$ \gamma = \gamma_{1} + i \gamma_{2},~~ D = {\bf N} ( D_{1} + i D_{2} ),~~
N_{3} = {\bf N} D_{3} / \gamma_{\parallel}. \eqno (7. 14) $$
By neglecting the rapidly varying terms, one obtains the
classical optical Bloch equations in the "rotating wave approximation" :
$$ {d {\bf S} \over dt} + ( \gamma_{\perp} + i \Delta) {\bf S} = i \chi N, $$
$$ {dN \over dt } + \gamma_{\parallel} (N - N_{3} ) = {i \over 2} ( {\bf S}
\chi^{*} - {\bf S}^{*} \chi ). \eqno (7.15) $$
In order to include the effects of these terms we consider solutions of the
form:
$$ E = {\bf E}_{0} + {1 \over 2} ( {\bf E}_{1} e^{-i \omega t} + {\bf E}^{*}
_{1}e^{i \omega t} ) + {1 \over 2} ( {\bf E}_{2} e^{-2i \omega t} + {\bf E}^{*}
_{2} e^{2i \omega t}) + ..., $$
$$ S ={\bf S}_{0} + {1 \over 2} ( {\bf S}_{1} e^{-i \omega t} + {\bf S}^{*}_{1}
e^{i \omega t} ) + {1 \over 2}({\bf S}_{2} e^{-2i \omega t} + {\bf S}^{*}_{2}
e^{2i \omega t}) + ..., $$
$$ N = {\bf N}_{0} + {1 \over 2} ({\bf N}_{1} e^{-i \omega t} + {\bf N}^{*}_{1}
e^{i \omega t}) + {1 \over 2} ({\bf N}_{2} e^{-2i \omega t} + {\bf N}^{*}_{2}
e^{2i \omega t}) + ..., \eqno (7.16) $$
where $ {\bf E}_{0} , {\bf E}_{1}, {\bf E}_{2}, ... , {\bf S}_{0}, {\bf S}_{1},
{\bf S}_{2}, ... , {\bf N}_{0}, {\bf N}_{1}, {\bf N}_{2}, ... $
are slowly varying functions which are called "amplitudes". For the
amplitudes of zeroth and first order we get the following
equations:
$$ {d {\bf S}_{1} \over dt} + (\gamma^{'}_{\perp} + i \Delta) {\bf S}_{1} =
- \gamma_{1} {\bf N}_{1}, $$
$$ \gamma_{a} {\bf S}_{1} + { \gamma_{1} - i (\gamma_{2} - 2 \chi_{0}) \over 2}
{\bf N} + i \chi_{1} {\bf N}_{0} = 0, $$
$${d{\bf N}_{0}\over dt}+ \gamma_{\parallel} ({\bf N}_{0} - N_{3}) + \gamma_{1}
{\bf S}_{0}={1\over 2}( {\bf S}_{1} \chi^{*}_{1} - {\bf S}^{*}_{1}\chi_{1}), $$
$$ {d {\bf S}_{0} \over dt} + \gamma^{'}_{\perp} {\bf S}_{0} + \gamma_{1} ({\sl
N}_{0} - N_{1} ) = 0, \eqno (7.17) $$
where $ \chi_{0} =  \mu{\bf E}_{0}/\hbar,~~\chi_{1} = \mu
{\bf E}_{1}/\hbar,~~N_{1} = {\bf N} D_{1} / \gamma_{1} $ .

In these equations the new parameter $ \gamma_{1} \not= 0 $
leads to a coupling of the polarization variable ${\bf S}_{1}$
with the electric field variable $ \chi_{1,}$ i.e. to an interaction
of the atom with the electric field.
In this case one obtains the polarization equation
$$ {d {\bf S}_{1} \over dt}+ \left ( {\gamma^{"}_{\perp} - i \gamma^{'}_{\perp}
(\gamma_{2}-2\chi_{0})/\gamma_1\over 1-i(\gamma_{2}- 2 \chi_{0} ) / \gamma_{1}}
+ i \Delta \right) {\bf S}_{1} = {2 i \chi_{1} {\bf N}_{0}\over 1-i( \gamma_{2}
- 2 \chi_{0} ) / \gamma_{1} }, \eqno (7.18) $$
which can be compared with the conventional Bloch polarization equation (7.15).
We see that this equation describes new physical phenomena. For instance,
in
the steady state, with the notations
$$ \Gamma = {\gamma_{2} - 2 \chi_{0} \over \gamma_{1} },~~ \zeta = {\gamma^{'}
_{\perp} \over \gamma^{"}_{\perp}} \Gamma,~~ \gamma ^{'}_{\parallel} = \gamma_{
\parallel} (1-{\gamma^{2}_{1} \over \gamma^{'}_{\perp} \gamma_{\parallel}}) ,~~
\chi^{2}_{s} = {\gamma^{'}_{\parallel} \gamma^{"}_{\perp} \over 2}, $$
$$ N^{e} = {N_{3} - {\gamma_{1} N_{1} \over \gamma^{'}_{\perp}} \over 1 - {
\gamma^{2} \over \gamma^{'}_{\perp} \gamma_{\parallel}}} ,~~ \delta = {\Delta
\over \gamma^{"}_{\perp}},~~\varepsilon = {\chi_{1} \over \chi_{s}},
\eqno (7.19) $$
one obtains the polarization amplitude
$$ {\bf S}_{1}=2N^{e}{\chi_{s} \over \gamma^{"}_{\perp}} \cdot {[\delta - \zeta
+ i (1 + \Gamma \delta)] \varepsilon \over (1 + \Gamma \delta)^{2} + (\delta -
\zeta)^{2} + (1 + \Gamma \delta) \vert \varepsilon \vert^{2}}. \eqno (7.20) $$
For the slowly varying amplitudes $ \chi_{1+}$ and $ \chi_{1-} $ of the field
variables,
$$ \chi_{1} = \chi_{1+} e^{ikz} + \chi_{1-} e^{-ikz}, \eqno (7.21) $$
the Maxwell equations take the form [96]:
$$ {d \chi_{1+} \over dz} = i {\mu g \over c \hbar} {\bf S}_{1+}, $$
$$ {d \chi_{1-} \over dz} = - i {\mu g \over c \hbar} {\bf S}_{1-}, $$
$$ \chi_{0} =-{\mu^{2}\over \varepsilon_{0} \hbar} {\bf S}_{0}, \eqno (7.22) $$
where $ g = \omega \mu/ 2 \varepsilon_{0} $ is the coupling coefficients
and ${\bf S}_{1+}, {\bf S}_{1-}$ are the Fourier transforms of the polarization
$ {\bf S}_{1} $ as a function of the coordinate $z$:
$${\bf S}_{1+}={k\over 2\pi} {\int^{ \pi}_{-\pi} {\bf S}_{1} (z) e^{-ikz} dz},~
{\bf S}_{1-} = {k \over 2\pi}{\int^{ \pi}_{- \pi} {\bf S}_{1} (z)
e^{ikz} dz}. \eqno (7.23) $$ Eqs. (7.22) with the expression (7.20)
describe the propagation of an electromagnetic plane wave, taking
into account the self-reflection effect [97]. Neglecting the
counter-propagating wave due to the self-reflection, one obtains the
absorption coefficient
$$\alpha= - {1 \over \vert \varepsilon \vert } {d \vert \varepsilon \vert \over
dz} = {\alpha_{0} \over 2} { 1 + \Gamma \delta \over (1 + \Gamma \delta)^{2} +
(\delta - \zeta)^{2} + (1 + \Gamma \delta) \vert \varepsilon \vert^{2}}
\eqno (7.24) $$
and the coresponding dephasing
$${d\theta\over dz} = - {\delta - \zeta \over 1 + \Gamma \delta} \alpha, \eqno
(7.25) $$
where $ \alpha_{0} = 4 \mu g N^{e} / e \hbar \gamma^{\parallel}_{\perp} $ .

According to Eq. (7.24) the absorption coefficient $\alpha $ becomes
negative when the electric field intensity $\vert \varepsilon
\vert^{2} $ is sufficiently small for values of the atomic detuning
$\delta$ with $1+\Gamma \delta < 0.$ In that case the
electromagnetic wave is amplified, taking energy from the
environment. At the same time, the difference between the population
of the ground state and the population of the excited state,
$$ N_{0} = {N^{e} \over { 1 + { 1 + \Gamma \delta \over{(\delta - \zeta)^{2}
+ (1 + \Gamma \delta)^{2}}} {{\vert \chi_{1} \vert^{2} \over
\chi^{2}_{s}}}}} \eqno (7.26) $$ becomes larger than its equilibrium
value $ N^{e} $. When the ensemble of atoms has a smaller
temperature than the environment, the energy passes from the
environment to the atomic system and finally to the electromagnetic
field. The absorption coefficient asymmetry with the atomic detuning
$\delta$, can be observed in the experimental data of Sandle and
Gallagher [27,98]. The laser radiation absorption by Na atoms in a
buffer gas of Ar atoms has been studied by McCartan and Farr [99].
They varied the pressure of the buffer gas and measured the spectral
line width $w$ and the resonance frequency shift $ \Delta \omega_{0}
$. Using the photoluminescence method [100], we obtain the following
expression of the spectral line width
$$ W = 2 \pi \Delta \omega^{2} { \alpha \over \int \alpha d\omega},\eqno (7.27)
$$
while the resonance frequency shift becomes
$$ \Delta \omega_{0} = \gamma^{"}_{\perp} \xi. \eqno (7.28) $$
Integrating the expression (7.24) one obtains
$$ W = 2 \gamma^{"}_{\perp} (1 + \Gamma \xi). \eqno (7.29) $$
By taking into account (7.28), this expression becomes
$$ W = 2 \Delta \omega_{0} {1 + \Gamma \xi \over \xi} \simeq 2 \Delta \omega
_{0}. \eqno (7.30) $$ Consequently the spectral line width $W$ is
proportional to the line frequency shift $\Delta\omega_0$ which is
in agreement with the experimental results of McCartan and Farr
[99].

{\bf 8. Summary}

The Lindblad theory provides a selfconsistent treatment of damping as a
possible
extension of quantum mechanics to open systems. In the present review first
we studied the damped quantum oscillator by using the Schr\"odinger and
Heisenberg representations. According to this theory we have calculated the
damping of the expectation values of coordinate and momentum and the variances
as functions of time. The resulting time dependence of the expectation values
yields an exponential damping.
Then we have shown that the quasiprobability distributions can be used to solve
the problem of dissipation for the harmonic oscillator. From the master
equation of the damped quantum oscillator we have derived the corresponding
Fokker-Planck equations in the Glauber $P$, the antinormal ordering $Q$ and
the Wigner $W$ representations and have made a comparative study of these
quasiprobability distributions. The Fokker-Planck equations we
obtained describe an Ornstein-Uhlenbeck process. We have proven that the
variances found from
the Fokker-Planck equations in these representations are the same. We have
solved these equations in the steady state and showed that the Glauber $P$
function (when it exists), the $Q$ and the Wigner $W$ functions are
two-dimensional Gaussians with different widths.
We have also calculated the time evolution of the density matrix. For this
purpose we applied the method of the generating function of the density
matrix. In this case the density matrix can be obtained by taking partial
derivatives of
the generating  function. The generating function depends on a set of
time-dependent coefficients which are calculated as solutions of linear
differential equations of first order. Depending on the initial conditions
for these coefficients, the density matrix evolves differently in time.
When the asymptotic state is a Gibbs state in the case of a thermal bath,
a Bose-Einstein distribution results as density matrix. Also for the case
that the initial density matrix is chosen as a Glauber packet, a simple
analytical expression for the density matrix has been derived.
The density matrix can be used in various physical applications where a
bosonic degree of freedom moving in a harmonic oscillator potential is
damped. For example, one needs to determine nondiagonal transition elements
of the density matrix, for an oscillator perturbed by a weak electromagnetic
field in addition to its coupling to a heat bath. The density matrix can also
be derived from the solution of the Fokker-Planck equation for the coherent
state  representation.

The theory was applied to the damping of the charge equilibration
in deep inelastic collisions of heavy ions. The comparison of theoretical
results with experimental data shows that the overdamped solution succeeded
to describe these experimental data.

We have also shown the Hamiltonian of the
proton and neutron asymmetry degrees
of freedom in deep inelastic collisions as an example for two coupled and
damped oscillators. The usual limitation of the Lindblad theory is that
the damping time is long compared with the characteristic time of the
oscillators. This condition is not too well satisfied in deep inelastic
collisions of nuclei, where the time scale is of the order of the relaxation
time. Therefore, in these applications we consider the Lindblad theory as an
axiomatic procedure for describing the dissipation processes and accept its
parameters as free quantities, fitted to the
experimental data.

We obtained simple models for the dissipation of the angular momentum, by
using the generators of the Lorentz group as openess operators of the system.
For given values of the diffusion and friction coefficients we obtained an
exponential damping of the angular momentum, in agreement with the situation
encountered in heavy ion collisions. If the opening of the system fulfils
rotational symmetry and the Hamiltonian is a positive function of $L^2$ and
$N^2,$ (square of generators of Lorentz group), then the Lindblad master
equation describes a physical system coupled
to a reservoir of infinite temperature, which increases the energy and angular
momentum exponentially.

We calculated the tunneling spectrum as a function of some barrier
characteristics: the tunneling operator, the overlap integral, the transition
elements of the coordinate and momentum. Besides Gamow's tunneling process with
energy conservation, additional terms with energy transfer to or form
environment are obtained. For low values of
temperature, the whole spectrum is situated at energies smaller than the
$Q$-value; for higher enough values of temperature, transitions at energies
higher than the $Q$-value appear. Generally, we shown that dissipation
stimulates
the tunneling process and leads to an exotic decay law.

Using Lindblad's theory of open systems, for the resonant atom-field
interaction,
we found new optical equations. We showned that the interaction with the
environment
consists not only in a decay, but also in a coupling of the atomic observables.
This coupling has experimental evidence in the absobtion spectrum of the laser
radiation and in optical bistability.

At the same time, this phenomenon leads to a remarkable physical effect of
energy
transfer from the dissipative environment, which is cooling, to the coherent
electromagnetic field propagating through it.

Recently we assist to a revival of interest in quantum brownian
motion as a paradigm of quantum open systems. There are many
motivations. The possibility of preparing systems in macroscopic
quantum states led to the problems of dissipation in tunneling and
of loss of quantum coherence (decoherence). These problems are
intimately related to the issue of quantum-to-classical transition.
All of them point the necessity of a better understanding of open
quantum systems and all requires the extension of the model of
quantum brownian motion. Our results allow such extensions and also
explain some earlier observations. For a first comment of the result
(3.4),{\it iii)} see Sec. 4.2 of Ref. [101]. The fact that the
Schr\"odinger equation corresponding to the Lindblad equation is
nonlinear was obtained in Ref. [17] and was quoted in Ref. [102].
The approach used in Ref. [17] to generate the Dekker master
equation was applied in Ref. [93] to generate a family of master
equations for local quantum dissipation. New developments of this
kind were also obtained in Ref. [103].

{\bf References}

\item{1.}
R. W. Hasse, J. Math. Phys. {\bf 16} (1975) 2005

\item{2.}
E. B. Davies, Quantum Theory of Open Systems, Academic Press, New York, 1976

\item{3.}
J. Messer, Acta Phys. Austriaca {\bf 58} (1979) 75

\item{4.}
H. Spohn, Rev. Mod. Phys. {\bf 52} (1980) 569

\item{5.}
H. Dekker, Phys. Rep. {\bf 80} (1981) 1

\item{6.}
P. Exner, Open Quantum Systems and Feynman Integrals, Reidel, Dordrecht, 1985

\item{7.}
K. H. Li, Phys. Rep. {\bf 134} (1986) 1

\item{8.}
R.Haake, Springer Tracts in Mod. Phys. {\bf 66} (1973) 98

\item{9.}
R. S. Ingarden and A. Kossakowski, Ann. Phys. (N.Y.) {\bf 89} (1975) 451

\item{10.}
V. Gorini and A. Kossakovski, J. Math. Phys. {\bf 17} (1976) 1298

\item{11.}
A. Kossakowski, Rep. Math. Phys. {\bf 3} (1972) 247

\item{12.}
A. Kossakowski, Bull. Acad. Polon. Sci. Math. Astron. Phys. {\bf 20} (1972)
1021

\item{13.}
R. S. Ingarden, Acta Phys. Polonica A {\bf 43} (1973) 1

\item{14.}
G. Lindblad, Commun. Math. Phys. {\bf 48} (1976) 119

\item{15.}
V. Gorini, A. Frigerio, M. Verri, A. Kossakowski and E. C. G. Sudarshan,
Rep. Math. Phys. {\bf 13} (1978) 149

\item{16.}
G. Lindblad, Rep. Math. Phys. {\bf 10} (1976) 393

\item{17.}
A. Sandulescu and H. Scutaru, Ann. Phys. (N.Y.) {\bf 173} (1987) 277

\item{18.}
A. Isar, A. Sandulescu and W. Scheid, J. Phys. G - Nucl. Part. Phys. {\bf 17}
(1991) 385

\item{19.}
A. Isar, W. Scheid and A. Sandulescu, J. Math. Phys. {\bf 32} (1991) 2128

\item{20.}
A. Isar, A. Sandulescu and W. Scheid, J. Math. Phys. {\bf 34} (1993) 3887

\item{21.}
S. Jang, Nucl. Phys. A {\bf 499} (1989) 250

\item{22.}
A. Pop, A. Sandulescu, H. Scutaru and W. Greiner, Z. Phys. A {\bf 329} (1988)
357

\item{23.}
A. Sandulescu, H. Scutaru and W. Scheid, J. Phys. A - Math. Gen. {\bf 20}
(1987) 2121

\item{24.}
A. Isar, A. Sandulescu, H. Scutaru and W. Scheid, Nuovo Cimento A {\bf 103}
(1990) 413

\item{25.}
A. Isar, A. Sandulescu and W. Scheid, Int. J. Mod. Phys. A {\bf 5}
(1990) 1773

\item{26.}
E. Stefanescu, A. Sandulescu and W. Greiner, Int. J. Mod. Phys. E {\bf 2}
(1993) 233

\item{27.}
M. Razavi and A. Pimpale, Phys. Rep. {\bf 168} (1988) 305

\item{28.}
H. Dekker, Phys. Rev. A {\bf 38} (1988) 6351

\item{29.}
A. Sandulescu and E. Stefanescu, Physica A {\bf 161} (1989) 525

\item{30.}
G. C. Emch, Algebraic Methods in Statistical Mechanics and Quantum
Field Theory, Wiley, New York, 1972

\item{31.}
V. Gorini, A. Kossakowski and E. C. G. Sudarshan, J. Math. Phys. {\bf 17}
(1976) 821

\item{32.}
P. Talkner, Ann. Phys. (N.Y.) {\bf 167} (1986) 390

\item{33.}
H. Dekker, Phys. Rev. A {\bf 16} (1977) 2126

\item{34.}
H. Dekker, Phys. Lett. A {\bf 74} (1979) 15

\item{35.}
H. Dekker, Phys. Lett. A {\bf 80} (1980) 369

\item{36.}
H. Dekker and M. C. Valsakumar, Phys. Lett. A {\bf 104} (1984) 67

\item{37.}
E. Merzbacher, Quantum Mechanics, Wiley, New York, 1970

\item{38.}
H. Hofmann and P. J. Siemens, Nucl. Phys. A {\bf 275} (1977) 464

\item{39.}
H. Hofmann, C. Gr\'egoire, R. Lucas and C. Ng\^o, Z. Phys. A  {\bf 293} (1979)
229

\item{40.}
R. W. Hasse, Nucl. Phys. A {\bf 318} (1979) 480

\item{41.}
E. M. Spina and H. A. Weidenm\"uller, Nucl. Phys. A {\bf 425} (1984) 354

\item{42.}
C. W. Gardiner and M. J. Collet, Phys. Rev. A {\bf 31} (1985) 3761

\item{43.}
T. A. B. Kennedy and D. F. Walls, Phys. Rev. A {\bf 37} (1988) 152

\item{44.}
G. S. Agarwal, Phys. Rev. {\bf 178} (1969) 2025

\item{45.}
G. S. Agarwal, Phys. Rev. A {\bf 4} (1971) 739

\item{46.}
S. Dattagupta, Phys. Rev. A {\bf 30} (1984) 1525

\item{47.}
C. M. Savage and D. F. Walls, Phys. Rev. A {\bf 32} (1985) 2316

\item{48.}
N. Lu, S. Y. Zhu and G. S. Agarwal, Phys. Rev. A {\bf 40} (1989) 258

\item{49.}
S. Jang and C. Yannouleas, Nucl. Phys. A {\bf 460} (1986) 201

\item{50.}
S.Chaturvedi, P.D.Drummond and D.F.Walls, J. Phys. A - Math. Gen. {\bf 10}
(1977) L187

\item{51.}
P. D. Drummond and C. W. Gardiner, J. Phys. A - Math. Gen. {\bf 13} (1980) 2353

\item{52.}
P. D. Drummond, C. W. Gardiner and D. F. Walls, Phys. Rev. A {\bf 24} (1981)
914

\item{53.}
M. Hillery, R. F. O 'Connell, M. O. Scully and E. P. Wigner, Phys. Rep.
{\bf 106} (1984) 121

\item{54.}
E. P. Wigner, Phys. Rev. {\bf 40} (1932) 749

\item{55.}
E. J. Glauber, Phys. Rev. {\bf 131} (1963) 2766

\item{56.}
E. J. Glauber, Phys. Rev. Lett. {\bf 10} (1963) 84

\item{57.}
E. C. G. Sudarshan, Phys. Rev. Lett. {\bf 10} (1963) 277

\item{58.}
W. Weidlich, H. Risken and H. Haken, Z. Phys. {\bf 204} (1967) 223

\item{59.}
M. Lax and W. H. Louisell, IEEE J. Q. Electron. {\bf 3} (1967) 47

\item{60.}
K. E. Cahill and R. J. Glauber, Phys. Rev. A {\bf 177} (1969) 1882

\item{61.}
R. J. Glauber, in Laser Handbook, ed. by F. T. Arecchi and E. O.
Schultz-Dubois, North-Holland, Amsterdam, 1972

\item{62.}
J. R. Klauder and E. C. G. Sudarshan, Fundamentals of Quantum Optics, Benjamin,
New York, 1968

\item{63.}
K. E. Cahill, Phys. Rev. {\bf 180} (1969) 1239

\item{64.}
K. E. Cahill, Phys. Rev. {\bf 180} (1969) 1244

\item{65.}
W. H. Louisell, Quantum Statistical Properties of Radiation, Wiley, New York,
1973

\item{66.}
H. Haken, Rev. Mod. Phys. {\bf 47} (1975) 67

\item{67.}
C. W. Gardiner, Handbook of Stochastic Methods, Springer, Berlin, 1982

\item{68.}
G. E. Uhlenbeck and L. S. Ornstein, Phys. Rev. {\bf 36} (1930) 823

\item{69.}
M. C. Wang and G. E. Uhlenbeck, Rev. Mod. Phys. {\bf 17} (1945) 323

\item{70.}
H. Weyl, The Theory of Groups and Quantum Mechanics, Dover, New York, 1931

\item{71.}
H. Freiesleben and J. B. Kratz, Phys. Rep. {\bf 106} (1984) 1

\item{72.}
W.U.Schr\"oder and J.R.Huizenga, Treatise on Heavy-Ion Science, vol.2,
Plenum, New York, 1984

\item{73.}
J.A.Maruhn, W.Greiner and W.Scheid, Heavy Ion Collisions, vol.2,
North-Holland, Amsterdam, l980

\item{74.}
A. Sandulescu, M. Petrovici, A. Pop, M. S. Popa, T. Hahn, K. H.
Ziegenhain and W. Greiner, J. Phys. G. - Nucl. Part. Phys. {\bf 7} (1981) L55

\item{75.}
E. S. Hernandez, W. D. Myers, J. Randrup and B. Remaud, Nucl. Phys. A
{\bf 361} (1981) 483

\item{76.}
H. Breuer, A. C. Mignerey, V. E. Viola, K. L. Wolf, J. R. Birkelund, D.
Hilscher, J. R. Huizenga, W. V. Schr\"oder and W. W. Wilde, Phys. Rev. C
{\bf 28} (1983) 1080

\item{77.}
G. Wolschin, Nukleonika {\bf 22} (1977) 1165

\item{78.}
G. Wolschin, Fizika {\bf 9} Suppl. 4 (1977) 415

\item{79.}
C. Riedel, G. Wolschin and W. N\"orenberg, Z. Phys. A {\bf 290} (1979) 47

\item{80.}
R. K. Gupta, M. M\"unchow, A. Sandulescu and W. Scheid, J. Phys. G. {\bf 10}
(1984) 209

\item{81.}
D. H. E. Gross and K. H. Hartmann, Phys. Rev. C {\bf 24} (1981) 2526

\item{82.}
W. V. Schr\"oder, J. R. Huizenga and J. Randrup, Phys. Lett. B {\bf 98} (1981)
355

\item{83.}
J. R. Birkelund, H. Freiesleben, J. R. Huizehga, W. V. Schr\"oder, W. W.
Wilcke, K. L. Wolf, J. P. Unik and V. E. Viola, Phys. Rev. C {\bf 26} (1982)
1984

\item{84.}
A. C. Merchant and W. N\"orenberg, Z. Phys. A {\bf 308} (1982) 315

\item{85.}
V. V. Dodonov and O. V. Manko, Physica A {\bf 130} (1985) 353

\item{86.}
A. A. Belavin, B. Ya. Zel'dovich, A. M. Perelomov and V. S. Popov, Z. Eksp.
Teor. Fiz. {\bf 56} (1969) 264

\item{87.}
Wu-Ki Tung, Group Theory in Physics,  World Scientific, Singapore, 1985

\item{88.}
A. Isar, A. Sandulescu and W. Scheid, Rev. Roum. Phys. {\bf 35} (1990) 13

\item{89.}
G. Gamow, Z. Phys. {\bf 51} (1928) 204

\item{90.}
A. O. Caldeira and A. J. Leggett, Ann. Phys. (N.Y.) {\bf 149} (1983) 374

\item{91.}
K. Fujikawa et al., Phys. Rev. Lett. {\bf 68} (1992) 1093

\item{92.}
E. G. Harris, Phys. Rev. A {\bf 48} (1993) 995

\item{93.}
M. R. Gallis, Phys. Rev. A {\bf 48} (1993) 1028

\item{94.}
D. M. Brink, private comm.

\item{95.}
R. P. Feynman, F. L. Vernon Jr. and R. W. Hellwarth, J. Appl. Phys. {\bf 28}
(1957) 49

\item{96.}
H. J. Carmichael, Optica Acta {\bf 27} (1980) 147

\item{97.}
L. Roso-Franco, J. Opt. Soc. Amer. {\bf 4} (1987) 1878

\item{98.}
W. J. Sandle and A. Gallagher, Phys. Rev. A {\bf 24} (1981) 2017

\item{99.}
D. G. McCartan and J. M. Farr, J. Phys. B {\bf 9} (1976) 985

\item{100.}
R. H. Chatham, A. Gallagher and E. L. Lewis, J. Phys. B {\bf 13} (1980) L7

\item{101.}
V. V. Dodonov, V. I. Man'ko, O. V. Man'ko, Quantum nonstationary
oscillator, Trudy FIAN, vol. 191, Nauka, Moskow, 1989

\item{102.}
A. Iannussis and R. Mignani, Physica A {\bf 152} (1988) 469

\item{103.}
N. V. Antonenko, S. P. Ivanova, R. V. Jolos and W. Scheid, J. Phys.
G: Nucl. Part. Phys. {\bf 20} (1994) 1447

\vfill
\eject
\bye